%% file: main.tex
\documentclass[aps,preprint,showpacs,superscriptaddress,letterpaper,showkeys,superscriptaddress,nofootinbib,floatfix]{revtex4-1} 

\usepackage{graphicx}  
\usepackage{dcolumn}   
\usepackage{bm}        
\usepackage{adjustbox}
\usepackage{amssymb}   
\usepackage[caption=false]{subfig}
\usepackage{xcolor}
\usepackage{subfig}
\usepackage{amsmath}
\usepackage{subfiles}
\usepackage{placeins}
\usepackage{array}
\usepackage[normalem]{ulem}

\newcommand{\minerva}{MINERvA}

\newcommand{\Wexp}{\ensuremath{W_{\rm\scriptstyle exp}}}
\newcommand{\Ehad}{\ensuremath{E_{\rm\scriptstyle had}}}
\newcommand{\gevsqcsq}{GeV$^2/c^2$}
\newcommand{\gevcsq}{GeV$/c^2$}

\newcommand{\pagePmu}{2}
\newcommand{\pagePmuZ}{8}
\newcommand{\pagePmuT}{7}
\newcommand{\pageThmu}{1}
\newcommand{\pageQ}{3}
\newcommand{\pageW}{4}
\newcommand{\pageTpi}{6}
\newcommand{\pageThpi}{5}

\newcommand{\oneplot}{0.45\textwidth}

\newcommand{\kevinComment}[1]{{\noindent \bf\color{purple} \small [KSM: #1]}}

\newcommand{\aaronComment}[1]{{\noindent \bf\color{orange} \small [AMB: #1]}}

\newcommand{\lauraComment}[1]{{\noindent \bf\color{teal} \small [LJF: #1]}}

\newcommand{\firstReviewAdd}[1]{{\bf\color{blue} \protect #1}}
\newcommand{\firstReviewStrike}[1]{{\bf\color{blue}\protect \sout{#1}}}
\newcommand{\firstReviewComment}[1]{{\noindent \bf\color{blue} \small [DAH: #1]}}

\newcommand{\secondReviewAdd}[1]{{\bf\color{purple} \protect #1}}
\newcommand{\secondReviewStrike}[1]{{\bf\color{purple}\protect \sout{#1}}}
\newcommand{\secondReviewComment}[1]{{\noindent \bf\color{purple} \small [DAH: #1]}}

\newcommand{\firstRefereeAdd}[1]{{\bf\color{green} \protect #1}}
\newcommand{\firstRefereeStrike}[1]{{\bf\color{green}\protect \sout{#1}}}

\newcommand{\paperComment}[1]{{\noindent \bf\color{red}\small [Committee: #1]}}
\newcommand{\inProgress}[1]{{\noindent \bf\color{red}\small [Work to complete: #1]}}

\newcommand{\showSections}[0]{0}

\renewcommand{\kevinComment}[1]{}
\renewcommand{\lauraComment}[1]{}
\renewcommand{\aaronComment}[1]{}
\renewcommand{\paperComment}[1]{}
\renewcommand{\firstReviewComment}[1]{}
\renewcommand{\secondReviewComment}[1]{}

\renewcommand{\firstReviewAdd}[1]{\protect #1}
\renewcommand{\firstReviewStrike}[1]{}
\renewcommand{\inProgress}[1]{}
\renewcommand{\secondReviewAdd}[1]{\protect #1}
\renewcommand{\secondReviewStrike}[1]{}
\renewcommand{\firstRefereeAdd}[1]{\protect #1}
\renewcommand{\firstRefereeStrike}[1]{}

\newcommand{\supp}{the supplemental material}

\newif\ifPRLandSupp
\PRLandSupptrue

\newif\ifColorFigures
\ColorFigurestrue

\makeatletter
\renewcommand{\p@subsection}{}
\makeatother

\begin{document}





\title{Simultaneous measurement of 
\firstReviewStrike{muon neutrino}
\firstReviewAdd{$\nu_\mu$} charged-current 
\firstReviewAdd{ single} $\pi^+$ production in 
\firstReviewStrike{carbon, water, iron, lead, and plastic}
\firstReviewAdd{CH, C, $\mathbf{H_2O}$, Fe, and Pb} targets
\firstReviewAdd{in MINERvA}}
\newcommand{\bkgdtype}{_MENU1PI}


\newcommand{\Rutgers}{Rutgers, The State University of New Jersey, Piscataway, New Jersey 08854, USA}
\newcommand{\Hampton}{Hampton University, Dept. of Physics, Hampton, VA 23668, USA}
\newcommand{\Dortmund}{Institute of Physics, Dortmund University, 44221, Germany }
\newcommand{\Otterbein}{Department of Physics, Otterbein University, 1 South Grove Street, Westerville, OH, 43081 USA}
\newcommand{\JMU}{James Madison University, Harrisonburg, Virginia 22807, USA}
\newcommand{\Florida}{University of Florida, Department of Physics, Gainesville, FL 32611}
\newcommand{\UCIrvine}{Department of Physics and Astronomy, University of California, Irvine, Irvine, California 92697-4575, USA}
\newcommand{\CBPF}{Centro Brasileiro de Pesquisas F\'{i}sicas, Rua Dr. Xavier Sigaud 150, Urca, Rio de Janeiro, Rio de Janeiro, 22290-180, Brazil}
\newcommand{\PUCP}{Secci\'{o}n F\'{i}sica, Departamento de Ciencias, Pontificia Universidad Cat\'{o}lica del Per\'{u}, Apartado 1761, Lima, Per\'{u}}
\newcommand{\INRM}{Institute for Nuclear Research of the Russian Academy of Sciences, 117312 Moscow, Russia}
\newcommand{\Jlab}{Jefferson Lab, 12000 Jefferson Avenue, Newport News, VA 23606, USA}
\newcommand{\Pittsburgh}{Department of Physics and Astronomy, University of Pittsburgh, Pittsburgh, Pennsylvania 15260, USA}
\newcommand{\Guanajuato}{Campus Le\'{o}n y Campus Guanajuato, Universidad de Guanajuato, Lascurain de Retana No. 5, Colonia Centro, Guanajuato 36000, Guanajuato M\'{e}xico.}
\newcommand{\Athens}{Department of Physics, University of Athens, GR-15771 Athens, Greece}
\newcommand{\Tufts}{Physics Department, Tufts University, Medford, Massachusetts 02155, USA}
\newcommand{\WM}{Department of Physics, William \& Mary, Williamsburg, Virginia 23187, USA}
\newcommand{\FNAL}{Fermi National Accelerator Laboratory, Batavia, Illinois 60510, USA}
\newcommand{\Purdue}{Department of Chemistry and Physics, Purdue University Calumet, Hammond, Indiana 46323, USA}
\newcommand{\MCLA}{Massachusetts College of Liberal Arts, 375 Church Street, North Adams, MA 01247}
\newcommand{\UMD}{Department of Physics, University of Minnesota -- Duluth, Duluth, Minnesota 55812, USA}
\newcommand{\Northwestern}{Northwestern University, Evanston, Illinois 60208}
\newcommand{\UNI}{Facultad de Ciencias, Universidad Nacional de Ingenier\'{i}a, Apartado 31139, Lima, Per\'{u}}
\newcommand{\Rochester}{Department of Physics and Astronomy, University of Rochester, Rochester, New York 14627 USA}
\newcommand{\Austin}{Department of Physics, University of Texas, 1 University Station, Austin, Texas 78712, USA}
\newcommand{\USM}{Departamento de F\'{i}sica, Universidad T\'{e}cnica Federico Santa Mar\'{i}a, Avenida Espa\~{n}a 1680 Casilla 110-V, Valpara\'{i}so, Chile}
\newcommand{\Geneva}{University of Geneva, 1211 Geneva 4, Switzerland}
\newcommand{\Chicago}{Enrico Fermi Institute, University of Chicago, Chicago, IL 60637 USA}
\newcommand{\hired}{}
\newcommand{\OregonState}{Department of Physics, Oregon State University, Corvallis, Oregon 97331, USA}
\newcommand{\oxford}{Oxford University, Department of Physics, Oxford, OX1 3PJ United Kingdom}
\newcommand{\umiss}{University of Mississippi, Oxford, Mississippi 38677, USA}
\newcommand{\upenn}{Department of Physics and Astronomy, University of Pennsylvania, Philadelphia, PA 19104}
\newcommand{\AMU}{AMU Campus, Aligarh, Uttar Pradesh 202001, India}
\newcommand{\wroclaw}{University of Wroclaw, plac Uniwersytecki 1, 50-137 Wroa\l{}aw, Poland}
\newcommand{\Mohali}{Department of Physical Sciences, IISER Mohali, Knowledge City, SAS Nagar, Mohali - 140306, Punjab, India}
\newcommand{\CINVESTAV}{Departamento de Fisica Col. San Pedro Zacatenco, 07360 Mexico, DF, Av. Instituto PolitÃ©cnico Nacional, Mexico}
\newcommand{\york}{York University, Department of Physics and Astronomy, Toronto, Ontario, M3J 1P3 Canada}
\newcommand{\ND}{Department of Physics, University of Notre Dame, Notre Dame, Indiana 46556, USA}
\newcommand{\ICL}{The Blackett Laboratory,  Imperial College London,  London SW7 2BW, United Kingdom}
\newcommand{\warwick}{Coventry CV4 7AL, UK}

\
\newcommand{\mascencioThanks}{Now at Iowa State University, Ames, IA 50011, USA}
\newcommand{\ricfregianThanks}{now at Department of Physics and Astronomy, University of California at Davis, Davis, CA 95616, USA}
\newcommand{\finerThanks}{Now at Los Alamos National Laboratory, Los Alamos, New Mexico 87545, USA}
\newcommand{\kleykampThanks}{now at Department of Physics and Astronomy, University of Mississippi, Oxford, MS 38677}
\newcommand{\bamThanks}{Now at University of Minnesota, Minneapolis, Minnesota 55455, USA}
\newcommand{\byaeggyThanks}{Now at Department of Physics, University of Cincinnati,  Cincinnati, Ohio 45221, USA}

%
%
\author{A.~Bercellie}                     \affiliation{\Rochester}
\author{K.A.~Kroma-Wiley}                 \affiliation{\upenn}  \affiliation{\Rochester}
\author{S.~Akhter}                        \affiliation{\AMU}
\author{Z.~Ahmad~Dar}                    \affiliation{\WM}  \affiliation{\AMU}
\author{F.~Akbar}                         \affiliation{\AMU}
\author{V.~Ansari}                        \affiliation{\AMU}
\author{M.~V.~Ascencio}\thanks{\mascencioThanks}  \affiliation{\PUCP}
\author{M.~Sajjad~Athar}                  \affiliation{\AMU}
\author{L.~Bellantoni}                    \affiliation{\FNAL}
\author{M.~Betancourt}                    \affiliation{\FNAL}
\author{A.~Bodek}                         \affiliation{\Rochester}
\author{J.~L.~Bonilla}                    \affiliation{\Guanajuato}
\author{A.~Bravar}                        \affiliation{\Geneva}
\author{H.~Budd}                          \affiliation{\Rochester}
\author{G.~Caceres}\thanks{\ricfregianThanks}  \affiliation{\CBPF}
\author{T.~Cai}                           \affiliation{\Rochester}
\author{G.A.~D\'{i}az~}                   \affiliation{\Rochester}
\author{H.~da~Motta}                      \affiliation{\CBPF}
\author{S.A.~Dytman}                      \affiliation{\Pittsburgh}
\author{J.~Felix}                         \affiliation{\Guanajuato}
\author{L.~Fields}                        \affiliation{\ND}
\author{A.~Filkins}                       \affiliation{\WM}
\author{R.~Fine}\thanks{\finerThanks}     \affiliation{\Rochester}
\author{A.M.~Gago}                        \affiliation{\PUCP}
\author{H.~Gallagher}                     \affiliation{\Tufts}
\author{P.K.Gaur}                         \affiliation{\AMU}
\author{A.~Ghosh}                         \affiliation{\USM}  \affiliation{\CBPF}
\author{S.M.~Gilligan}                    \affiliation{\OregonState}
\author{R.~Gran}                          \affiliation{\UMD}
\author{E.Granados}                       \affiliation{\Guanajuato}
\author{D.A.~Harris}                      \affiliation{\york}  \affiliation{\FNAL}
\author{D.~Jena}                          \affiliation{\FNAL}
\author{S.~Jena}                          \affiliation{\Mohali}
\author{J.~Kleykamp}\thanks{\kleykampThanks}  \affiliation{\Rochester}
\author{A.~Klustov\'{a}}                  \affiliation{\ICL}
\author{M.~Kordosky}                      \affiliation{\WM}
\author{D.~Last}                          \affiliation{\upenn}
\author{T.~Le}                            \affiliation{\Tufts}  \affiliation{\Rutgers}
\author{A.~Lozano}                        \affiliation{\CBPF}
\author{X.-G.~Lu}                         \affiliation{\warwick}  \affiliation{\oxford}
\author{I.~Mahbub}                        \affiliation{\UMD}
\author{E.~Maher}
\affiliation{\MCLA}
\author{S.~Manly}                         \affiliation{\Rochester}
\author{W.A.~Mann}                        \affiliation{\Tufts}
\author{C.~Mauger}                        \affiliation{\upenn}
\author{K.S.~McFarland}                   \affiliation{\Rochester}
\author{B.~Messerly}\thanks{\bamThanks}   \affiliation{\Pittsburgh}
\author{J.~Miller}                        \affiliation{\USM}
\author{O.~Moreno}                        \affiliation{\WM}  \affiliation{\Guanajuato}
\author{J.G.~Morf\'{i}n}                  \affiliation{\FNAL}
\author{D.~Naples}                        \affiliation{\Pittsburgh}
\author{J.K.~Nelson}                      \affiliation{\WM}
\author{C.~Nguyen}                        \affiliation{\Florida}
\author{A.~Olivier}                       \affiliation{\Rochester}
\author{V.~Paolone}                       \affiliation{\Pittsburgh}
\author{G.N.~Perdue}                      \affiliation{\FNAL}  \affiliation{\Rochester}
\author{K.-J.~Plows}                      \affiliation{\oxford}
\author{M.A.~Ram\'{i}rez}                 \affiliation{\upenn}  \affiliation{\Guanajuato}
\author{R.D.~Ransome}                     \affiliation{\Rutgers}
\author{H.~Ray}                           \affiliation{\Florida}
\author{D.~Ruterbories}                   \affiliation{\Rochester}
\author{H.~Schellman}                     \affiliation{\OregonState}
\author{C.J.~Solano~Salinas}              \affiliation{\UNI}
\author{H.~Su}                            \affiliation{\Pittsburgh}
\author{M.~Sultana}                       \affiliation{\Rochester}
\author{V.S.~Syrotenko}                   \affiliation{\Tufts}
\author{B.~Utt}                           \affiliation{\UMD}
\author{E.~Valencia}                      \affiliation{\WM}  \affiliation{\Guanajuato}
\author{N.H.~Vaughan}                     \affiliation{\OregonState}
\author{A.V.~Waldron}                     \affiliation{\ICL}
\author{B.~Yaeggy}\thanks{\byaeggyThanks}  \affiliation{\USM}
\author{L.~Zazueta}                       \affiliation{\WM}

\collaboration{The MINERvA Collaboration}\ \noaffiliation

\date{\today}
\begin{abstract}
\firstReviewStrike{Single charged pion production is a common interaction at accelerator-based neutrino oscillation experiments.  The MINERvA experiment has measured semi-inclusive differential cross sections $\nu_\mu A\to \mu^- \pi^+ X$ on targets of scintillator, carbon, water, iron, and lead in a neutrino beam with $\left< E_\nu\right> \approx 6$~GeV.  We observe a strong suppression at low $Q^2$ and an enhancement at low $T_\pi$ relative to models in both the light and heavy nuclear targets.  This result has much higher statistics than previous measurements and includes the first comparison of $\pi^+$ production from different nuclei in a single measurement.}
\firstReviewAdd{
Neutrino-induced charged-current single $\pi^+$ production in the $\Delta(1232)$ resonance region is of considerable interest to accelerator-based neutrino oscillation experiments.  In this work, high statistics differential cross sections are reported for the semi-exclusive reaction 
$\nu_\mu A \to \mu^- \pi^+ +$ nucleon(s) on scintillator, carbon, water, iron, and lead targets recorded by MINERvA using a wide-band $\nu_\mu$  beam with $\left< E_\nu \right> \approx 6$~GeV.   Suppression of the cross section at low $Q^2$ and enhancement of low $T_\pi$  are observed in both light and heavy nuclear targets compared to phenomenological models used in current neutrino interaction generators.  The cross-sections per nucleon for iron and lead compared to CH across the kinematic variables probed are 0.8 and 0.5 respectively, a scaling which is also not predicted by current generators. 
}
\end{abstract}

\maketitle

\if\showSections1 \kevinComment{Will remove section headings with \textbackslash showSections switch above.  As of first version to the committee the paper looks to be about 1/3 or 1/2 of a column too long.} \fi 

\kevinComment{For Aaron to add to supplemental material: fluxes in all targets}

\if\showSections1 \section{Motivation} \fi

Charged pion production is an important process for accelerator-based oscillation experiments.  For low energy experiments, T2K~\cite{T2K:2011qtm},  T2HK~\cite{Hyper-Kamiokande:2018ofw}, and SBN~\cite{MicroBooNE:2015bmn}, this process is both a significant background to quasielastic scattering and is used as a signal reaction when the pion is identified.  For higher energy experiments, NOvA~\cite{NOvA:2004blv} and DUNE~\cite{DUNE:2015lol}, the process is a large fraction of the signal\firstRefereeStrike{ reactions}.  \firstReviewStrike{Neutrino energy reconstruction in these experiments relies on relating the visible energy from the $\pi^\pm$ in the detector to its larger total energy.}
\firstReviewAdd{ Neutrino oscillation experiments rely on an accurate model of pion production} \firstReviewStrike{is one of the necessary ingredients }
to evaluate neutrino energy for events with charged pions. 
\firstReviewAdd{ Although NOvA, T2K, and T2HK utilize relatively low-A nuclear media such as hydrocarbons or water, SBN and DUNE experiments rely on argon (A = 40) targets.   In order for these 
\firstRefereeStrike{latter} experiments to make use of the wealth of neutrino interaction data obtained with light nuclei, a knowledge of neutrino cross-section scaling as a function of target A is required. This work \firstRefereeStrike{sheds first light on}\firstRefereeAdd{tests} scaling behavior by simultaneously \firstRefereeStrike{mapping}\firstRefereeAdd{measuring} the differential cross sections of $\nu_\mu + A \to \mu^- + \pi^+ +$ nucleons as the target nucleon number is changed from carbon (A=12) to lead (A= 208).   We also present \firstReviewAdd{cross section} ratios \firstRefereeStrike{of the iron and lead targets} to scintillator, where many systematic uncertainties cancel, allowing a precise measurement of the nuclear dependence.  }


Previous measurements of charged-current pion production~\cite{AguilarArevalo:2010bm,Adamson:2014pgc, Eberly:2014mra,Aliaga:2015wva,McGivern:2016bwh,Altinok:2017xua} have found disagreements between data and models \firstRefereeStrike{in the event rate} at both low $Q^2$ (the negative of the square of the four momentum
transfer from the lepton) and in pion kinetic energy.  Simulations of this process must model both the primary neutrino-nucleon interaction \firstRefereeStrike{as well as}\firstRefereeAdd{and} a variety of nuclear effects, including \firstReviewStrike{additional} interactions of produced hadrons within the nucleus. This convolution of primary production and nuclear modification makes it difficult to isolate the \firstRefereeAdd{model} features \firstRefereeStrike{of these models that are }responsible for the disagreements\firstRefereeStrike{ with data}.  Measurements on multiple nuclei can help \firstRefereeStrike{untangle}\firstRefereeAdd{separate} nuclear effects from other aspects of the interaction. 
\firstReviewStrike{
In this letter, the MINERvA collaboration presents the first simultaneous measurement of charged-current pion production on many nuclear targets -- scintillator, carbon, water, iron, and lead.  We also present \firstReviewAdd{cross section} ratios of each of the latter targets to scintillator, where many systematic uncertainties cancel, allowing a precise measurement of the nuclear dependence.  }

\if\showSections1 \section{Signal Definition} \fi

Signal reactions for this \firstReviewStrike{result} \firstReviewAdd{measurement} are $\nu_\mu$ charged-current interactions which produce a single negatively charged muon and a single positively charged pion. Any number of baryons may be \secondReviewStrike{ejected from the nucleus}\secondReviewAdd{in the final state}; however, no other mesons may be produced.  \firstRefereeStrike{In order to ensure high reconstruction efficiency} \firstRefereeAdd{To match \minerva's acceptance}, the muon is restricted in momentum ($p_\mu$) and angle with respect to the neutrino beam ($\theta_\mu$), such that $1.5<p_\mu<20$ GeV/c and $\theta_\mu<13^\circ$\firstRefereeAdd{.}\firstRefereeStrike{, while the p}\firstRefereeAdd{  P}ion\firstRefereeStrike{s must have} kinetic energy ($T_\pi$) \firstRefereeStrike{such that}\firstRefereeAdd{is restricted to} $35<T_\pi< 350$ MeV.  

To \firstRefereeStrike{focus on}\firstRefereeAdd{enhance} reactions \firstRefereeStrike{in the}\firstRefereeAdd{involving a} $\Delta$ \firstRefereeAdd{baryon }resonance\firstRefereeStrike{ region}, a selection is \firstRefereeStrike{also} made on the invariant hadronic mass of the final state.  We define $\Wexp=M^2+2M\Ehad-Q^2$, where $\Ehad=E_\nu-E_\mu$, $M$ is the average of the proton and neutron masses.  $\Wexp$ is the invariant mass of the hadronic final state under the assumption that 
\firstReviewAdd{the} target nucleon is at rest.  \Wexp\ is required to be less than $1.4$ \gevcsq.

This analysis provides measurements in eight kinematic variables: the magnitude of the muon momentum and its longitudinal and transverse components ($p_\mu$, $p_{\mu,||}$, and $p_{\mu,T}$), the angle of the muon ($\theta_\mu$), the kinetic energy and angle of the charged pion with respect to the neutrino beam ($T_\pi$ and $\theta_\pi$),  $Q^2$, and \Wexp\ as defined by the kinematics of the final state particles.

\if\showSections1 \section{Minerva Detector and Beam} \fi
The\firstReviewAdd{se} measurements \firstReviewStrike{presented in this paper} use event samples collected at the MINERvA detector using the medium energy NuMI beam at \firstRefereeStrike{the Fermi National \firstReviewAdd{Accelerator} Laboratory}\firstRefereeAdd{Fermilab}~\cite{Aliaga:2013uqz}.  To create the neutrino beam, 120~GeV protons impact a \firstReviewStrike{fixed} graphite target and produce pions and kaons.  Two magnetic horns focus \firstRefereeStrike{and direct} these charged particles into a pipe where they decay into primarily muon neutrinos. A GEANT4 simulation of the NuMI beamline~\cite{Aliaga:2016oaz} predicts the neutrino flux using constraints from hadron production data.  Previous MINERvA measurements of \firstRefereeStrike{the} neutrino elastic scattering on atomic electrons, $\nu e^- \rightarrow \nu e^-$, constrain the normalization of the neutrino flux, reducing the uncertainty of the flux between 2 and 20 GeV from 7.8\% to 3.9\%~\cite{Valencia:2019mkf,MINERvA:2022vmb}. Another MINERvA measurement of inverse muon decay, $\nu_\mu e^{-}\rightarrow \mu^-\nu_e$, constrains the flux at higher neutrino energies~\cite{MINERvA:2021dhf}. 
\firstReviewAdd{These data represent an \firstRefereeStrike{total} exposure of $10.6\times 10^{20}$ protons on target, half of which were taken with the water target \firstRefereeStrike{full and half with the target empty}\firstRefereeAdd{filled}.}

The MINERvA detector consists of a central polystyrene-based scintillator tracker with an upstream nuclear target region and downstream electromagnetic and hadronic calorimeters.  The nuclear target region contains five planes of passive material comprised of carbon, water, iron, and lead targets. \firstReviewAdd{The scintillator tracking planes are 95\% CH by weight.} 
The passive targets are interspersed between regions of active tracking volumes. The finely segmented tracking volumes of the MINERvA detector consist of hexagonal planes of \firstReviewAdd{nestled triangular scintillator strips with a pitch of 1.7~cm} \firstReviewStrike{1.7 cm wide scintillator strips}, allowing for spatial reconstruction with a resolution of 3~mm~\cite{Aliaga:2013uqz}.  The MINOS Near Detector, located 2~m downstream of the MINERvA detector\firstRefereeStrike{, serves as a muon spectrometer for exiting muons, allowing for the reconstruction of muon charge and momentum} \firstRefereeAdd{measures charge and momentum of exiting muons}~\cite{MINOS:2008hdf}. 

\if\showSections1 \section{Simulation} \fi
The simulation of the MINERvA detector utilizes GEANT4 version 4.9.4p2~\cite{Agostinelli:2002hh} with the QGSP\_BERT physics list ~\cite{Bertini:1968,Bertini:1971,Kaidalov:1982} to model the detector response. Calibrations using through-going muons provide the absolute energy scale~\cite{Aliaga:2013uqz}. Measurements on a scaled down version of the MINERvA detector in a charged particle test beam set the hadronic energy response~\cite{Aliaga:2015aqe}.  Overlaying data onto simulated events captures the effects of overlapping activity due to additional beam interactions.  

A modified version of the GENIE v2.12.6 event generator, \firstRefereeAdd{denoted ``MnvTune v4.3.1",} supplies the \firstRefereeAdd{neutrino interaction} simulation \firstRefereeStrike{of the signal and background neutrino interactions}~\cite{Andreopoulos:2009rq}. 
For inelastic events with the invariant hadronic mass $W<1.7$ GeV/$c^2$, resonance pion production \firstReviewStrike{uses} \firstReviewAdd{assumes the} Rein-Seghal model~\cite{Rein:1980wg} with an axial mass of $M_{A}^{RES} = 1.12$ GeV/$c^2$~
\firstReviewAdd{\cite{Kuzmin:2006dh}}.  Deep inelastic scattering \firstRefereeStrike{(DIS)} relies on the Bodek-Yang model~\cite{Bodek:2004pc} tuned to agree with external measurements of pion production and total cross sections.  Coherent pion production is simulated with the Rein-Seghal~\cite{Rein:1982pf,Rein:2006di} model with corrections for \firstReviewStrike{non-zero} muon mass.  The nuclear medium is modeled as a relativistic Fermi gas (\firstReviewAdd{with Fermi momentum} $p_F\sim 250$~MeV/$c$)~\cite{Moniz:1971mt} with an added Bodek-Ritchie high momentum tail~\cite{Bodek:1980ar,Bodek:1981wr}.  Simulation of hadron final state interactions within the nucleus are predicted by the INTRANUKE-hA package~\cite{Andreopoulos:2009rq}.

\if\showSections1 \section{Minerva Tunes} \fi

The GENIE simulation has been tuned \firstRefereeStrike{in order} to better reproduce MINERvA data and provide more accurate signal and background models\firstRefereeStrike{ resulting in a model known as ``MnvTune v4.3.1"}.  Modifications to the quasielastic process are described in Ref.~\cite{Carneiro:2019jds,Rodrigues:2015hik}. \secondReviewAdd{Pion production through baryon resonances is modified to match the $D_2$ bubble chamber data as in Ref.~\cite{Rodrigues:2016xjj}.}   Coherent pion production is reweighted in both the energy of the pion $E_\pi$ and $\theta_\pi$ to agree with a recent measurement of coherent pion production \firstRefereeStrike{in the medium energy NuMI beam}~\cite{AlexCoherent}.  \firstRefereeStrike{Additionally,} The normalization of coherent pion production is increased by 43.7\% \firstRefereeStrike{in order} to account for coherent interactions on hydrogen\firstRefereeAdd{,} known as diffractive pion production, based on the Kopeliovic  model~\cite{Kopeliovich:2012tu}.  \secondReviewAdd{The changes above comprise ``MnvTune v4.2.1''.}

\secondReviewStrike{Pion production through baryon resonances is modified to match the $D_2$ bubble chamber data as in Ref.~\cite{Rodrigues:2016xjj}.} However, \firstRefereeStrike{the study of events in the scintillator target found a significant disagreement} \firstRefereeAdd{signal events in the scintillator disagree} with the prediction of this model\firstRefereeStrike{; t}.  The model \firstRefereeAdd{for all targets except hydrogen} was corrected \firstRefereeAdd{by matching the simulated cross section versus $Q^2$ to the data.  This results in improved estimates of}\firstRefereeStrike{ to produce accurate} efficiencies and \firstRefereeStrike{subtractions of background distributions}\firstRefereeAdd{backgrounds}.  \firstRefereeStrike{For all targets except hydrogen, this correction was made by matching the reconstructed $Q^2$ distribution in the scintillator tracker to the simulation}.  The correction \firstRefereeStrike{significantly} decreases the single pion production for $Q^2<0.1$~\gevsqcsq~with a slow logarithmic increase for higher $Q^2$.  Details of this tune are given in 
\ifPRLandSupp
\supp\firstRefereeStrike{, section \ref{section:LowQ2Tune}}.
\else
\supp.
\fi
\firstRefereeStrike{\secondReviewAdd{This last tune completes ``MnvTune v4.3.1".}}
\firstRefereeAdd{This modification, combined with those of MnvTune v4.2.1, 
 forms MnvTune v4.3.1.}


\if\showSections1 \section{Event Selection} \fi

\begin{figure}
    \centering
    \captionsetup[subfigure]{labelformat=empty,justification=centering}    
    \ifColorFigures
       \includegraphics[width=\oneplot]{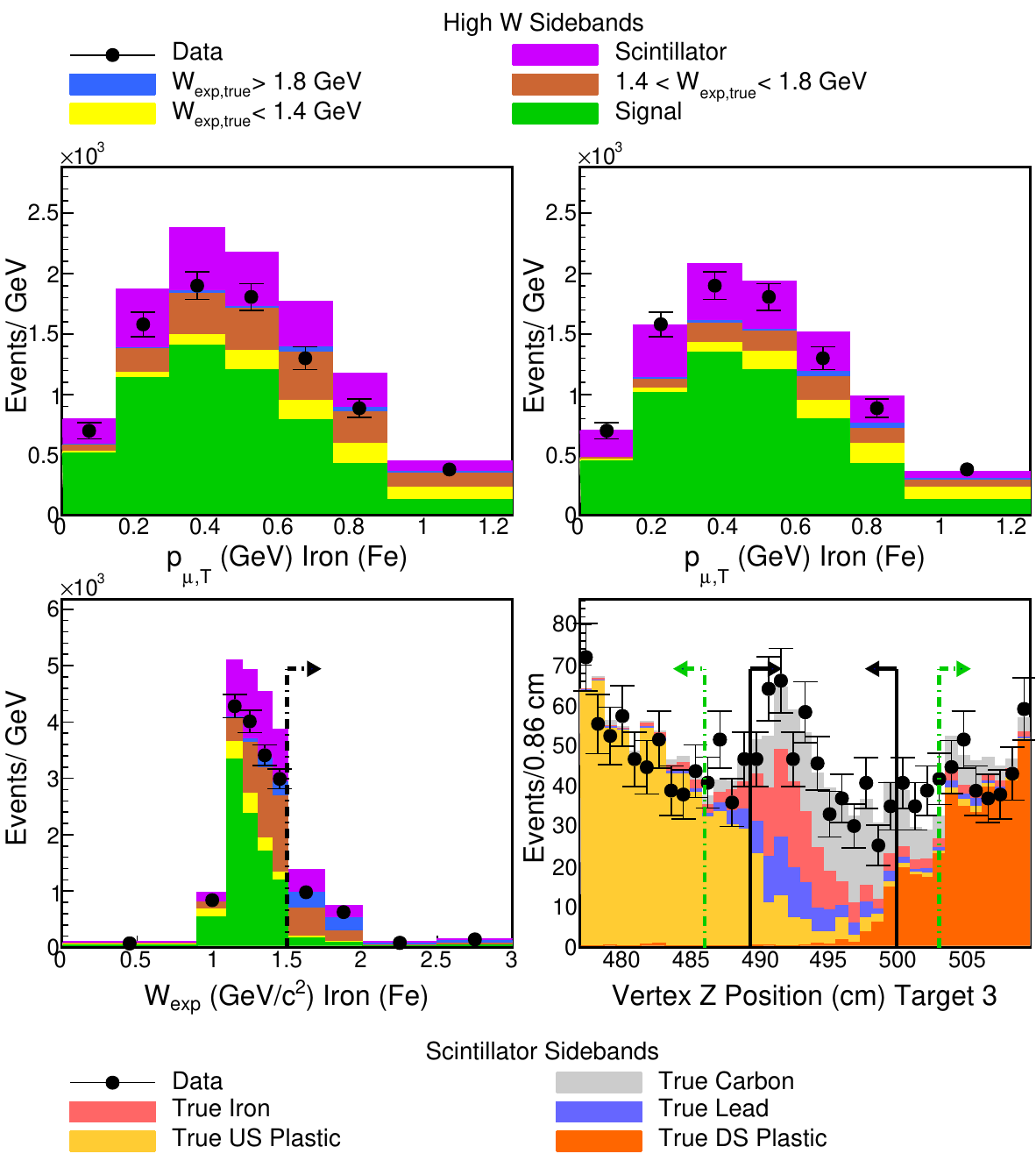}
    \else
      \includegraphics[width=\oneplot]{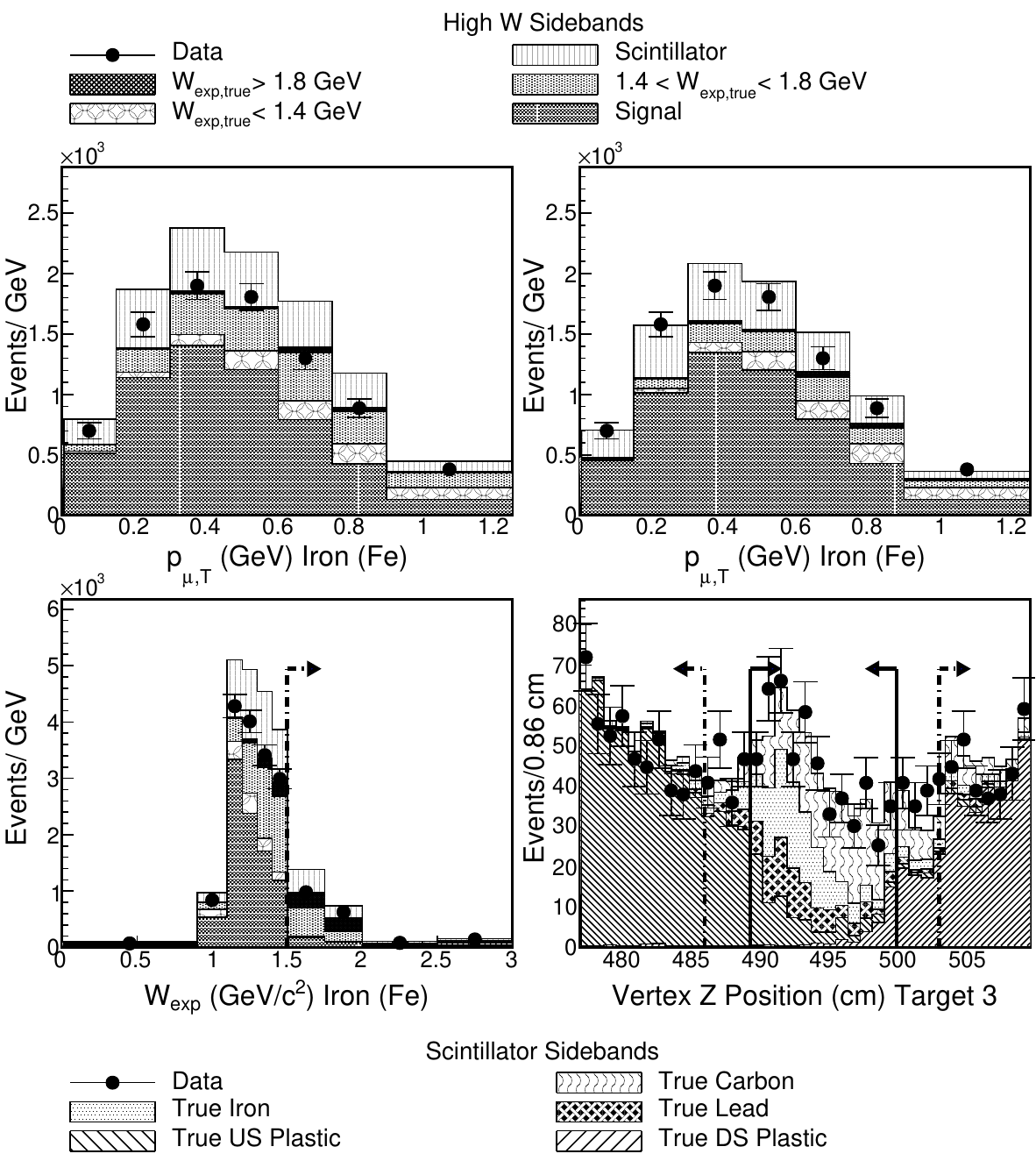}
    \fi  
    \caption{$p_{\mu,T}$ distributions in iron for the signal regions before (top left) and after (top right) the simulation has been constrained with background estimates from data, the high $W$ sideband in $W_{exp}$ in iron (bottom left) and the \firstReviewStrike{plastic} \firstReviewAdd{scintillator} sidebands in vertex z of target 3 (bottom left).  
    The background in the bottom plots have not been constrained. The solid (dashed) arrows in the lower right plot delineate the signal (sideband) regions.
    \firstRefereeAdd{Events labeled $W_{exp,true}<1.4 GeV$ are those that pass the signal $W_{exp,true}$ selection but fail other elements of the signal definition.  } 
    \kevinComment{The plot are very small.  Less whitespace; legends to bottom?}
     \lauraComment{Plot labels could also move to caption if necessary.  Axis labels need to be bigger. }
     \paperComment{can’t read legend–improve text size and location of the legend; put text inside the figure to enlarge the figure by saving some space}
    }
    \label{fig:signalAndSidebands}
\end{figure}

\begin{figure}
    \centering
    \ifColorFigures
       \includegraphics[width=\oneplot]{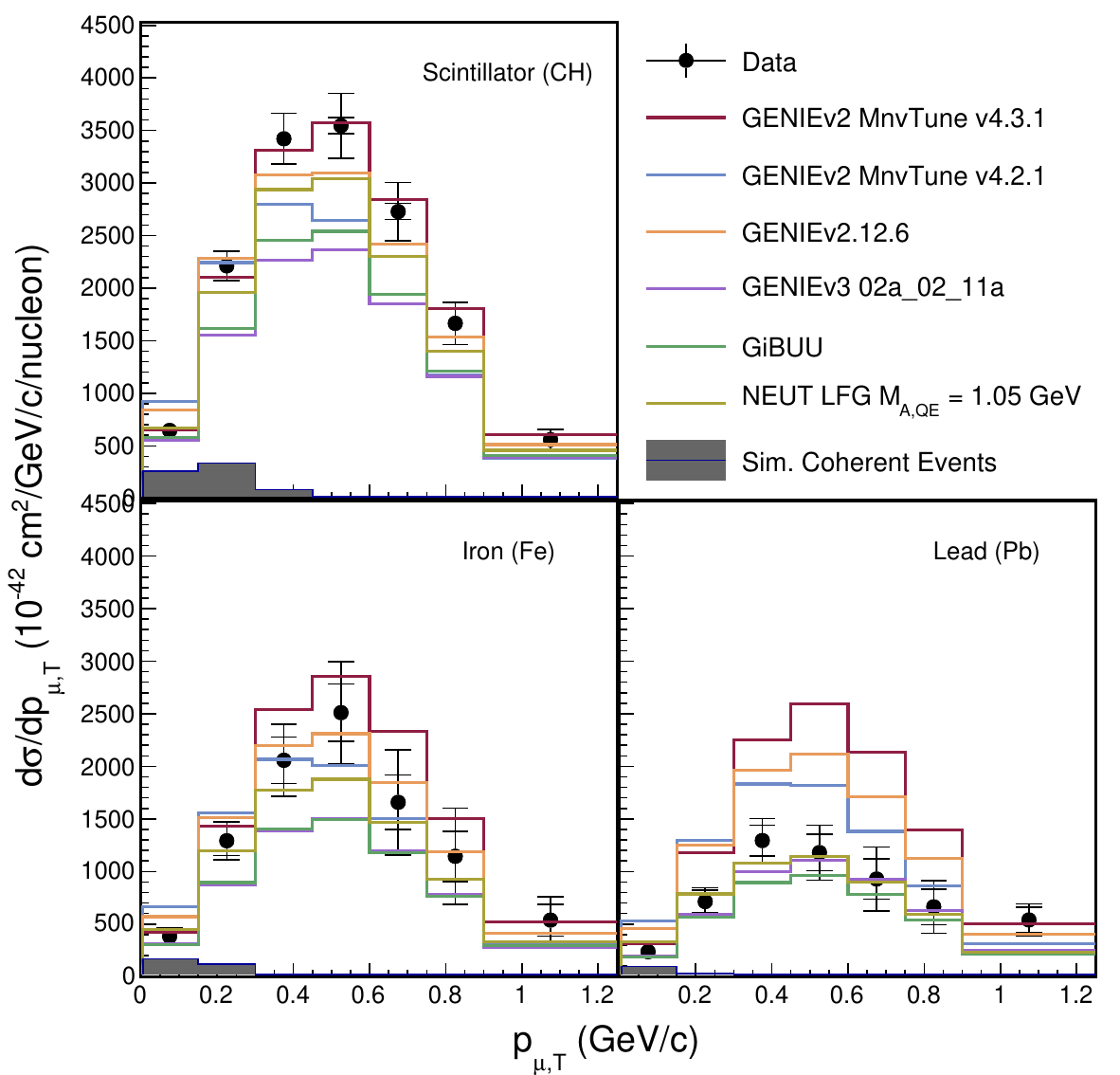}
    \else
      \includegraphics[width=\oneplot]{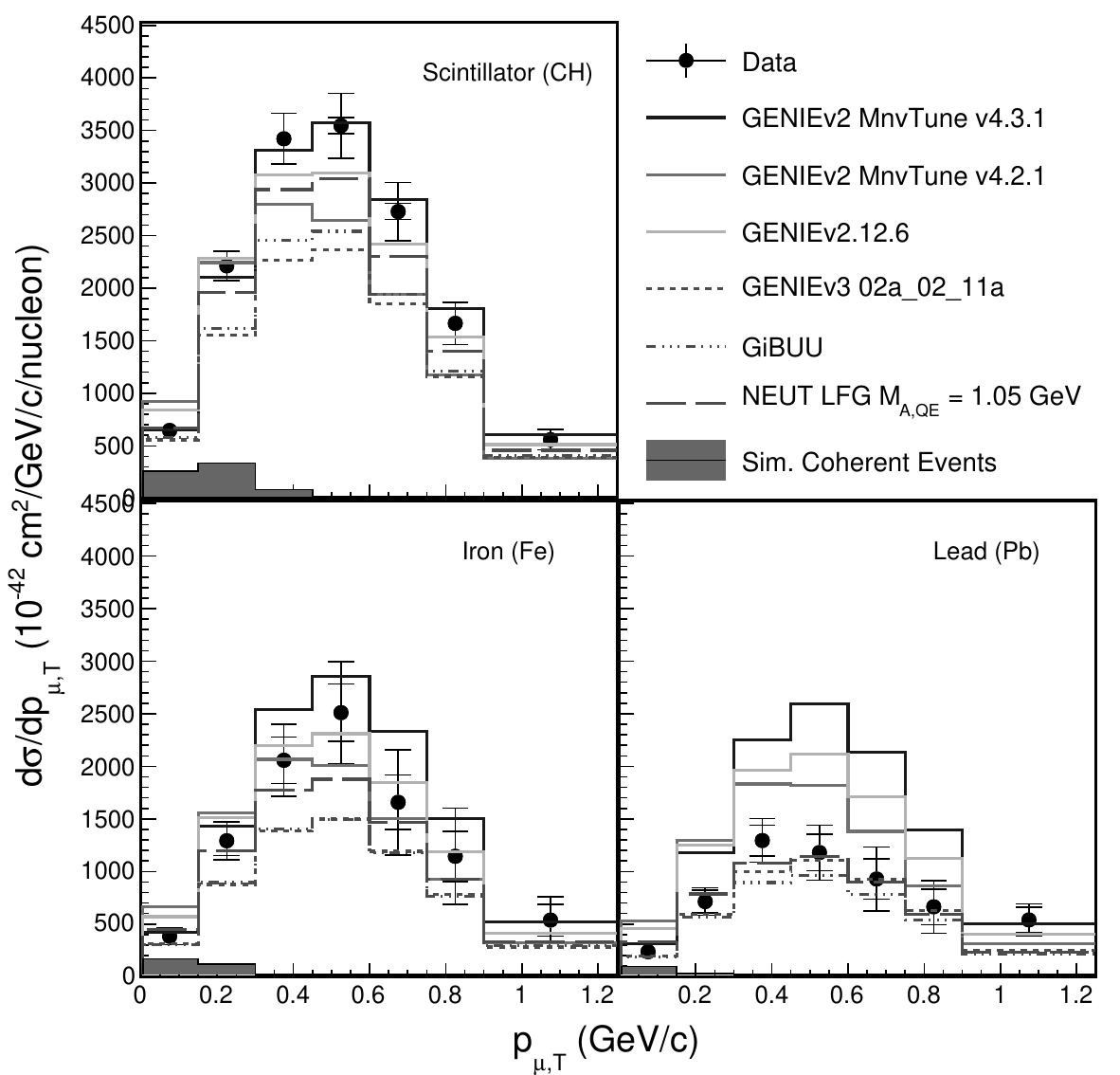}
    \fi  
    \caption{Differential cross section $\frac{d\sigma}{dp_{\mu,T}}$ \firstReviewAdd{measured on} 
    \firstReviewStrike{in}  scintillator, iron, and lead 
    \firstReviewAdd{(solid points) compared to GENIE, GIBUU and NEUT generators, all of which are discrepant with one if not all nuclei}.
    \inProgress{Will add a table with chi2 for each models in the Supplemental material}  
    }
    \label{fig:absolutePtmuXsec}
\end{figure}
\begin{figure}
    \centering
    \ifColorFigures
       \includegraphics[width=\oneplot]{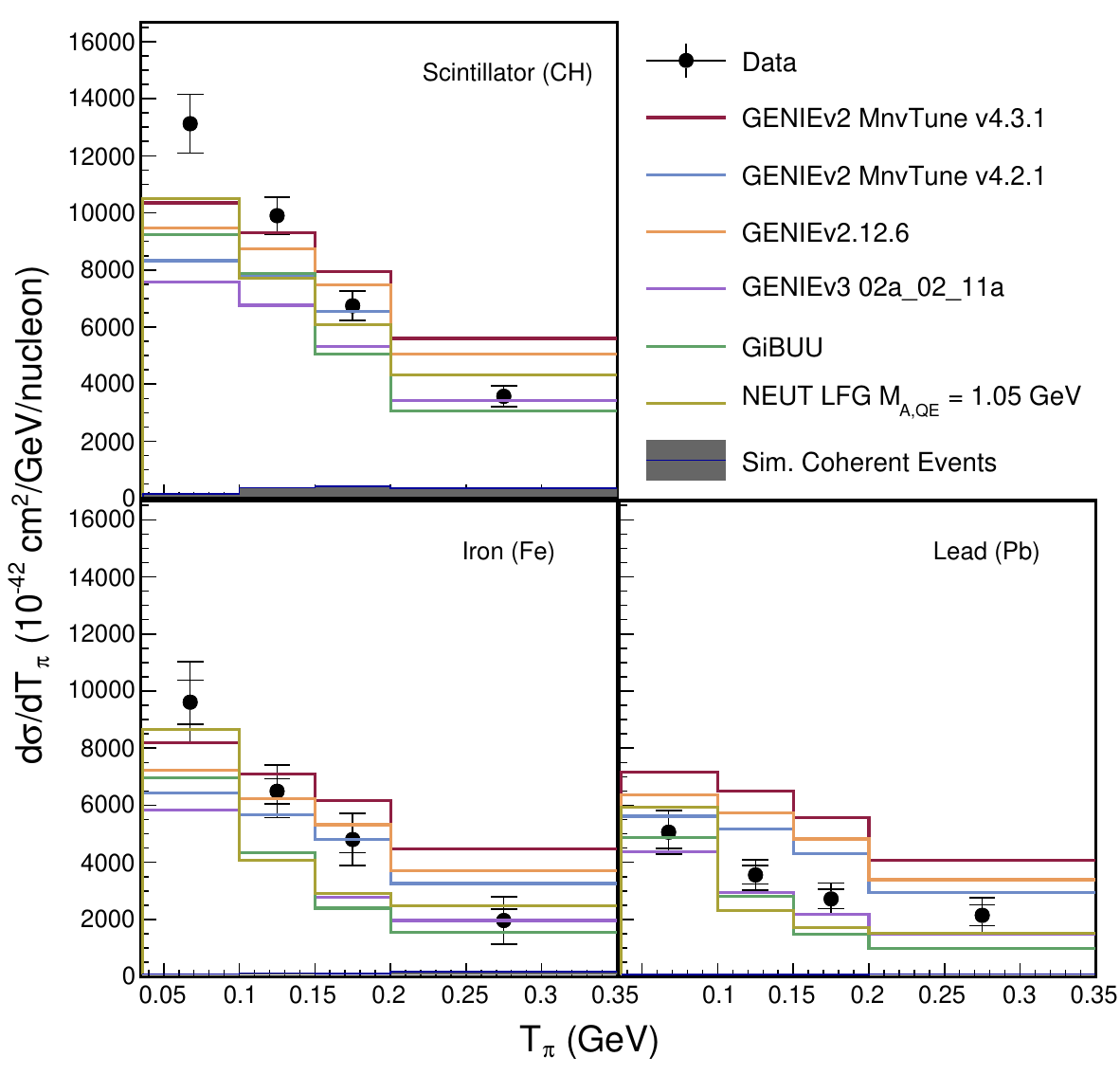}
    \else
      \includegraphics[width=\oneplot]{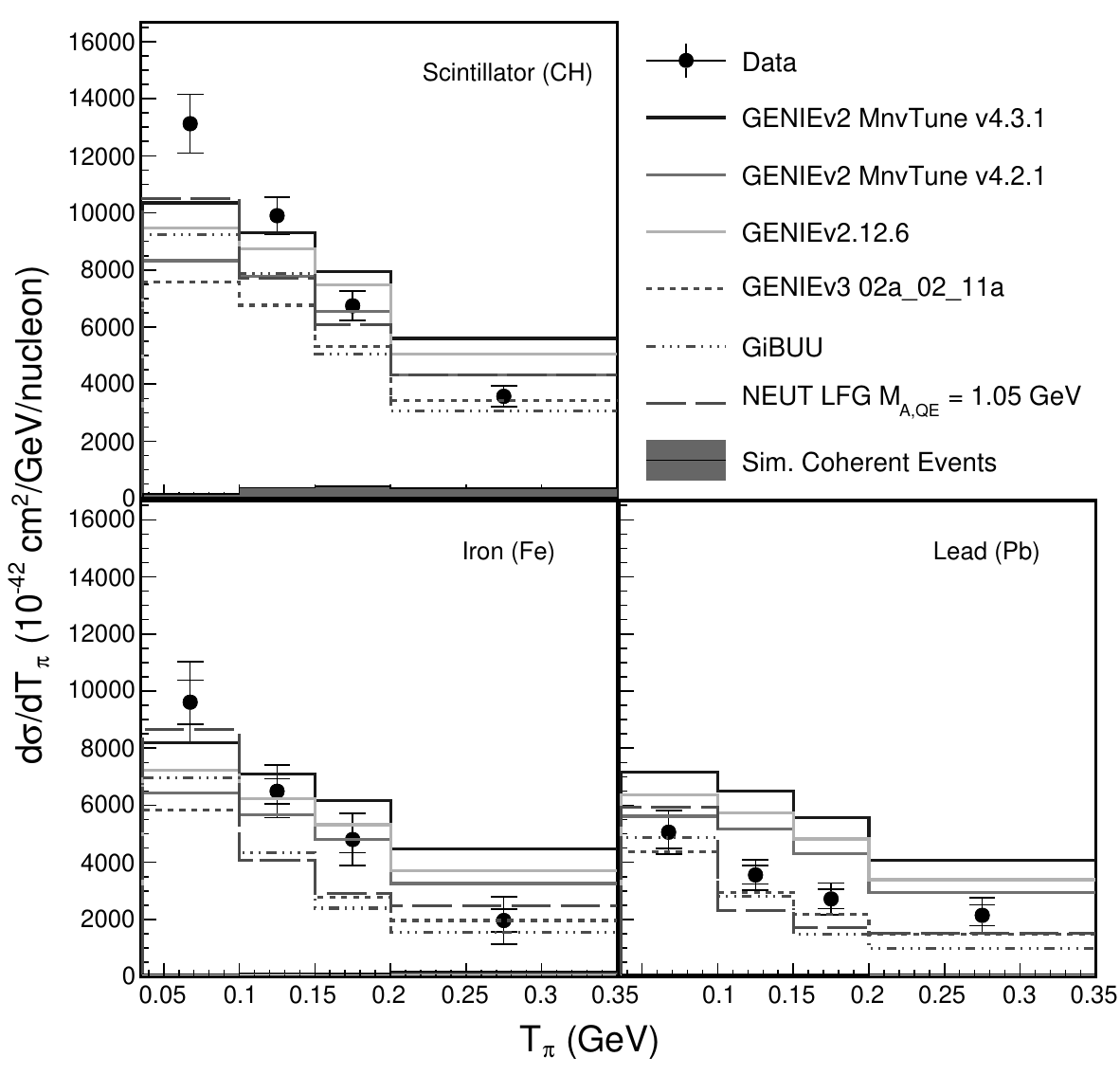}
    \fi  

    \caption{Differential cross section $\frac{d\sigma}{dT_\pi}$ in scintillator, iron, and lead \firstReviewAdd{ (solid points) compared to GENIE, GIBUU and NEUT generators, all of which are discrepant}. }    
    \label{fig:absoluteTpiXsec}
\end{figure}

\begin{figure}
    \centering
    \captionsetup[subfigure]{labelformat=empty,justification=centering}
     \ifColorFigures
       \includegraphics[width=\oneplot]{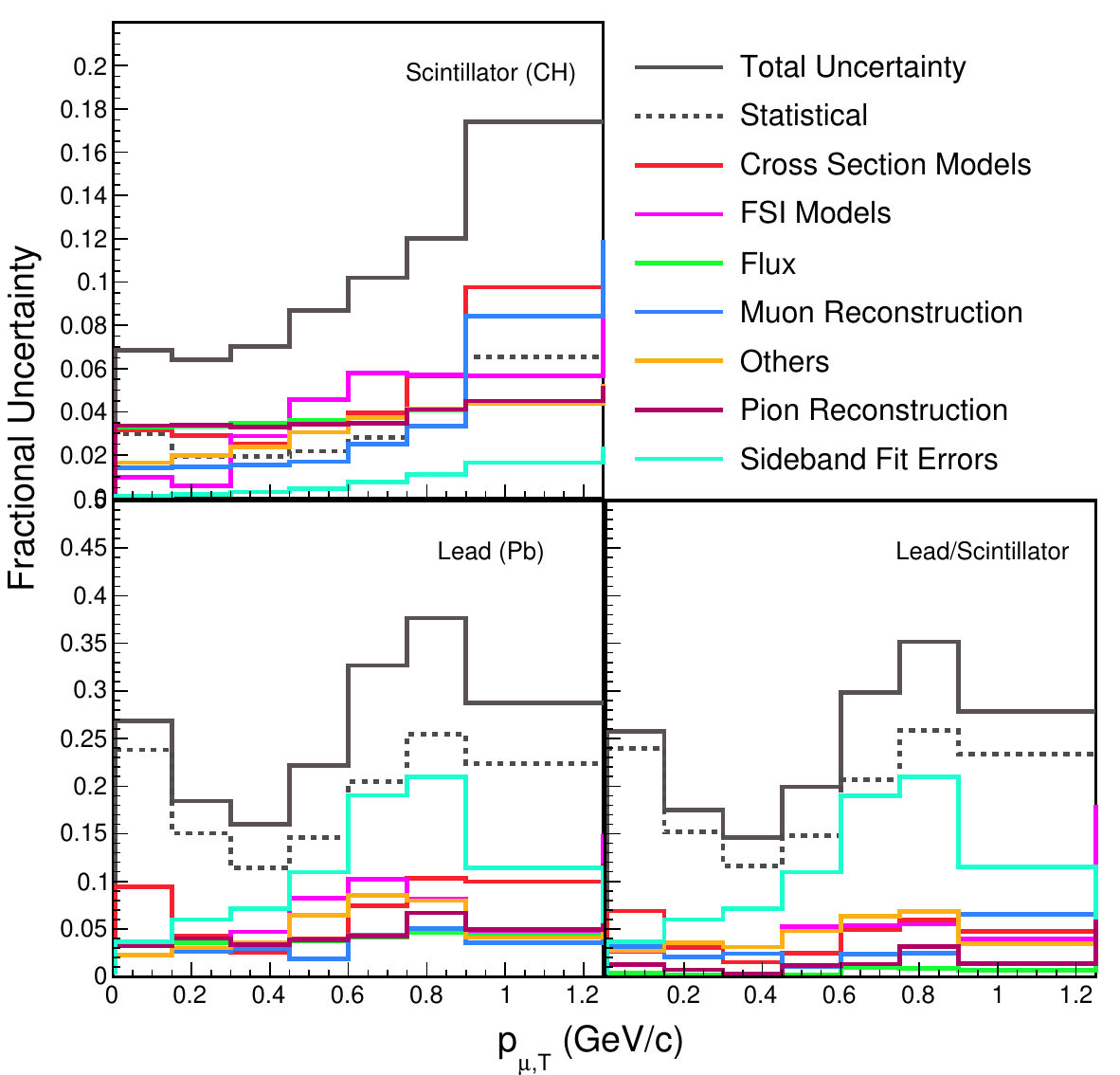}
    \else
      \includegraphics[width=\oneplot]{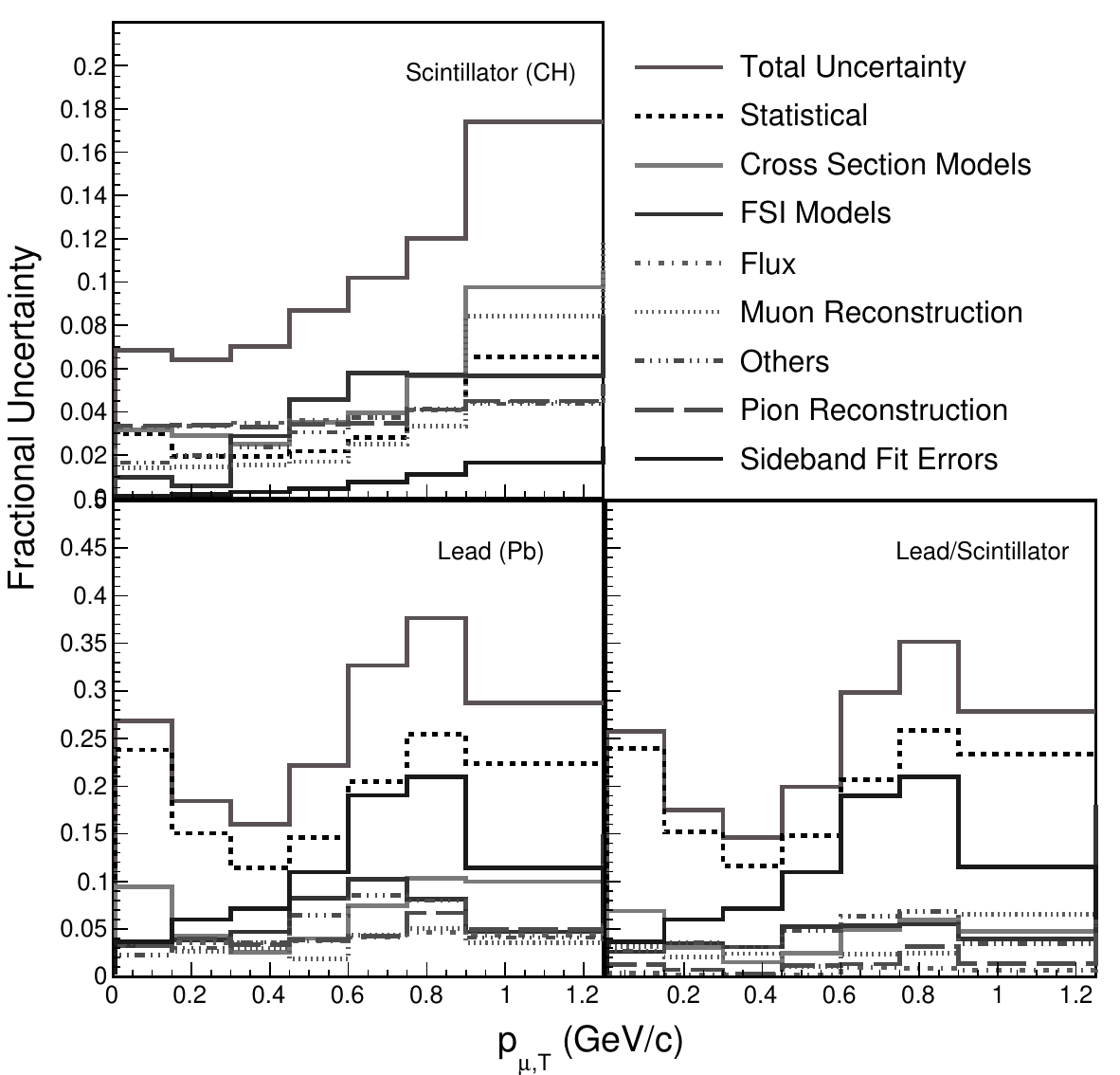}
    \fi  

    \caption{Systematic uncertainties on the differential cross section $\frac{d\sigma}{dp_{\mu,T}}$ in lead and scintillator and uncertainties of the ratio of $\left(\frac{d\sigma_{Pb}}{dp_{\mu,T}}\right)/\left(\frac{d\sigma_{CH}}{dp_{\mu,T}}\right)$.}
    
    \label{fig:absolutePtmuXsecSystIllustration}
\end{figure}

\begin{figure}
    \centering
    \ifColorFigures
       \includegraphics[width=\oneplot]{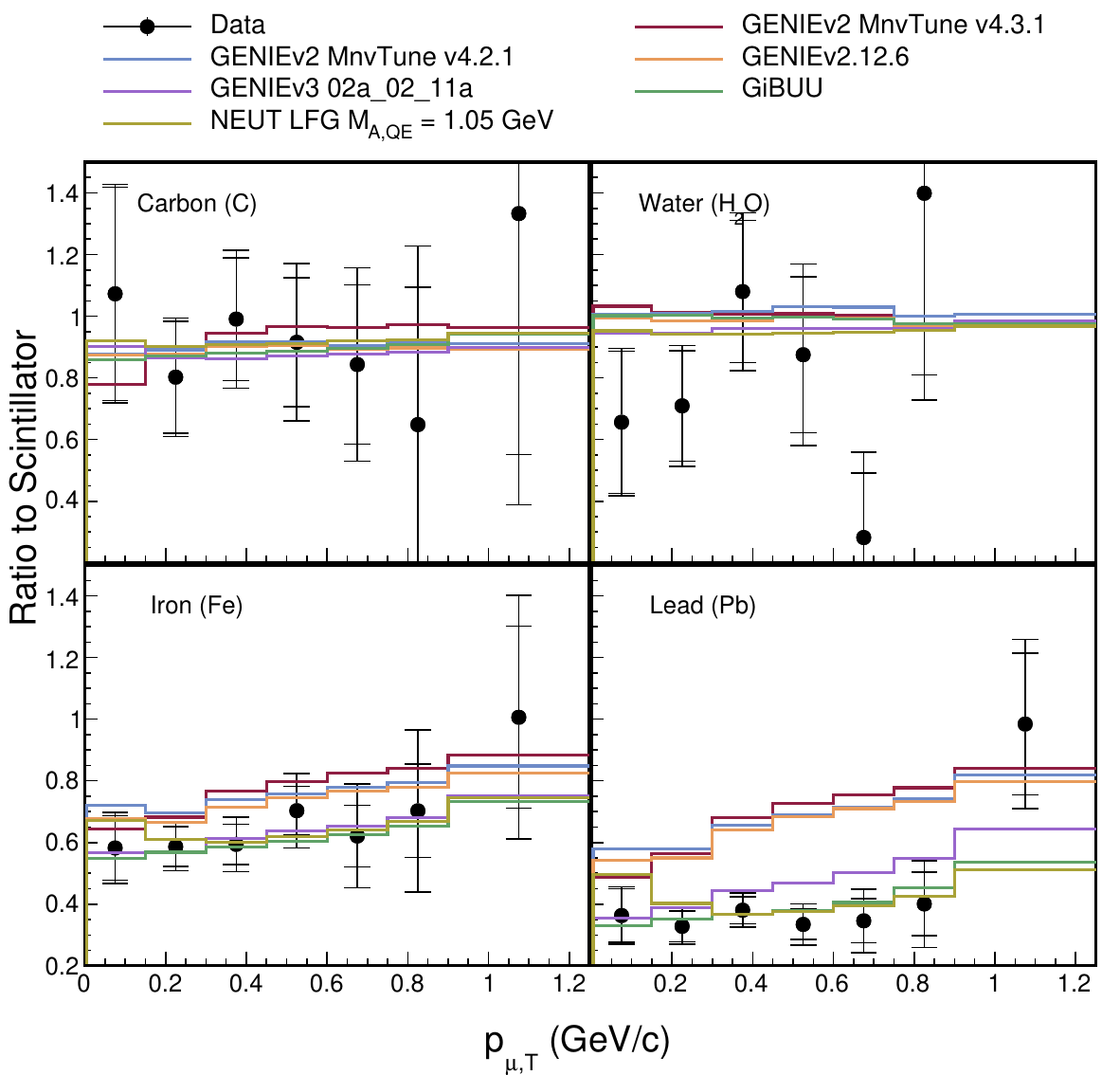}
    \else
      \includegraphics[width=\oneplot]{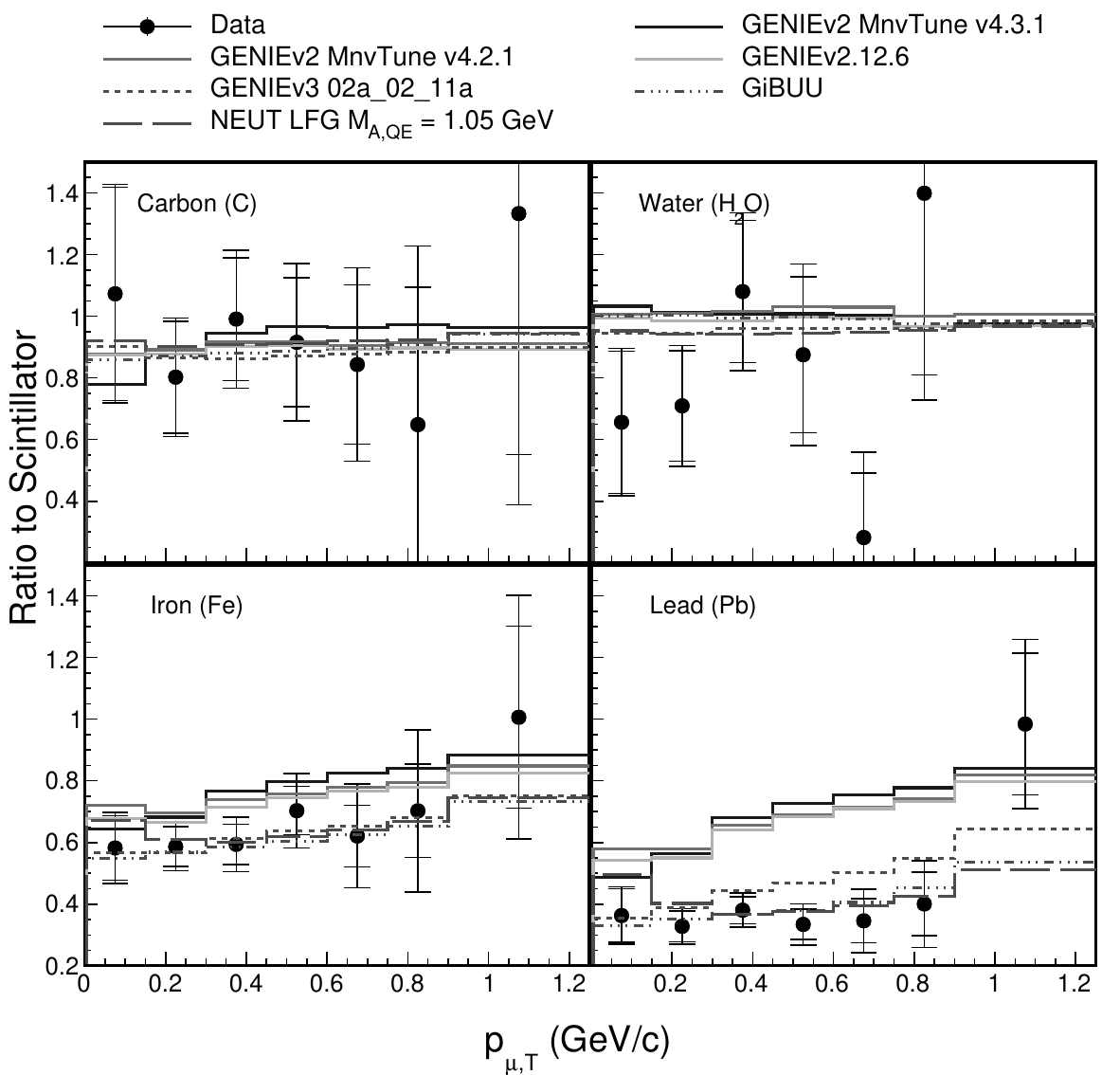}
    \fi  
    \caption{Cross section ratios $(\frac{d\sigma_A}{dp_{\mu,T}})/(\frac{d\sigma_{CH}}{dp_{\mu,T}})$ for carbon, water, iron, and lead \firstReviewAdd{ (solid points), as compared to GENIE, NEUT, and GIBUU.}}    
    \label{fig:PtmuXsecRatios}
\end{figure}
\begin{figure}
    \centering
    \ifColorFigures
       \includegraphics[width=\oneplot]{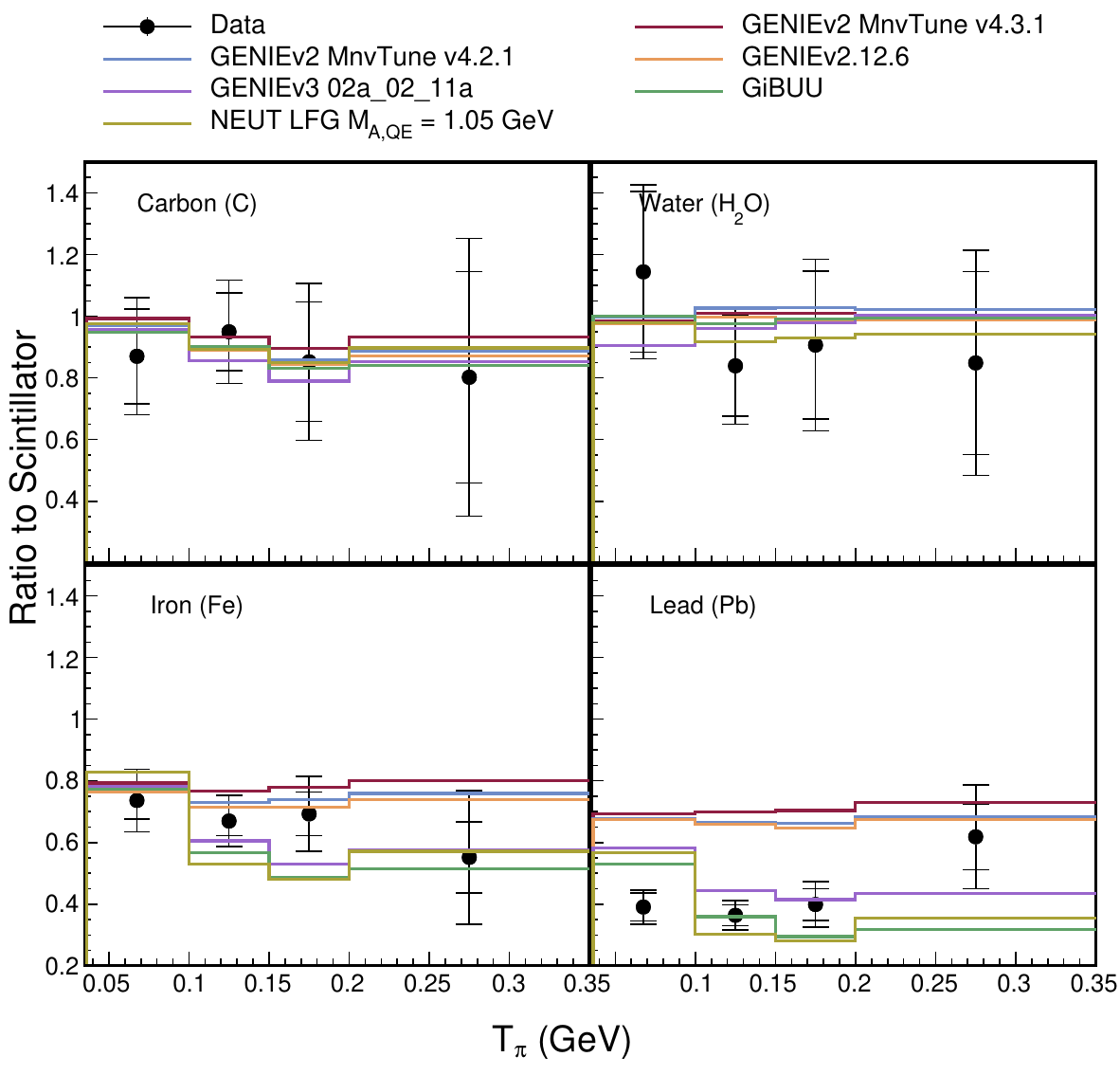}
    \else
      \includegraphics[width=\oneplot]{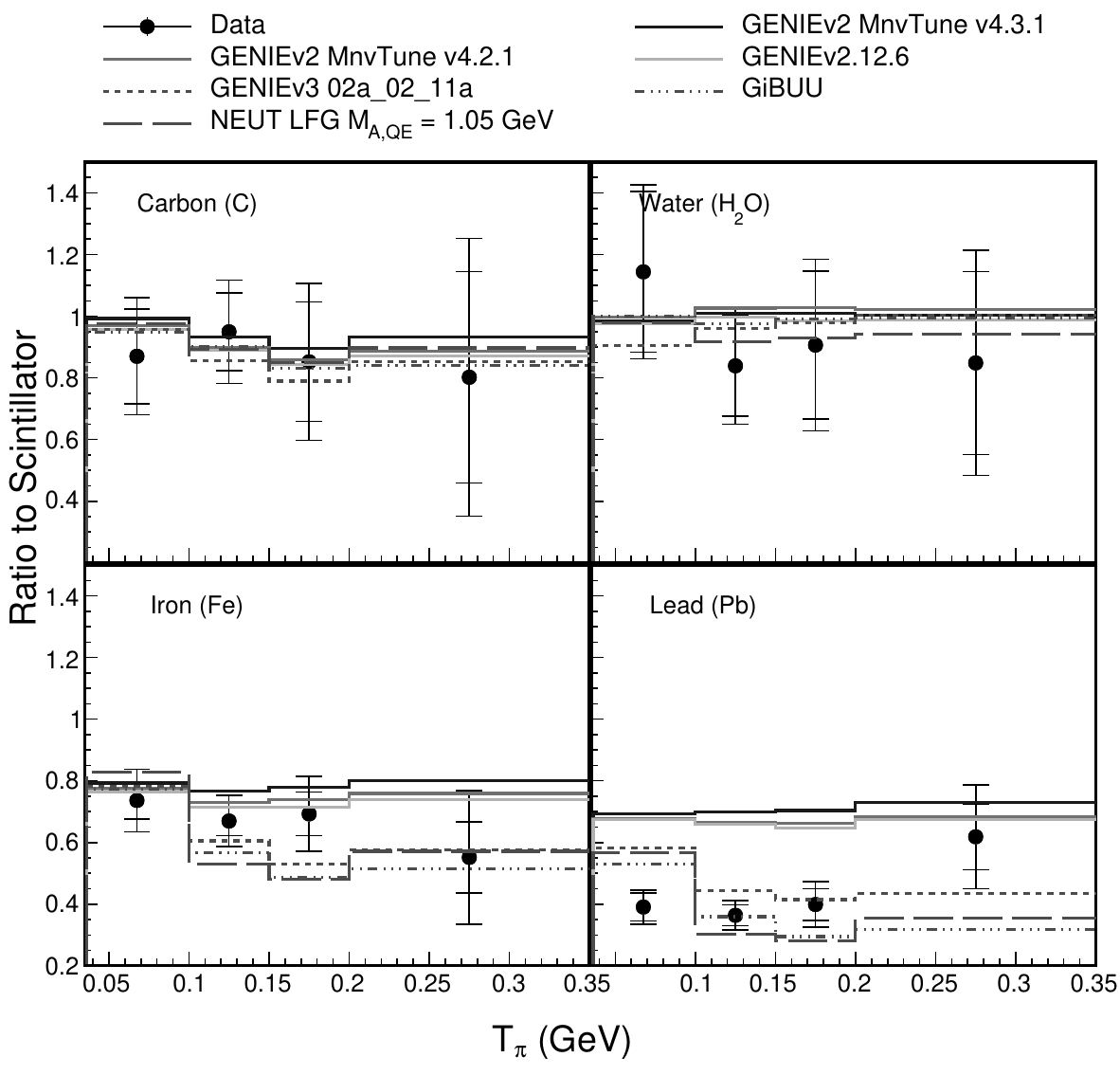}
    \fi  

    \caption{Cross section ratios $(\frac{d\sigma_A}{dT_{\pi^+}})/(\frac{d\sigma_{CH}}{dT_{\pi^+}})$ for carbon, water, iron, and lead \firstReviewAdd{ (solid points) compared to predictions from GENIE, NEUT, and GIBUU.  The predictions from GIBUU and NEUT agree better with the measurements than those from GENIE.} }    
    \label{fig:TpiXsecRatios}
\end{figure}

\kevinComment{RATIO FIGURES: illustrate "universality" of Q2 fudge, add a model to the ratio where fudge is only applied to C and CH.  Maybe also look into adding the correction to hydrogen as another alternate model (but probably it's the same size and opposite direction as the difference between 4.2.1 and 4.3.1, so not the highest priority).}

Selected event\firstRefereeAdd{s}\firstReviewStrike{in the analysis sample} are required to have at least two tracked particles that start in the correct target.  One track must be identified as a negatively charged muon by the MINOS Near Detector.  One \firstRefereeStrike{of the} remaining track\firstRefereeStrike{s} must be identified as a non-interacting charged pion by matching a Michel electron to the endpoint of the track and having longitudinal energy deposition ($dE/dx$) consistent with a non-interacting pion.  All \firstRefereeStrike{other} remaining tracks must not have \firstRefereeStrike{a} $dE/dx$ consistent with a pion.  
The charged pion tracking efficiency decreases at lower momentum and at angles perpendicular to the detector axis.  The efficiency to reconstruct the Michel electron from the $\mu^+$ was predicted from the simulation and \firstRefereeStrike{is} validated in the scintillator tracker \firstRefereeStrike{by studies of} \firstRefereeAdd{using} stopping muons produced \firstRefereeStrike{in neutrino interactions} in the \firstReviewAdd{rock upstream of the detector}~\cite{BercellieThesis}. 
The $\Wexp<1.4$~\gevcsq ~requirement strongly reduces the number of multipion events. Both $\Wexp$ and $Q^2$ are computed using the visible calorimetric energy in the detector to estimate $\Ehad$ \secondReviewAdd{\cite{BercellieThesis}}.  
\firstReviewAdd{After all cuts there are 33,231 events in the tracker, and 1403 (1033) events in the iron (lead), and 295 (291) events in the carbon (water).  }

\if\showSections1 \section{Background constraints} \fi

The two primary backgrounds \firstRefereeStrike{to this signal} are pion production events with $\Wexp>1.4$~\gevcsq~and pion production events that appear to originate from the passive nuclear targets but are actually produced in the adjacent scintillator.  First, simulated events with reconstructed $\Wexp>1.4$~\gevcsq~ in the scintillator target are weighted to match data using two scale factors, one for events with true $\Wexp$ between $1.4$ and {$1.8$~\gevcsq} and one for events with $\Wexp$ above $1.8$~\gevcsq.  \firstRefereeAdd{True $\Wexp$ is computed in the same manner as reconstructed $\Wexp$ but uses true simulated values.}  \firstRefereeStrike{At this stage, in order t}\firstRefereeAdd{T}o improve the background prediction from events in scintillator adjacent to the passive target, pions with low $T_\pi$ or $\theta_\pi > 90^\circ$ also receive additional weight\firstRefereeStrike{ correction}s to improve \firstRefereeStrike{their} agreement with the scintillator target data.  
\firstReviewStrike{Second,}
\firstReviewAdd{ Next,} simulated events in each target are 
\firstReviewStrike{then} weighted to correct the prediction of adjacent scintillator backgrounds using  events observed in data near the passive targets.  Finally, a weight is applied for events with high reconstructed $\Wexp$, in the same manner as above, for each target material.  Fig.~\ref{fig:signalAndSidebands} shows a sample of data used in the background constraints, before and after constraints.  The full set of weights are given in \ifPRLandSupp
\supp\firstRefereeStrike{, Sec.\ref{section:SBTune}}.
\else
\supp.
\fi


\if\showSections1 \section{Differential cross sections and ratios} \fi

Kinematic smearing due to the detector in the background-subtracted distributions \firstReviewStrike{are} \firstReviewAdd{is} removed with iterative D'Agostini unfolding~\cite{D'Agostini:1994zf,DAgostini:2010xxxxx} as implemented in the RooUnfold~\cite{Adye:2011gm} framework. The number of iterations was chosen by studying the fidelity of unfolding randomly thrown pseudodata samples generated from alternate physics models. 
\firstReviewStrike{Corrections for detector acceptance or selection efficiencies are applied to each distribution, which are $\sim4.5\%$ for events on scintillator and $\sim1\%$ for events on passive materials.} 
\firstReviewAdd{The event distributions are corrected for selection efficiency and acceptance, which is 4.5\% (1\%) on average in the tracker (passive targets).  Efficiencies are available in the supplement.  }
The differential cross sections are obtained by normalizing the resulting event rates by the integrated neutrino flux and \firstRefereeStrike{the} number of target nucleons.  

One complication of forming ratios of cross sections between the passive targets and the tracker is that the fluxes have small differences due to the distributions of mass relative to the beam axis, as shown in \firstReviewAdd{\supp}\firstReviewStrike{Fig.~\ref{fig:fluxRatios}}.  This is corrected by measuring the \firstReviewStrike{plastic} cross section \firstReviewAdd{on scintillator} in several regions of the detector, each integrated over a slightly different neutrino energy distribution.  Linear combinations of these regional cross sections are used to form a cross section integrated over a flux\firstReviewStrike{identical to that of} \firstReviewAdd{ that matches \firstRefereeStrike{the one seen by}} the relevant target.  These cross sections are used to form the ratio of target-to-tracker cross-section ratios~\cite{BercellieThesis,KleykampThesis}.


Figure~\ref{fig:absolutePtmuXsec} shows the differential cross section $\frac{d\sigma}{dp_{\mu,T}}$ \firstReviewAdd{ on} materials with highest statistics (iron, lead, and scintillator). Lower statistics measurements on water and carbon are available in the supplement.  \firstReviewStrike{The masses of all the passive targets are small enough that } 
\firstReviewAdd{The statistical uncertainty dominates in all measurements, except in the scintillator.  }
\firstReviewStrike{ The masses of the carbon and water target are smaller than the lead and iron, and thus the cross sections on the former targets have particularly large uncertainties.}  
\firstReviewStrike{ An illustration of the }
\firstReviewAdd{ An example of the } uncertainties 
\firstReviewStrike{in}
\firstReviewAdd{on} the cross sections 
\firstReviewStrike{ is} 
\firstReviewAdd{are} shown in Fig.~\ref{fig:absolutePtmuXsecSystIllustration}; uncertainties in the \firstRefereeStrike{assumed input} cross section models which enter via efficiency and unfolding are the largest, but are \firstReviewAdd{lower than the statistical uncertainties in all targets but the scintillator.} 
\firstReviewStrike{ typically comparable or less than statistical errors, except in the highest statistics bin of the scintillator. } Statistical uncertainties in the sideband constraints are also significant for the passive targets.  Systematic uncertainties are smaller in the cross section ratios, which benefit from partial cancellation of the flux, cross section model, and reconstruction uncertainties.  

Ratios of cross sections to that in scintillator are shown for $\frac{d\sigma}{dp_{\mu,T}}$ and $\frac{d\sigma}{dT_\pi}$ in Figs.~\ref{fig:PtmuXsecRatios} and \ref{fig:TpiXsecRatios}.  
\firstReviewStrike{These results show \firstReviewStrike{ nearly universal behavior over the range of nuclei studied: that is,}
\firstReviewAdd{that} the cross section ratios are reasonably well described by a \firstReviewStrike{ universal } scale factor of unity in the carbon and water, and by a constant 
$\approx 0.8\secondReviewAdd{\pm 0.1}$ in iron and $\approx 0.5\secondReviewAdd{\pm 0.1}$ in lead
with no large modifications to the shape in $p_{\mu,T}$ or $T_\pi$
The ratios are largely also consistent with \firstReviewStrike{ this universal scaling observation} the same scaling.
}
\firstReviewAdd{The cross-section ratios \firstRefereeAdd{of} carbon or water to scintillator can be characterized by a scale factor of unity, while ratios between iron or lead to scintillator are constant at ~0.8 and ~0.5 respectively, with no large modifications to distribution shapes for either $p_{\mu,T}$ or $T_\pi$.}     
Cross sections and cross section ratios to scintillator in other kinematic variables are tabulated and shown in 
\ifPRLandSupp
\supp\firstRefereeStrike{, sections \ref{section:AbsXsec} and \ref{section:RatioXsec}}.
\else
\supp.
\fi
\firstReviewAdd{Ratios in the other variables  also exhibit this 
scaling behavior.

None of the six generator models investigated capture the evolution of absolute cross section.  
The generators NEUT\cite{Hayato:2009zz} and GIBUU\cite{Gallmeister:2016dnq} do correctly predict the cross section ratios between Fe and Pb and scintillator, as shown in Figs.~\ref{fig:PtmuXsecRatios} and \ref{fig:TpiXsecRatios}, while the GENIE predictions do not agree.  The GENIE models use pion- and nucleon-nucleus scattering data to implement single-step absorption and other scattering processes (INTRANUKE-hA)~\cite{Andreopoulos:2009rq}, whereas NEUT, NuWRO and GIBUU employ different microscopic transport algorithms to simulate pion and nucleon intranuclear scattering.}

As previously noted, the scintillator cross section showed a large discrepancy with the MINERvA-tuned GENIE model, which is corrected with a weight as a function of $Q^2$.  In Fig.~\ref{fig:absolutePtmuXsec}, the differential cross section as a function of the related observable $p_{\mu,T}$ $\left(Q^2\approx{p_{\mu,T}}^2\left[ 1+{\cal O}(\frac{\Ehad}{E_\mu})\right] \right)$ shows poor agreement both with untuned GENIE models (GENIEv2.12.6) and with tuned models without this weighting function (GENIEv2 MnvTune v4.2.1).  The \firstRefereeAdd{shape of the} data is in better agreement with the weighted model (GENIEv2 MnvTune v4.3.1) in all targets, even though the weight is derived using only the scintillator measurement.  \firstRefereeAdd{The absolute normalization of the cross section is not well described by MnvTune v4.3.1 in Iron or Lead.}

Similarly, the ratios of $d\sigma/dp_{\mu,T}$ between the targets and scintillator are consistent with being independent of $p_{\mu,T}$.
A \firstRefereeStrike{notable} feature of both the GENIE v3 and GiBUU models is that they 
\firstReviewStrike{do appear to} predict the large suppression \firstRefereeStrike{in the} lead \firstRefereeStrike{target} relative to other targets.  However, as seen in Fig.~\ref{fig:TpiXsecRatios}, these models also predict a larger ratio at low $T_\pi$ in the heavy targets, a feature which is not seen in the data.

In summary, these data on single pion production on a wide variety of nuclei provide\firstReviewStrike{s} a new way to test nuclear models through their \firstRefereeAdd{nuclear }dependence\firstRefereeStrike{ on the size of the target nucleus}.  We observe a large modification to the predicted $Q^2$ distribution of single $\pi^+$ events \firstRefereeStrike{in scintillator, but it appears }that it can be consistently \firstRefereeStrike{applied to}\firstRefereeAdd{seen in} all nuclei.  These results \firstRefereeStrike{can} give guidance in how the \firstRefereeStrike{relative} wealth of data on scintillator targets can be applied to \firstRefereeStrike{reference} models for oxygen and argon for future neutrino experiments.

\begin{acknowledgements}

\paperComment{pulled version dated March 2022, v15, from DocDB.  https://minerva-docdb.fnal.gov/cgi-bin/sso/ShowDocument?docid=13929&version=15  Unclear how it is actually different...}

This document was prepared by members of the MINERvA Collaboration using the resources of the Fermi National Accelerator Laboratory (Fermilab), a U.S. Department of Energy, Office of Science, HEP User Facility. Fermilab is managed by Fermi Research Alliance, LLC (FRA), acting under Contract No. DE-AC02-07CH11359.
These resources included support for the MINERvA construction project, and support
for construction also
was granted by the United States National Science Foundation under
Award No. PHY-0619727 and by the University of Rochester. Support for
participating scientists was provided by NSF and DOE (USA); by CAPES
and CNPq (Brazil); by CoNaCyT (Mexico); by ANID PIA / APOYO AFB180002, CONICYT PIA ACT1413, and Fondecyt 3170845 and 11130133 (Chile); 
by CONCYTEC (Consejo Nacional de Ciencia, Tecnolog\'ia e Innovaci\'on Tecnol\'ogica), DGI-PUCP (Direcci\'on de Gesti\'on de la Investigaci\'on  - Pontificia Universidad Cat\'olica del Peru), and VRI-UNI (Vice-Rectorate for Research of National University of Engineering) (Peru); NCN Opus Grant No. 2016/21/B/ST2/01092 (Poland); by Science and Technology Facilities Council (UK); by EU Horizon 2020 Marie Skłodowska-Curie Action; by a Cottrell Postdoctoral Fellowship from the Research Corporation for Scientific Advancement; by an Imperial College London President's PhD Scholarship.  We thank the MINOS Collaboration for use of its near detector data. Finally, we thank the staff of
Fermilab for support of the beam line, the detector, and computing infrastructure.


\end{acknowledgements}

\bibliography{MINERvATargetPion}

\ifPRLandSupp

\clearpage
\onecolumngrid
\pagebreak
\section{Supplemental Material}
\renewcommand\thefigure{Supp.\arabic{figure}}
\setcounter{figure}{0}
\renewcommand\thetable{Supp.\Roman{table}}
\setcounter{table}{0}

\subsection{Flux Differences between Targets and Tracker before Reweighting of Tracker}
\FloatBarrier
\begin{figure}
    \centering
    \includegraphics[width=0.4\textwidth,page=4]{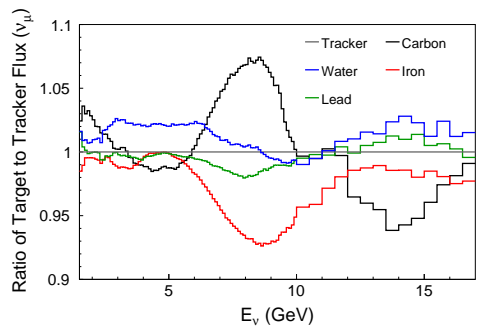}
    \caption{Ratio of neutrino fluxes in carbon, water, iron, and lead to scintillator as a function of energy before reweighting of the scintillator sample denominator.}
    \label{fig:fluxRatios}
\end{figure}
\pagebreak\clearpage

\subsection{Single $\pi^+$ $Q^2$ tune function}
\label{section:LowQ2Tune}
The disagreement between the initial tuned GENIE model described above and the scintillator data was addressed by applying a weighting function, $w(Q^2)$ to all non-coherent single pion production on nuclei other than hydrogen.   Coherent pion production and production on hydrogen are excluded from the correction because they are independently constrained by the results of Ref.~\cite{AlexCoherent} and Ref.~\cite{Rodrigues:2016xjj}, respectively. The predicted coherent and hydrogen cross section on scintillator in the data was subtracted from the measurement in the data, and compared to the same quantity in the simulation, and the ratio was fit to find $w(Q^2)$. 

This fit has the form shown in equation \ref{eq:LowQ2Eq}, where Erf is the Gaussian error function, and the fitted parameters are given in Tab.~\ref{tbl:LowQ2Fit}.  Figure \ref{fig:LowQ2Function} shows the cross section ratio as well as the fit.

\secondReviewAdd{These suppression at low $Q^2$ relative to high $Q^2$ is a feature that has been observed in different data sets, including MINOS on iron~\cite{MINOS:2014axb} and in other MINERvA pion measurements on scintillator~\cite{Stowell:2019zsh,Eberly:2014mra, McGivern:2016bwh, Altinok:2017xua}.  Models have some differences in prediction, although generally do not predict this large suppression.  The high $Q^2$ shape evolution of different versions and tunes of GENIE goes from GENIE\firstRefereeAdd{v2 MnvTune} v4.2.1 (steepest), to untuned GENIE v2.12.6 to GENIE v3 \firstRefereeAdd{02a\_02\_11a} to GENIE\firstRefereeAdd{v2 MnvTune} v4.3.1 (flattest) via changes to the axial form factor and, in the case of GENIE v3, also the vector form factor and \firstRefereeStrike{well as} FSI processes. Like the tunes of Refs.~\onlinecite{Stowell:2019zsh} and \onlinecite{MINOS:2014axb}, the \firstRefereeAdd{Mnv}Tune v4.3.1 adds the {\em ad hoc} additional shape at low $Q^2$ evident in these data.}

\FloatBarrier
\begin{align}
    w(Q^2) &= a+b\times \text{Max}\left[0,\text{Erf}\left[x_3 \ln\left(\frac{Q^2}{x_1}\right)\right]\right]
    \label{eq:LowQ2Eq}
\end{align}
\begin{table}
\centering
\begin{tabular}{c|c}
Parameter & Fitted Value\\  
\hline
$a$ & $0.475\pm 0.037$\\
\hline
$b$ & $1.837 \pm 0.537$\\
\hline
$x_1$ & $0.017\pm 0.003$\\
\hline
$x_3$ & $0.171\pm 0.069$\\
\hline
\hline
$\chi^2$/ndf & 1.58/11
\end{tabular}
\caption{Fitted parameters of the $Q^2$ model variation}
  \label{tbl:LowQ2Fit}
\end{table}
\begin{figure}
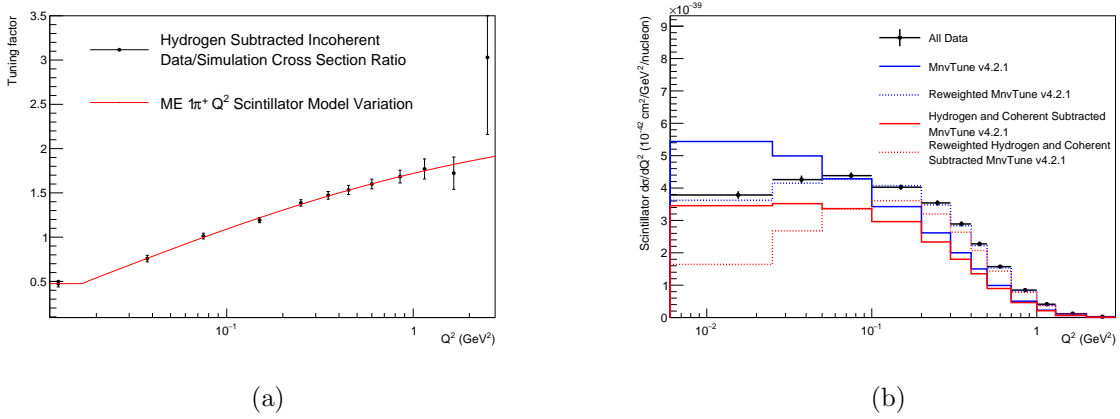

    \centering
    \captionsetup[subfigure]{justification=centering}
    \begin{tabular}{cc}
    \subfloat[][\firstRefereeStrike{Hydrogen Subtracted Incoherent Data/Simulation Cross Section Ratio and the $Q^2$ weighting function (FHC nu-e constraint)}]{\includegraphics[width=0.45\textwidth,page=10]{otherFigures/PlotLowQ2Fit_NewCoh_DataMCRatio.pdf}} &
    \hspace{1.5em}
    \subfloat[][\firstRefereeStrike{Unweighted and Reweighted Total and Hydrogen Subtracted Incoherent Simulated Cross Section Compared to Data Cross Section}
    ]
    {\includegraphics[width=0.45\textwidth,page=11]{otherFigures/PlotLowQ2Fit_NewCoh_DataMCRatio.pdf}} 
    \end{tabular}
    \firstRefereeAdd{\caption{(a) Hydrogen Subtracted Incoherent Data/Simulation Cross Section Ratio and the $Q^2$ weighting function (FHC nu-e constraint).  (b) Unweighted and Reweighted Total and Hydrogen Subtracted Incoherent Simulated Cross Section Compared to Data Cross Section.}}
    \label{fig:LowQ2Function}
\end{figure}
\pagebreak
\clearpage

\subsection{Sideband scale factors}
\label{section:SBTune}
\FloatBarrier
\setlength\tabcolsep{0.5em}
\begin{table}
\centering
\renewcommand{\arraystretch}{1.15}
\normalsize
\begin{tabular}{c|c|c|c}
Target & $1.4<W_\text{exp,true}<1.8$ GeV/$c^2$ & $1.8<W_\text{exp,true}$ GeV/$c^2$ & $\chi^2$/ndf\\
\hline
Carbon & $0.58 \pm 0.44$ & $1.71 \pm 0.71$ & 0.147/4\\
\hline
Water  & $1.53 \pm 0.30$ & $0.00 \pm 0.90$ & 18.956/4\\
\hline
Iron   & $0.48 \pm 0.25$ & $1.05 \pm 0.29$ & 1.705/4\\
\hline
Lead   & $0.09 \pm 0.16$ & $1.06 \pm 0.20$ & 0.324/4\\
\hline
Scintillator & $0.58 \pm 0.02$ & $0.80 \pm 0.02$ & 3.192/4
\end{tabular}
\caption{$W_{exp}$ sideband scales found from the fit in the high $W>1.4$ GeV sideband for all targets.   }
  \label{tbl:W_Sideband_Scales}
\end{table}

\begin{figure}
    \centering
    \begin{tabular}{cc}
      \subfloat[][Scintillator]{\includegraphics[width=\oneplot,page=1]{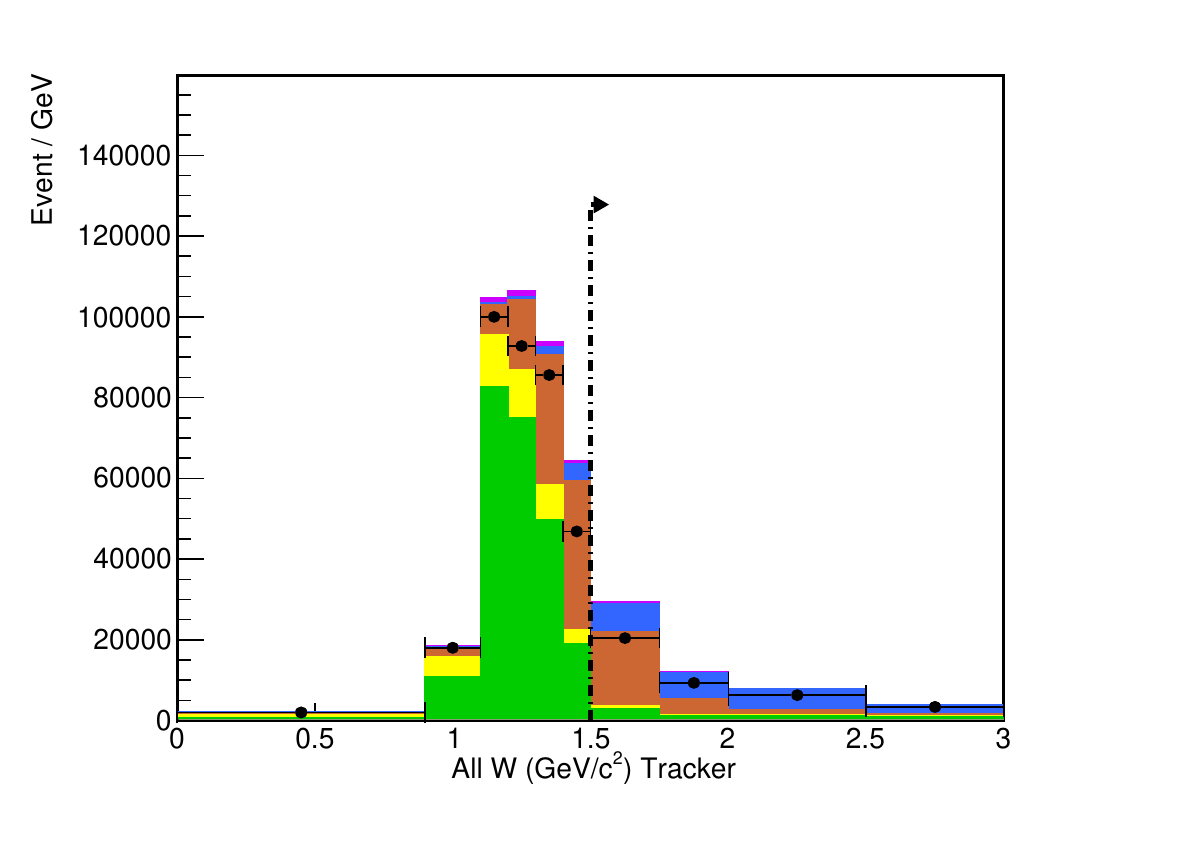}}&
      \includegraphics[height=0.3\textheight,page=1]{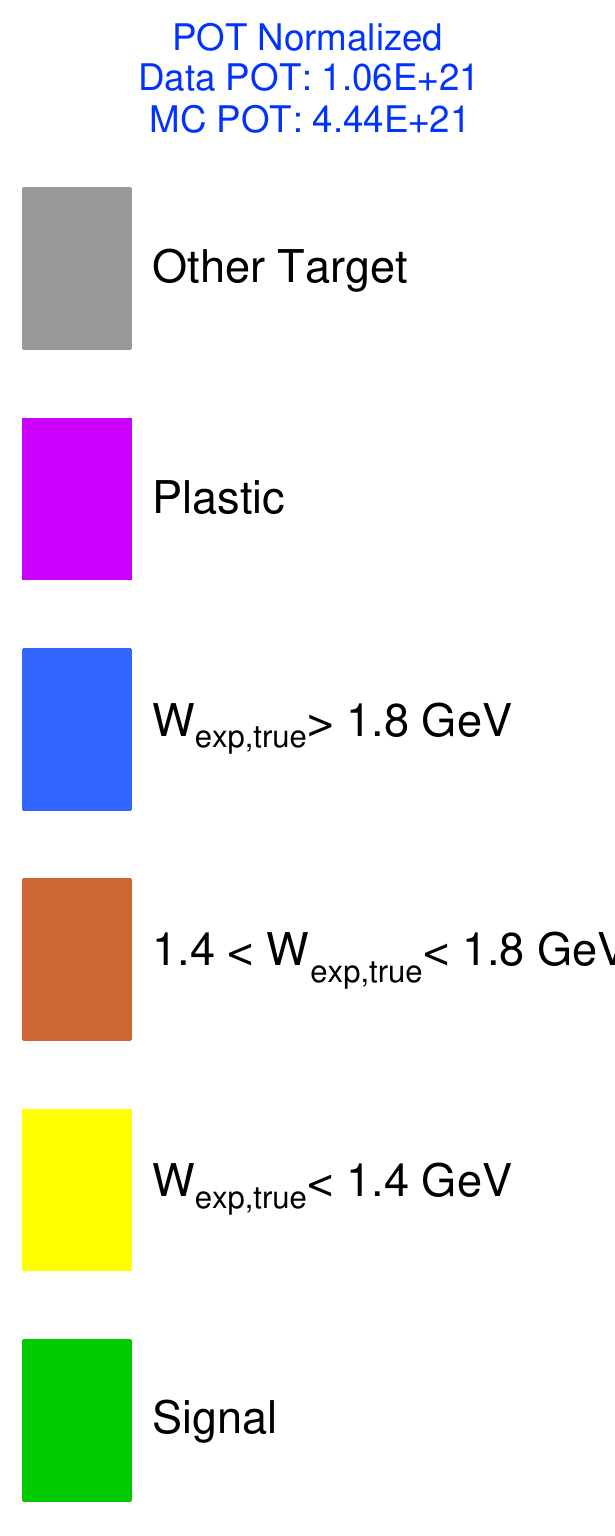}\\
      \subfloat[][Carbon]{\includegraphics[width=\oneplot,page=1]{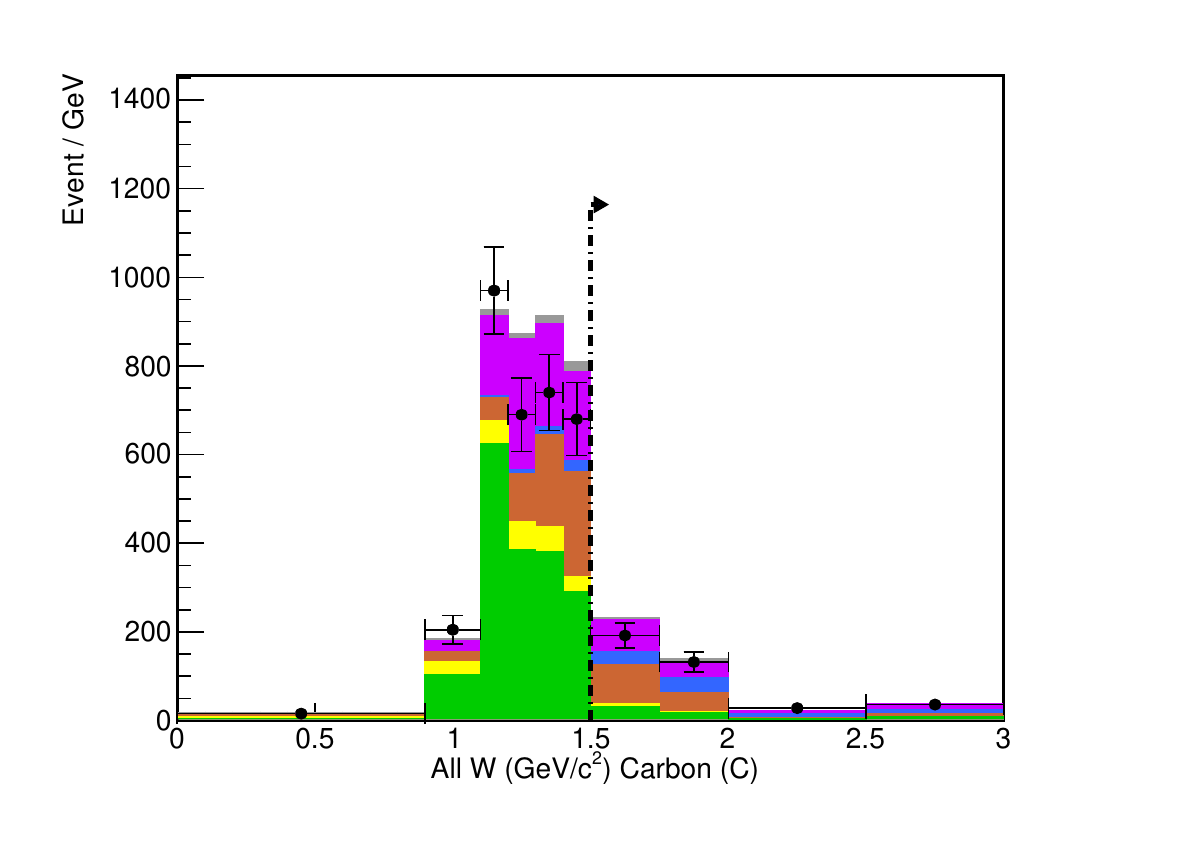}}&
      \subfloat[][Water]{\includegraphics[width=\oneplot,page=1]{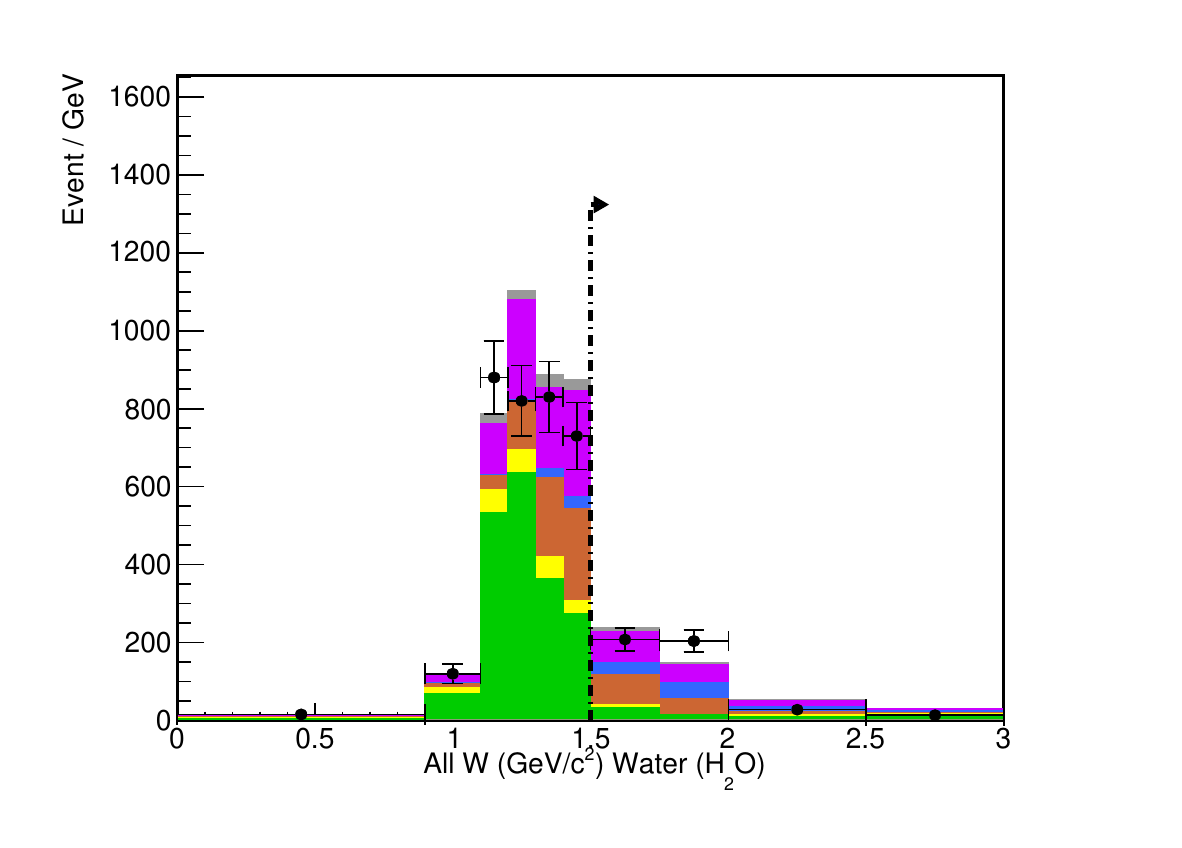}}\\
      \subfloat[][Iron]{\includegraphics[width=\oneplot,page=1]{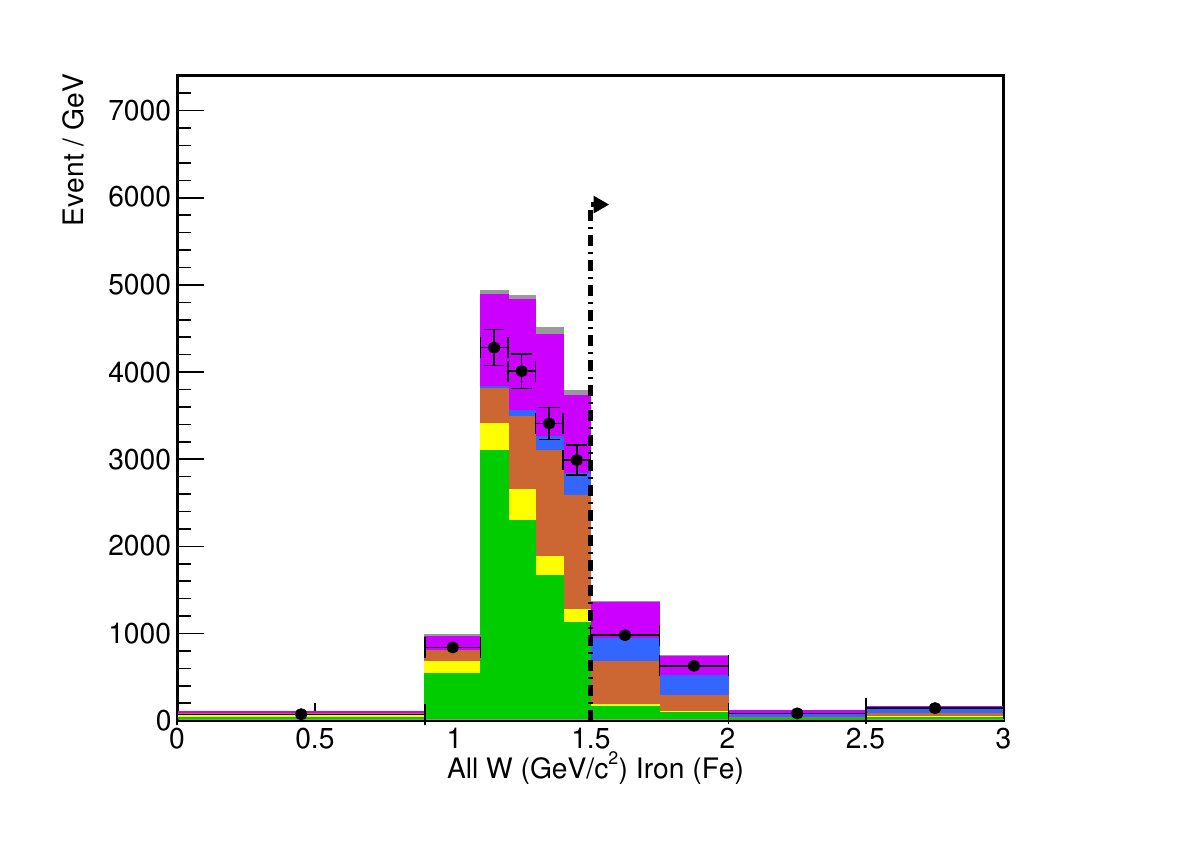}}&
      \subfloat[][Lead]{\includegraphics[width=\oneplot,page=1]{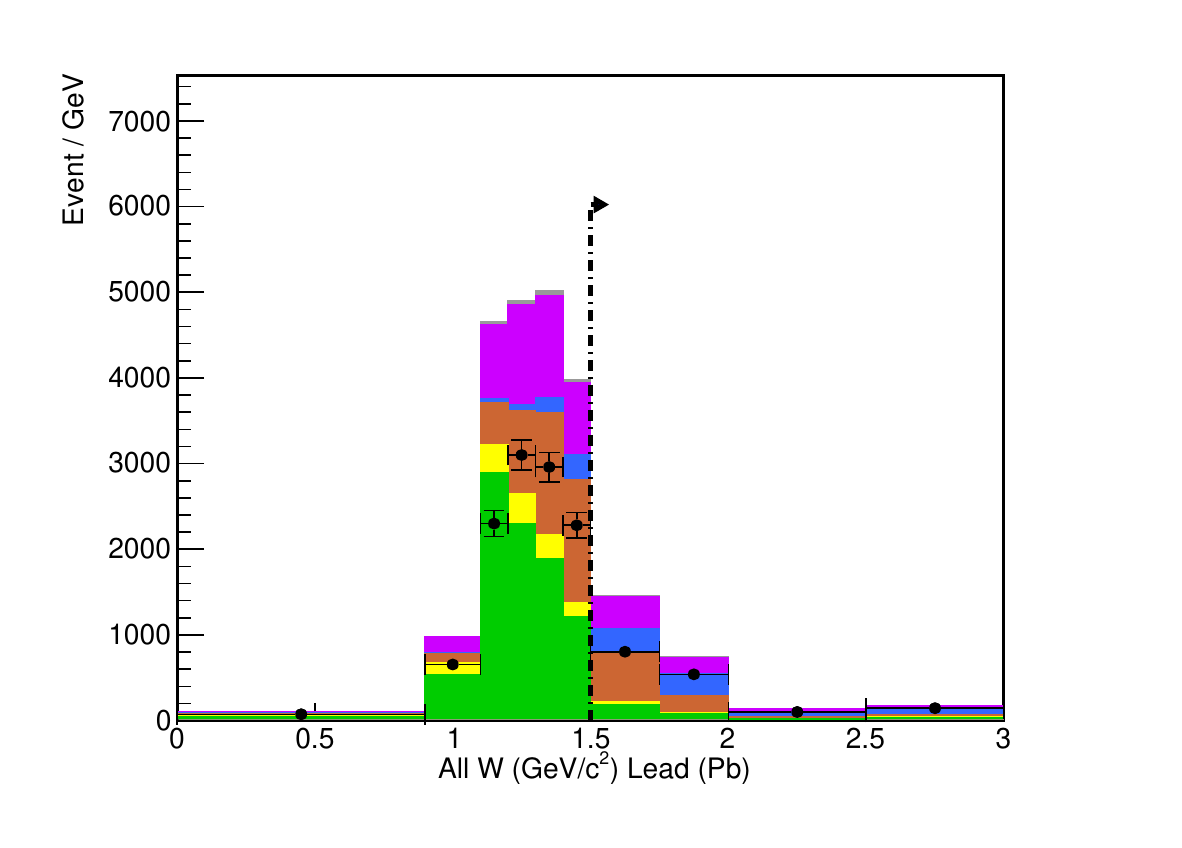}}
     \end{tabular}
    \caption{$W_{exp}$ distribution of selected events before constraining the simulation using background estimates from data.  The dashed arrow delineates the high $W_{exp}$ sideband region. }    
    \label{fig:WSBBeforeFit}
\end{figure}

\begin{figure}
    \centering
    \begin{tabular}{cc}
      \subfloat[][Scintillator]{\includegraphics[width=\oneplot,page=1]{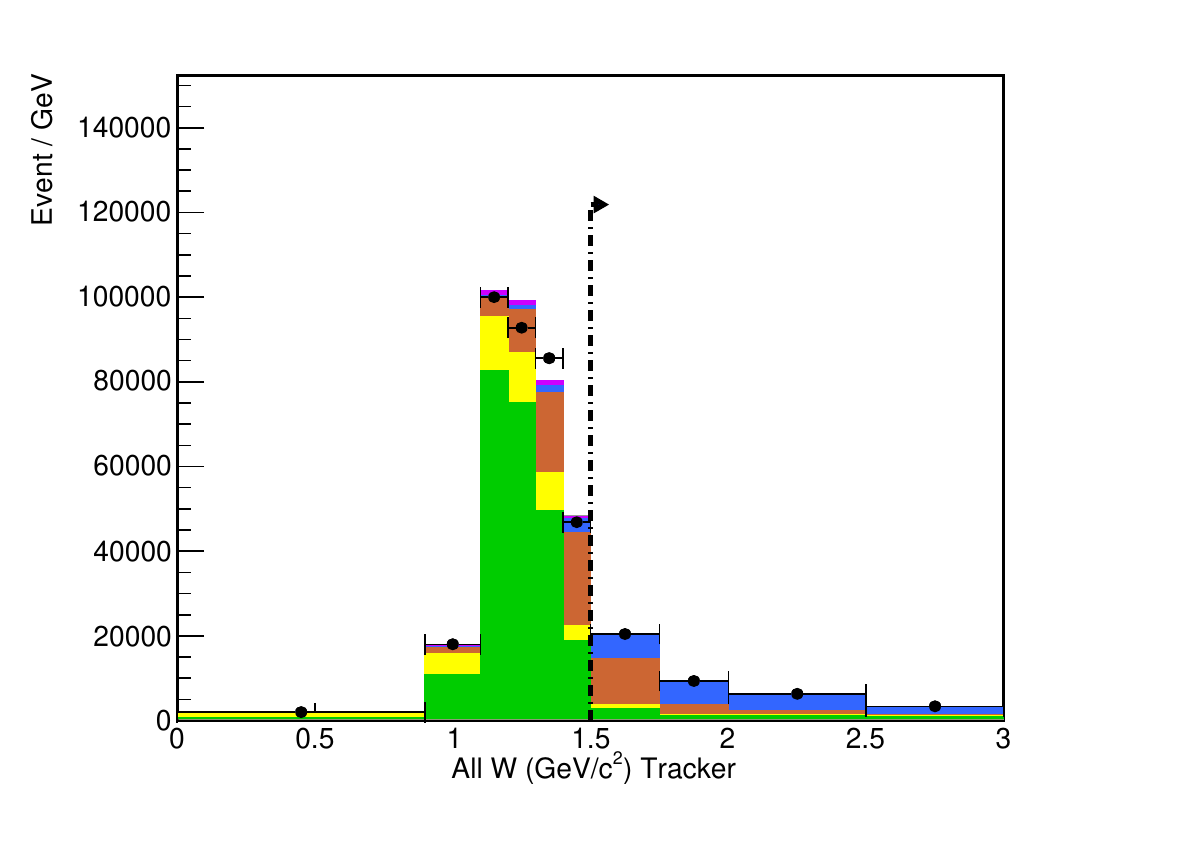}}&
      \includegraphics[height=0.3\textheight,page=1]{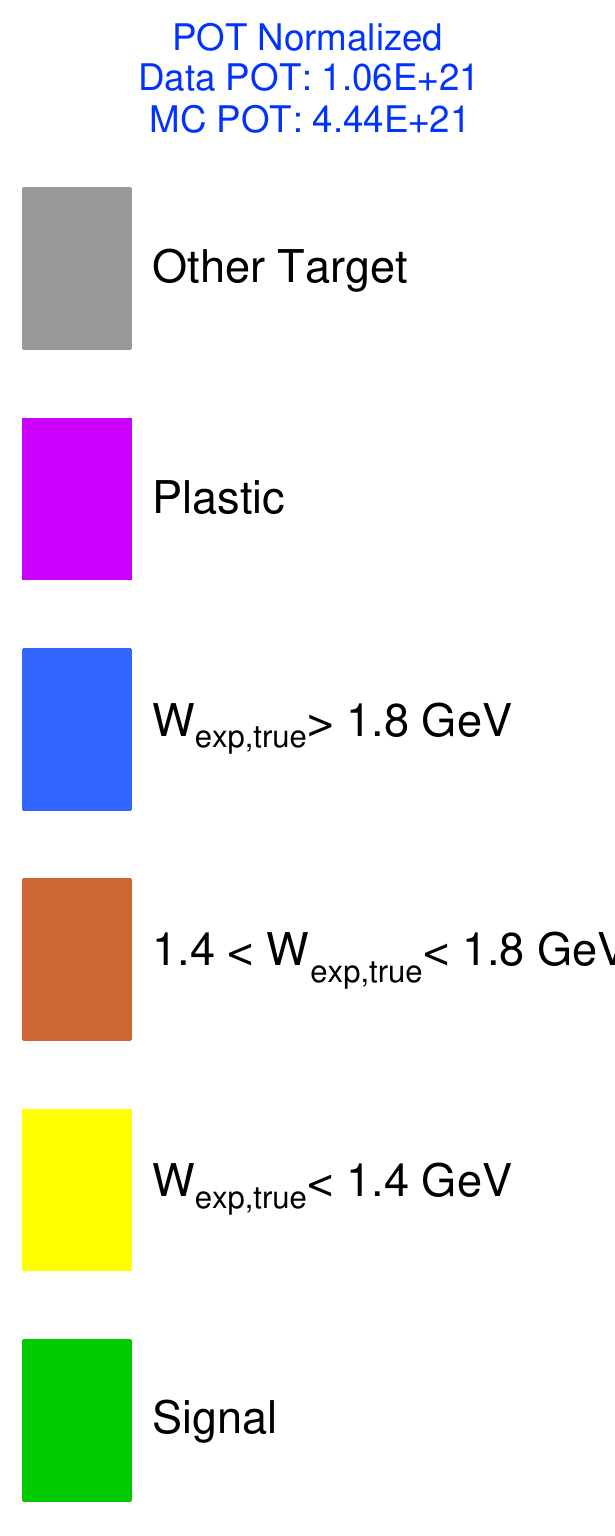}\\
      \subfloat[][Carbon]{\includegraphics[width=\oneplot,page=1]{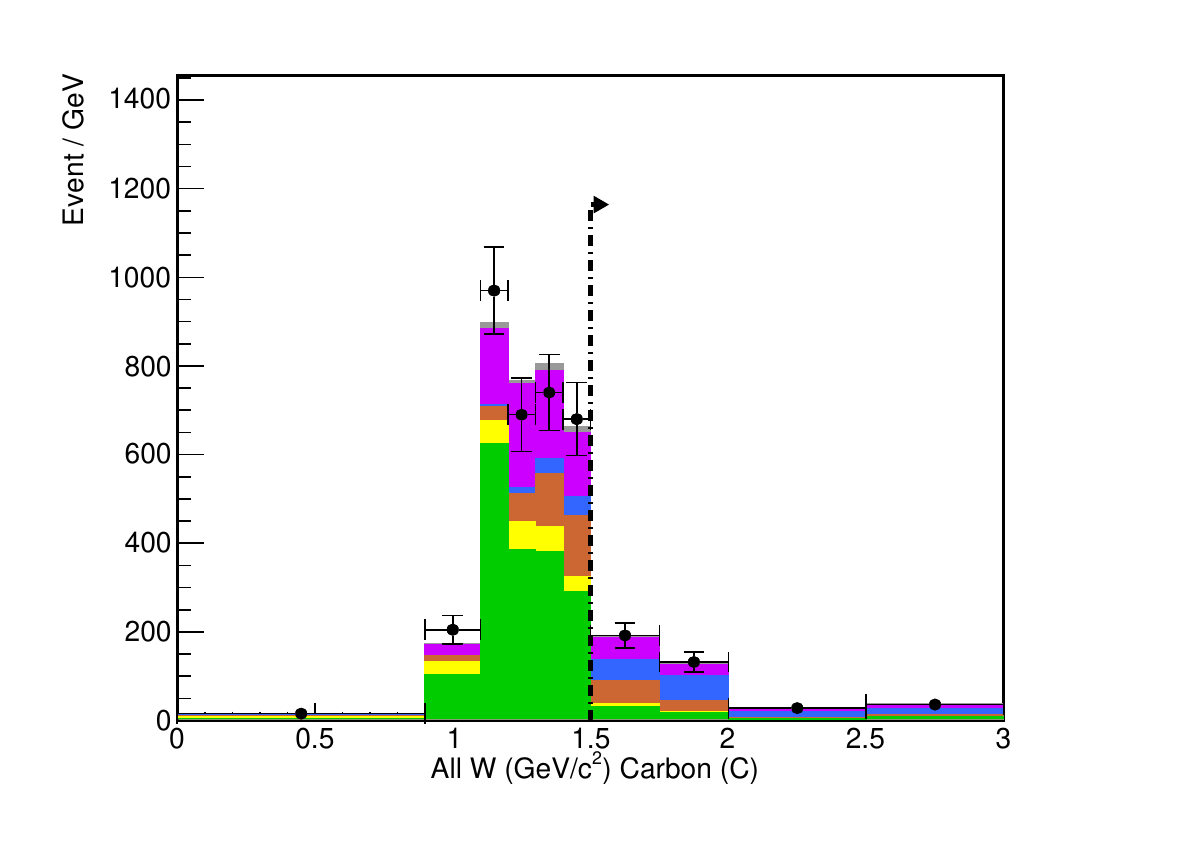}}&
      \subfloat[][Water]{\includegraphics[width=\oneplot,page=1]{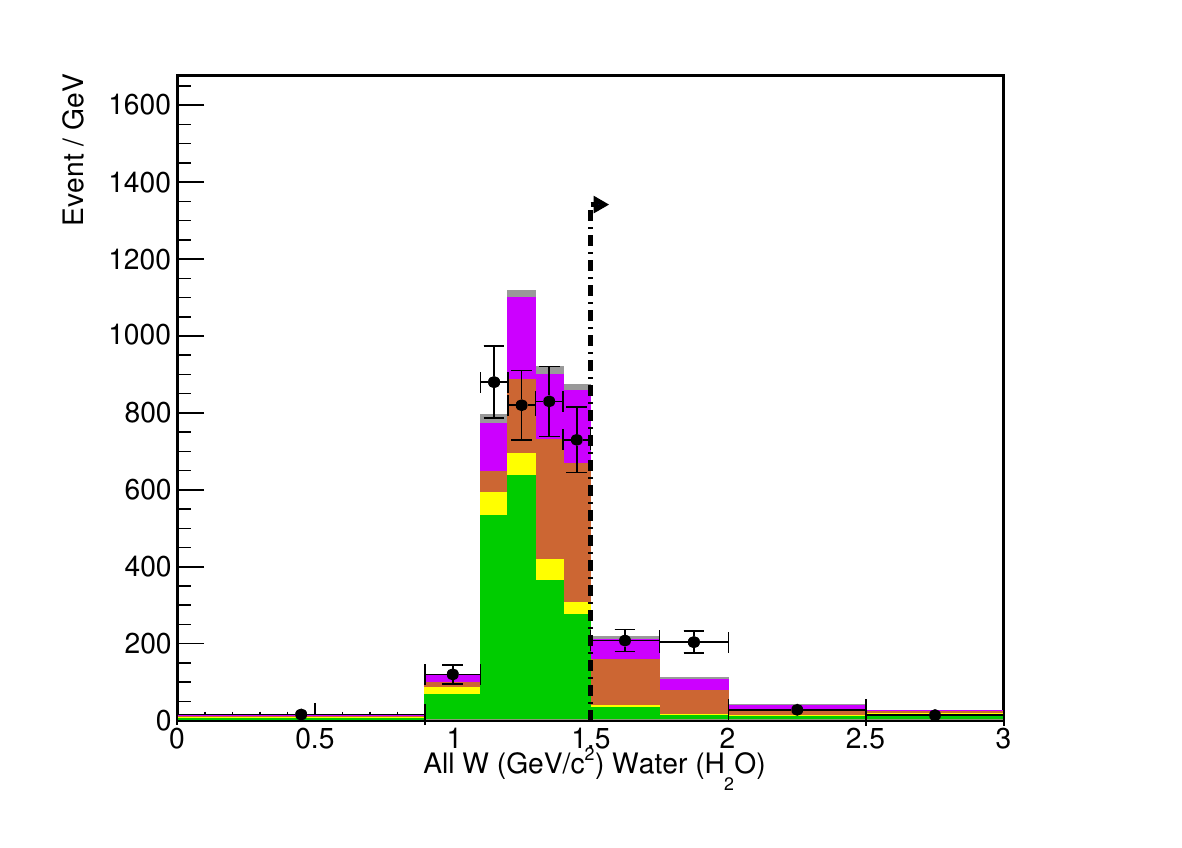}}\\
      \subfloat[][Iron]{\includegraphics[width=\oneplot,page=1]{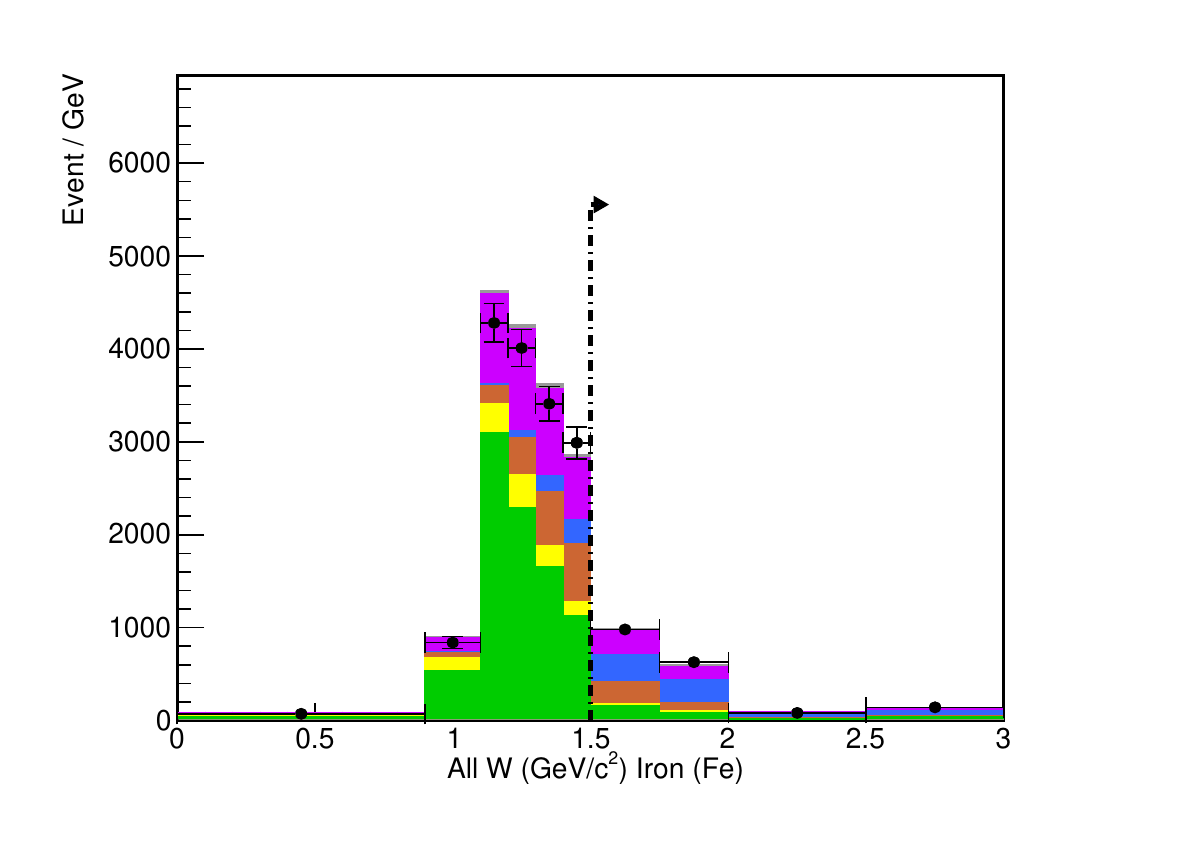}}&
      \subfloat[][Lead]{\includegraphics[width=\oneplot,page=1]{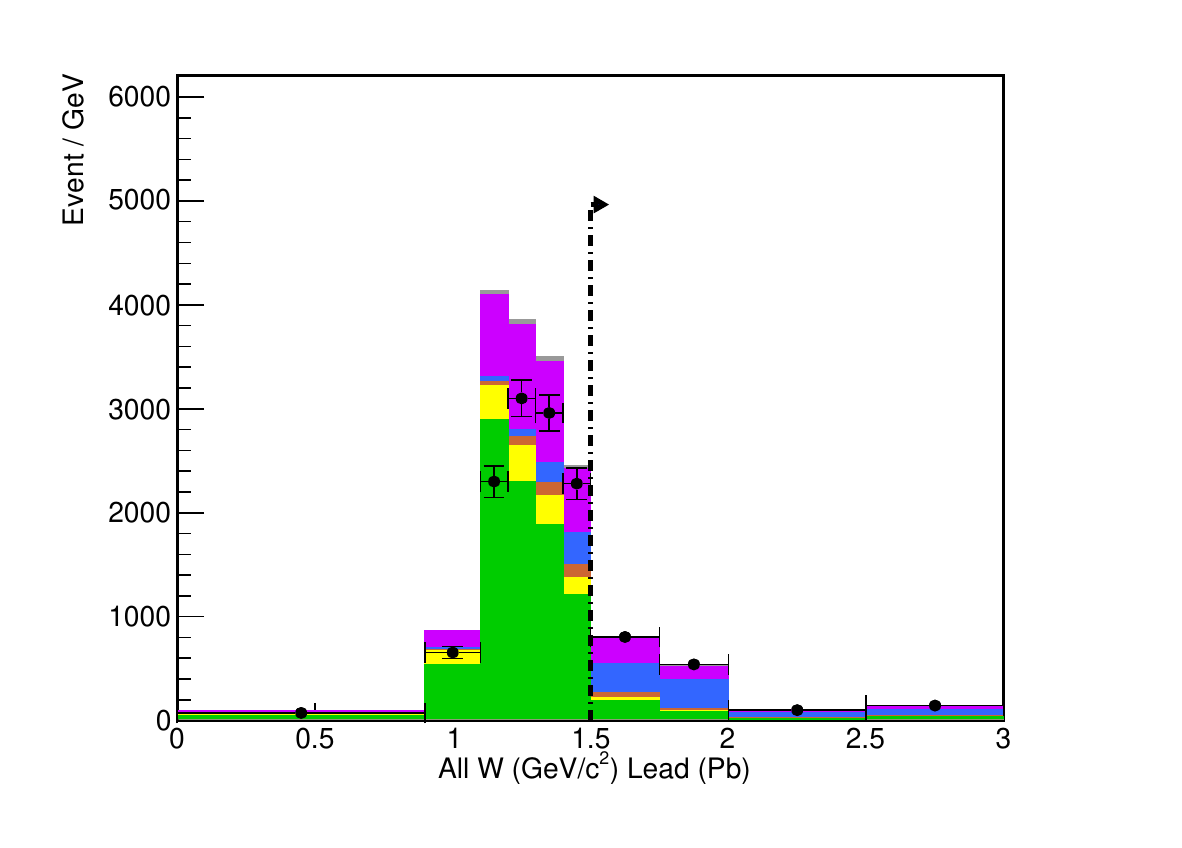}}
     \end{tabular}
    \caption{$W_{exp}$ distribution of selected events after constraining the simulation using background estimates from data.  The dashed arrow delineates the high $W_{exp}$ sideband region. }    
    \label{fig:WSBAfterFit}
\end{figure}

\begin{table}
\centering
\renewcommand{\arraystretch}{1.15}
\normalsize
\begin{tabular}{c|c|c|c|c}
Target & Upstream Scale & Upstream $\chi^2$/ndf & Downstream Scale & Downstream $\chi^2$/ndf\\
\hline
Target 1 & $0.69 \pm 0.13$ & 1.45/6  & $0.93 \pm 0.07$ & 2.76/6\\
\hline
Target 2 & $0.99 \pm 0.05$ & 3.90/7  & $0.88 \pm 0.07$ & 12.00/6\\
\hline
Target 3 & $0.92 \pm 0.04$ & 6.93/10 & $0.88 \pm 0.06$ & 4.03/7\\
\hline
Target 4 & $0.92 \pm 0.06$ & 4.18/5  & $0.83 \pm 0.05$ & 13.70/7\\
\hline
Target 5 & $0.92 \pm 0.05$ & 9.89/7  & $0.86 \pm 0.05$ & 11.89/7
\end{tabular}
\caption{Upstream and downstream scintillator target sideband scales for all nuclear targets}
  \label{tbl:Plas_Sideband_Scales}
\end{table}

\pagebreak
\clearpage
\subsection{Coherent Pion Production Weights}
\label{section:CohWeight}
\FloatBarrier
\begin{figure}
    \centering
    \includegraphics[width=0.9\textwidth,page=1]{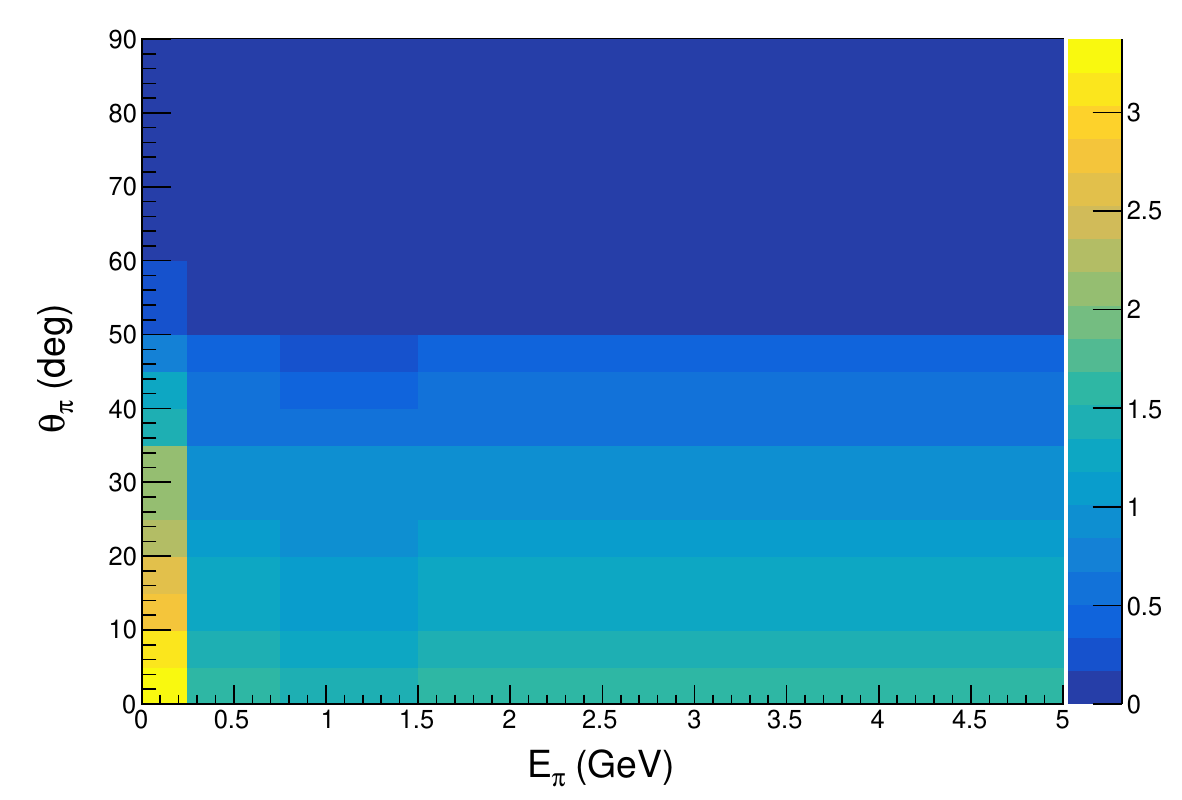}
    \caption{Plot of weights applied to coherent pion production events, given the kinematics of the coherent $\pi^+$, derived from ME MINERvA data~\cite{AlexCoherent}}
    \label{fig:CohWeightHist}
\end{figure}
\begin{table}
\centering
\renewcommand{\arraystretch}{1.15}
\begin{tabular}{c||c|c|c|c|c|c|c|c}
$E_\pi$ (GeV) & 0.0-0.25 & 0.25-0.5 & 0.5-0.75 & 0.75-1.0 & 1.0-1.5 & 1.5-2.0 & 2.0-3.0 & 3.0-5.0\\\cline{1-1}
$\theta_\pi$ (deg) &  & & & & & & \\
\hline
\hline
0-5& 3.372& 1.593& 1.593& 1.369& 1.360& 1.593& 1.593& 1.593\\
\hline
5-10& 3.083& 1.456& 1.456& 1.251& 1.243& 1.456& 1.456& 1.456\\
\hline
10-15& 2.742& 1.296& 1.296& 1.113& 1.106& 1.296& 1.296& 1.296\\
\hline
15-20& 2.577& 1.218& 1.218& 1.046& 1.039& 1.218& 1.218& 1.218\\
\hline
20-25& 2.345& 1.108& 1.108& 0.952& 0.945& 1.108& 1.108& 1.108\\
\hline
25-30& 2.116& 1.000& 1.000& 0.859& 0.853& 1.000& 1.000& 1.000\\
\hline
30-35& 2.116& 1.000& 1.000& 0.859& 0.853& 1.000& 1.000& 1.000\\
\hline
35-40& 1.422& 0.672& 0.672& 0.577& 0.573& 0.672& 0.672& 0.672\\
\hline
40-45& 1.238& 0.585& 0.585& 0.503& 0.499& 0.585& 0.585& 0.585\\
\hline
45-50& 0.713& 0.337& 0.337& 0.289& 0.288& 0.337& 0.337& 0.337\\
\hline
50-60& 0.263& 0.124& 0.124& 0.107& 0.106& 0.124& 0.124& 0.124\\
\hline
60-70& 0.049& 0.023& 0.023& 0.020& 0.020& 0.023& 0.023& 0.023\\
\hline
70-80& 0.049& 0.023& 0.023& 0.020& 0.020& 0.023& 0.023& 0.023\\
\hline
80-90& 0.049& 0.023& 0.023& 0.020& 0.020& 0.023& 0.023& 0.023\\
\end{tabular}
\caption{Tabulation of weights applied to coherent pion production events, given the kinematics of the coherent $\pi^+$, derived from ME MINERvA data~\cite{AlexCoherent}}
\end{table}

\pagebreak
\clearpage

\firstReviewAdd{
\subsection{Efficiencies}
\label{section:Efficiencies}
\FloatBarrier
\begin{figure}
    \centering
    \captionsetup[subfigure]{justification=centering}
    \begin{tabular}{cc}
    \subfloat[][Efficiency versus muon momentum]{\includegraphics[width=\oneplot,page=\pagePmu]{Efficiencies/MergeEffMig_MENU1PI_pi_channel_MinervaME1ABCDEFGLMNOP_MC_Tar_Grid.pdf}} &
    \hspace{1.5em}\subfloat[][Efficiency versus muon angle]{\includegraphics[width=\oneplot,page=\pageThmu]{Efficiencies/MergeEffMig_MENU1PI_pi_channel_MinervaME1ABCDEFGLMNOP_MC_Tar_Grid.pdf}}\\[4em]
    \hspace{1.5em}\subfloat[][Efficiency versus longitudinal muon momentum]{\includegraphics[width=\oneplot,page=\pagePmuZ]{Efficiencies/MergeEffMig_MENU1PI_pi_channel_MinervaME1ABCDEFGLMNOP_MC_Tar_Grid.pdf}}&
    \subfloat[][Efficiency versus transverse muon momentum]{\includegraphics[width=\oneplot,page=\pagePmuT]{Efficiencies/MergeEffMig_MENU1PI_pi_channel_MinervaME1ABCDEFGLMNOP_MC_Tar_Grid.pdf}}
    \end{tabular}
    \caption{Efficiencies for scintillator (top left), carbon (middle left), water (middle right), iron (bottom left) and lead (bottom right).  The shaded error bands show total uncertainties, including systematics and simulated statistics.  }
    \label{fig:SuppEfficiency_muon}
\end{figure}
\pagebreak
\clearpage
\begin{figure}
    \centering
    \captionsetup[subfigure]{justification=centering}
    \begin{tabular}{cc}
    \subfloat[][Efficiency versus $Q^2$]{\includegraphics[width=\oneplot,page=\pageQ]{Efficiencies/MergeEffMig_MENU1PI_pi_channel_MinervaME1ABCDEFGLMNOP_MC_Tar_Grid.pdf}} &
    \hspace{1.5em}\subfloat[][Efficiency versus $W$]{\includegraphics[width=\oneplot,page=\pageW]{Efficiencies/MergeEffMig_MENU1PI_pi_channel_MinervaME1ABCDEFGLMNOP_MC_Tar_Grid.pdf}}\\[4em]
    \hspace{1.5em}\subfloat[][Efficiency versus pion kinetic energy]{\includegraphics[width=\oneplot,page=\pageTpi]{Efficiencies/MergeEffMig_MENU1PI_pi_channel_MinervaME1ABCDEFGLMNOP_MC_Tar_Grid.pdf}}&
    \subfloat[][Efficiency versus pion angle]{\includegraphics[width=\oneplot,page=\pageThpi]{Efficiencies/MergeEffMig_MENU1PI_pi_channel_MinervaME1ABCDEFGLMNOP_MC_Tar_Grid.pdf}}
    \end{tabular}
    \caption{Efficiencies for scintillator (top left), carbon (middle left), water (middle right), iron (bottom left) and lead (bottom right).  The shaded error bands show total uncertainties, including systematics and simulated statistics.  }
    \label{fig:SuppEfficiency_other}
\end{figure}
\pagebreak
\clearpage
}

\subsection{Absolute Cross Section Plots}
\label{section:AbsXsec}
\FloatBarrier
\begin{figure}
    \centering
    \captionsetup[subfigure]{justification=centering}
    \begin{tabular}{cc}
    \subfloat[][Differential cross section $\frac{d\sigma}{dp_\mu}$ for \firstRefereeAdd{scintillator,} carbon, water, iron, and lead.]{\includegraphics[width=\oneplot,page=\pagePmu]{XsecAbs/diffXsec_new_jeffrey_flux_NewCoh_Nucleon_MinosMatched_pi_channel_MinervaME1ABCDEFGLMNOP_Data_Tar_Grid.pdf}} &
    \hspace{1.5em}\subfloat[][Differential cross section $\frac{d\sigma}{d\theta_\mu}$ for \firstRefereeAdd{scintillator,} carbon, water, iron, and lead.]{\includegraphics[width=\oneplot,page=\pageThmu]{XsecAbs/diffXsec_new_jeffrey_flux_NewCoh_Nucleon_MinosMatched_pi_channel_MinervaME1ABCDEFGLMNOP_Data_Tar_Grid.pdf}}\\[4em]
    \hspace{1.5em}\subfloat[][Differential cross section $\frac{d\sigma}{dp_{\mu,||}}$ for \firstRefereeAdd{scintillator,} carbon, water, iron, and lead.]{\includegraphics[width=\oneplot,page=\pagePmuZ]{XsecAbs/diffXsec_new_jeffrey_flux_NewCoh_Nucleon_MinosMatched_pi_channel_MinervaME1ABCDEFGLMNOP_Data_Tar_Grid.pdf}}&
    \subfloat[][Differential cross section $\frac{d\sigma}{dp_{\mu,T}}$ for \firstRefereeAdd{scintillator,} carbon, water, iron, and lead.]{\includegraphics[width=\oneplot,page=\pagePmuT]{XsecAbs/diffXsec_new_jeffrey_flux_NewCoh_Nucleon_MinosMatched_pi_channel_MinervaME1ABCDEFGLMNOP_Data_Tar_Grid.pdf}}
    \end{tabular}
           \caption{Differential cross sections versus muon variables.  }
    \label{fig:SuppAbsoluteMu}
\end{figure}
\pagebreak
\clearpage
\FloatBarrier
\begin{figure}
    \centering
    \captionsetup[subfigure]{justification=centering}
    \begin{tabular}{cc}
    \subfloat[][Differential cross section $\frac{d\sigma}{dQ^2}$ for \firstRefereeAdd{scintillator,} carbon, water, iron, and lead.]{\includegraphics[width=\oneplot,page=\pageQ]{XsecAbs/diffXsec_new_jeffrey_flux_NewCoh_Nucleon_MinosMatched_pi_channel_MinervaME1ABCDEFGLMNOP_Data_Tar_Grid.pdf}} &
    \hspace{1.5em}\subfloat[][Differential cross section $\frac{d\sigma}{dW_{exp}}$ for \firstRefereeAdd{scintillator,} carbon, water, iron, and lead.]{\includegraphics[width=\oneplot,page=\pageW]{XsecAbs/diffXsec_new_jeffrey_flux_NewCoh_Nucleon_MinosMatched_pi_channel_MinervaME1ABCDEFGLMNOP_Data_Tar_Grid.pdf}}\\[4em]
    \subfloat[][Differential cross section $\frac{d\sigma}{dT_\pi}$ for \firstRefereeAdd{scintillator,} carbon, water, iron, and lead.]{\includegraphics[width=\oneplot,page=\pageTpi]{XsecAbs/diffXsec_new_jeffrey_flux_NewCoh_Nucleon_MinosMatched_pi_channel_MinervaME1ABCDEFGLMNOP_Data_Tar_Grid.pdf}}&
    \hspace{1.5em}\subfloat[][Differential cross section $\frac{d\sigma}{d\theta_\pi}$ for \firstRefereeAdd{scintillator,} carbon, water, iron, and lead.]{\includegraphics[width=\oneplot,page=\pageThpi]{XsecAbs/diffXsec_new_jeffrey_flux_NewCoh_Nucleon_MinosMatched_pi_channel_MinervaME1ABCDEFGLMNOP_Data_Tar_Grid.pdf}}
    \end{tabular}
        \caption{Differential cross sections versus other variables.  }
    \label{fig:SuppAbsoluteRest}
\end{figure}
\pagebreak
\clearpage

\subsection{Absolute Cross Section Error Summaries}
\label{section:AbsXsecError}
\FloatBarrier
\begin{figure}
    \centering
    \captionsetup[subfigure]{justification=centering}
    \begin{tabular}{cc}
    \subfloat[][Fractional uncertainties of $\frac{d\sigma}{dp_\mu}$.]{\includegraphics[width=\oneplot,page=\pagePmu]{XsecAbs/diffXsec_new_jeffrey_flux_NewCoh_MENU1PI_Nucleon_MinosMatched_pi_channel_MinervaME1ABCDEFGLMNOP_Error_Grid.pdf}} &
    \hspace{1.5em}\subfloat[][Fractional uncertainties of $\frac{d\sigma}{d\theta_\mu}$.]{\includegraphics[width=\oneplot,page=\pageThmu]{XsecAbs/diffXsec_new_jeffrey_flux_NewCoh_MENU1PI_Nucleon_MinosMatched_pi_channel_MinervaME1ABCDEFGLMNOP_Error_Grid.pdf}}\\[4em]
    \hspace{1.5em}\subfloat[][Fractional uncertainties of $\frac{d\sigma}{dp_{\mu,||}}$.]{\includegraphics[width=\oneplot,page=\pagePmuZ]{XsecAbs/diffXsec_new_jeffrey_flux_NewCoh_MENU1PI_Nucleon_MinosMatched_pi_channel_MinervaME1ABCDEFGLMNOP_Error_Grid.pdf}}&
    \subfloat[][Fractional uncertainties of $\frac{d\sigma}{dp_{\mu,T}}$ .]{\includegraphics[width=\oneplot,page=\pagePmuT]{XsecAbs/diffXsec_new_jeffrey_flux_NewCoh_MENU1PI_Nucleon_MinosMatched_pi_channel_MinervaME1ABCDEFGLMNOP_Error_Grid.pdf}}
    \end{tabular}
    \caption{Systematic uncertainties of differential cross sections for scintillator (top left), carbon (middle left), water (middle right), iron (bottom left) and lead (bottom right)}
    \label{fig:SuppAbsoluteMuError}
\end{figure}
\pagebreak
\clearpage
\FloatBarrier
\begin{figure}
    \centering
    \captionsetup[subfigure]{justification=centering}
    \begin{tabular}{cc}
    \subfloat[][Fractional uncertainties of $\frac{d\sigma}{dQ^2}$.]{\includegraphics[width=\oneplot,page=\pageQ]{XsecAbs/diffXsec_new_jeffrey_flux_NewCoh_MENU1PI_Nucleon_MinosMatched_pi_channel_MinervaME1ABCDEFGLMNOP_Error_Grid.pdf}} &
    \hspace{1.5em}\subfloat[][Fractional Uncertainties $\frac{d\sigma}{dW_{exp}}$.]{\includegraphics[width=\oneplot,page=\pageW]{XsecAbs/diffXsec_new_jeffrey_flux_NewCoh_MENU1PI_Nucleon_MinosMatched_pi_channel_MinervaME1ABCDEFGLMNOP_Error_Grid.pdf}}\\[4em]
    \subfloat[][Fractional Uncertainties $\frac{d\sigma}{dT_\pi}$.]{\includegraphics[width=\oneplot,page=\pageTpi]{XsecAbs/diffXsec_new_jeffrey_flux_NewCoh_MENU1PI_Nucleon_MinosMatched_pi_channel_MinervaME1ABCDEFGLMNOP_Error_Grid.pdf}}&
    \hspace{1.5em}\subfloat[][Fractional Uncertainties $\frac{d\sigma}{d\theta_\pi}$.]{\includegraphics[width=\oneplot,page=\pageThpi]{XsecAbs/diffXsec_new_jeffrey_flux_NewCoh_MENU1PI_Nucleon_MinosMatched_pi_channel_MinervaME1ABCDEFGLMNOP_Error_Grid.pdf}}
    \end{tabular}
    \caption{Systematic uncertainties of differential cross sections for scintillator (top left), carbon (middle left), water (middle right), iron (bottom left) and lead (bottom right)}
    \label{fig:SuppAbsoluteRestError}
\end{figure}
\pagebreak
\clearpage

\subsection{Cross Section Ratio Plots}
\label{section:RatioXsec}
\FloatBarrier

\secondReviewAdd{One of the striking features in these comparisons is that there are large difference in the ratios between the GENIEv2 and GENIEv3 models for the iron and lead targets.  These are primarily from different the FSI effects in the generators. The increased pion scattering and absorption rates in heavy nuclei in GENIEv3 appear to better describe our data.}

\begin{figure}
    \centering
    \captionsetup[subfigure]{justification=centering}
    \begin{tabular}{cc}
    \subfloat[][Cross section ratios $\left(\frac{d\sigma_A}{dp_\mu}\right)/\left(\frac{d\sigma_{CH}}{dp_\mu}\right)$ for carbon, water, iron, and lead.]{\includegraphics[width=\oneplot,page=\pagePmu]{XsecRatios/ratioDiffXsec_new_jeffrey_flux_NewCoh_Nucleon_MinosMatched_pi_channel_MinervaME1ABCDEFGLMNOP_Data_Tar_Grid.pdf}} &
    \hspace{1.5em}\subfloat[][Cross section ratios $\left(\frac{d\sigma_A}{d\theta_\mu}\right)/\left(\frac{d\sigma_{CH}}{d\theta_\mu}\right)$ for carbon, water, iron, and lead.]{\includegraphics[width=\oneplot,page=\pageThmu]{XsecRatios/ratioDiffXsec_new_jeffrey_flux_NewCoh_Nucleon_MinosMatched_pi_channel_MinervaME1ABCDEFGLMNOP_Data_Tar_Grid.pdf}}\\[7em]
    \hspace{1.5em}\subfloat[][Cross section ratios $\left(\frac{d\sigma_A}{dp_{\mu,||}}\right)/\left(\frac{d\sigma_{CH}}{dp_{\mu,||}}\right)$ for carbon, water, iron, and lead.]{\includegraphics[width=\oneplot,page=\pagePmuZ]{XsecRatios/ratioDiffXsec_new_jeffrey_flux_NewCoh_Nucleon_MinosMatched_pi_channel_MinervaME1ABCDEFGLMNOP_Data_Tar_Grid.pdf}}&
    \subfloat[][Cross section ratios $\left(\frac{d\sigma_A}{dp_{\mu,T}}\right)/\left(\frac{d\sigma_{CH}}{dp_{\mu,T}}\right)$ for carbon, water, iron, and lead.]{\includegraphics[width=\oneplot,page=\pagePmuT]{XsecRatios/ratioDiffXsec_new_jeffrey_flux_NewCoh_Nucleon_MinosMatched_pi_channel_MinervaME1ABCDEFGLMNOP_Data_Tar_Grid.pdf}}
    \end{tabular}
        \caption{Cross section ratios versus muon variables.  }
    \label{fig:SuppRatioMu}
\end{figure}
\pagebreak
\clearpage
\FloatBarrier
\begin{figure}
    \centering
    \captionsetup[subfigure]{justification=centering}
    \begin{tabular}{cc}
    \subfloat[][Cross section ratios $\left(\frac{d\sigma_A}{dQ^2}\right)/\left(\frac{d\sigma_{CH}}{dQ^2}\right)$ for carbon, water, iron, and lead.]{\includegraphics[width=\oneplot,page=\pageQ]{XsecRatios/ratioDiffXsec_new_jeffrey_flux_NewCoh_Nucleon_MinosMatched_pi_channel_MinervaME1ABCDEFGLMNOP_Data_Tar_Grid.pdf}} &
    \hspace{1.5em}\subfloat[][Cross section ratios $\left(\frac{d\sigma_A}{dW_{exp}}\right)/\left(\frac{d\sigma_{CH}}{dW_{exp}}\right)$ for carbon, water, iron, and lead.]{\includegraphics[width=\oneplot,page=\pageW]{XsecRatios/ratioDiffXsec_new_jeffrey_flux_NewCoh_Nucleon_MinosMatched_pi_channel_MinervaME1ABCDEFGLMNOP_Data_Tar_Grid.pdf}}\\[7em]
    \subfloat[][Cross section ratios $\left(\frac{d\sigma_A}{dT_\pi}\right)/\left(\frac{d\sigma_{CH}}{dT_\pi}\right)$ for carbon, water, iron, and lead.]{\includegraphics[width=\oneplot,page=\pageTpi]{XsecRatios/ratioDiffXsec_new_jeffrey_flux_NewCoh_Nucleon_MinosMatched_pi_channel_MinervaME1ABCDEFGLMNOP_Data_Tar_Grid.pdf}}&
    \hspace{1.5em}\subfloat[][Cross section ratios $\left(\frac{d\sigma_A}{d\theta_\pi}\right)/\left(\frac{d\sigma_{CH}}{d\theta_\pi}\right)$ for carbon, water, iron, and lead.]{\includegraphics[width=\oneplot,page=\pageThpi]{XsecRatios/ratioDiffXsec_new_jeffrey_flux_NewCoh_Nucleon_MinosMatched_pi_channel_MinervaME1ABCDEFGLMNOP_Data_Tar_Grid.pdf}}
    \end{tabular}
    \caption{Cross section ratios versus other variables.  }
    
    \label{fig:SuppRatioRest}
\end{figure}
\pagebreak
\clearpage
\FloatBarrier
\subsection{Cross Section Ratio Error Summaries}
\label{section:RatioXsecError}
\begin{figure}
    \centering
    \captionsetup[subfigure]{justification=centering}
    \begin{tabular}{cc}
    \subfloat[][Fractional uncertainties $\left(\frac{d\sigma_A}{dp_\mu}\right)/\left(\frac{d\sigma_{CH}}{dp_\mu}\right)$.]{\includegraphics[width=\oneplot,page=\pagePmu]{XsecRatios/ratioDiffXsec_new_jeffrey_flux_NewCoh_MENU1PI_Nucleon_MinosMatched_pi_channel_MinervaME1ABCDEFGLMNOP_Error_Grid.pdf}} &
    \hspace{1.5em}\subfloat[][Fractional uncertainties $\left(\frac{d\sigma_A}{d\theta_\mu}\right)/\left(\frac{d\sigma_{CH}}{d\theta_\mu}\right)$.]{\includegraphics[width=\oneplot,page=\pageThmu]{XsecRatios/ratioDiffXsec_new_jeffrey_flux_NewCoh_MENU1PI_Nucleon_MinosMatched_pi_channel_MinervaME1ABCDEFGLMNOP_Error_Grid.pdf}}\\[7em]
    \hspace{1.5em}\subfloat[][Fractional uncertainties $\left(\frac{d\sigma_A}{dp_{\mu,||}}\right)/\left(\frac{d\sigma_{CH}}{dp_{\mu,||}}\right)$.]{\includegraphics[width=\oneplot,page=\pagePmuZ]{XsecRatios/ratioDiffXsec_new_jeffrey_flux_NewCoh_MENU1PI_Nucleon_MinosMatched_pi_channel_MinervaME1ABCDEFGLMNOP_Error_Grid.pdf}}&
    \subfloat[][Fractional uncertainties $\left(\frac{d\sigma_A}{dp_{\mu,T}}\right)/\left(\frac{d\sigma_{CH}}{dp_{\mu,T}}\right)$.]{\includegraphics[width=\oneplot,page=\pagePmuT]{XsecRatios/ratioDiffXsec_new_jeffrey_flux_NewCoh_MENU1PI_Nucleon_MinosMatched_pi_channel_MinervaME1ABCDEFGLMNOP_Error_Grid.pdf}}
    \end{tabular}
    \caption{Systematic uncertainties of cross section ratios for carbon/scintillator (top left), water/scintillator (top right), iron/scintillator (bottom left) and lead/scintillator (bottom right)}
    \label{fig:SuppRatioMuError}
\end{figure}
\pagebreak
\clearpage
\FloatBarrier
\begin{figure}
    \centering
    \captionsetup[subfigure]{justification=centering}
    \begin{tabular}{cc}
    \subfloat[][Fractional uncertainties $\left(\frac{d\sigma_A}{dQ^2}\right)/\left(\frac{d\sigma_{CH}}{dQ^2}\right)$.]{\includegraphics[width=\oneplot,page=\pageQ]{XsecRatios/ratioDiffXsec_new_jeffrey_flux_NewCoh_MENU1PI_Nucleon_MinosMatched_pi_channel_MinervaME1ABCDEFGLMNOP_Error_Grid.pdf}} &
    \hspace{1.5em}\subfloat[][Fractional uncertainties $\left(\frac{d\sigma_A}{dW_{exp}}\right)/\left(\frac{d\sigma_{CH}}{dW_{exp}}\right)$.]{\includegraphics[width=\oneplot,page=\pageW]{XsecRatios/ratioDiffXsec_new_jeffrey_flux_NewCoh_MENU1PI_Nucleon_MinosMatched_pi_channel_MinervaME1ABCDEFGLMNOP_Error_Grid.pdf}}\\[7em]
    \subfloat[][Fractional uncertainties $\left(\frac{d\sigma_A}{dT_\pi}\right)/\left(\frac{d\sigma_{CH}}{dT_\pi}\right)$.]{\includegraphics[width=\oneplot,page=\pageTpi]{XsecRatios/ratioDiffXsec_new_jeffrey_flux_NewCoh_MENU1PI_Nucleon_MinosMatched_pi_channel_MinervaME1ABCDEFGLMNOP_Error_Grid.pdf}}&
    \hspace{1.5em}\subfloat[][Fractional uncertainties $\left(\frac{d\sigma_A}{d\theta_\pi}\right)/\left(\frac{d\sigma_{CH}}{d\theta_\pi}\right)$.]{\includegraphics[width=\oneplot,page=\pageThpi]{XsecRatios/ratioDiffXsec_new_jeffrey_flux_NewCoh_MENU1PI_Nucleon_MinosMatched_pi_channel_MinervaME1ABCDEFGLMNOP_Error_Grid.pdf}}
    \end{tabular}
    \caption{Systematic uncertainties of cross section ratios for carbon/scintillator (top left), water/scintillator (top right), iron/scintillator (bottom left) and lead/scintillator (bottom right)}
    \label{fig:SuppRatioRestError}
\end{figure}
\pagebreak
\clearpage
\FloatBarrier

\onecolumngrid
\subsection{Cross Section Tables}
\FloatBarrier

\subfile{includeTex/PRL_Table_Var_Xsec}

\subsection{Tables of Ratios of cross sections}
\FloatBarrier
\subfile{includeTex/PRL_Table_Var_Ratio}
\fi 

\end{document}

%% file: includeTex/PRL_Table_Var_Xsec.tex
\setlength\tabcolsep{0.5em}
\begin{table}
  \centering
  \renewcommand{\arraystretch}{1.15}
  \caption{Measured cross section as function of $p_{\mu}$ on scintillator, in units of $10^{-42}$ $\text{cm}^2$/GeV/c/nucleon, and the absolute and fractional cross section uncertainties}
\begin{tabular}{c||c|c||c|c|c}
      Bin edges (GeV/c) &  $\sigma$  &  Abs. Unc.  &  Frac. Stat. Unc.  &  Frac. Sys. Unc.  &  Frac. Flux Unc. \\
      \hline
      \hline
      1.5-3.0 & 132.41 & 9.94 & 0.031 & 0.068 & 0.037\\
      3.0-4.0 & 339.29 & 25.45 & 0.022 & 0.072 & 0.037\\
      4.0-5.0 & 492.23 & 37.09 & 0.019 & 0.073 & 0.037\\
      5.0-6.0 & 521.41 & 39.22 & 0.017 & 0.073 & 0.037\\
      6.0-7.0 & 393.48 & 30.19 & 0.018 & 0.074 & 0.036\\
      7.0-8.5 & 177.43 & 14.31 & 0.022 & 0.078 & 0.035\\
      8.5-10.0 & 53.66 & 4.89 & 0.039 & 0.082 & 0.033\\
      10.0-20.0 & 9.51 & 0.90 & 0.046 & 0.083 & 0.035
    \end{tabular}
  \label{tbl:datanucleonplasticmuon_pa}
\end{table}
\begin{table}
  \centering
  \renewcommand{\arraystretch}{1.15}
  \caption{Statistical covariance matrix of measured cross section as function of $p_{\mu}$ on scintillator, in units of $10^{-84}$ $(\text{cm}^2$/GeV/c/nucleon$)^2$}
  \resizebox{\textwidth}{!}{
    \begin{tabular}{c||cccccccc}
      Bin edges (GeV/c) & 1.5-3.0 & 3.0-4.0 & 4.0-5.0 & 5.0-6.0 & 6.0-7.0 & 7.0-8.5 & 8.5-10.0 & 10.0-20.0\\
      \hline
      \hline
      1.5-3.0 & 16.658 & 4.105 & 0.935 & 0.338 & 0.170 & 0.049 & 0.020 & 0.002\\
      3.0-4.0 & 4.105 & 57.479 & 12.979 & 3.077 & 0.916 & 0.225 & 0.043 & 0.006\\
      4.0-5.0 & 0.935 & 12.979 & 83.056 & 19.009 & 4.249 & 0.707 & 0.145 & 0.016\\
      5.0-6.0 & 0.338 & 3.077 & 19.009 & 76.961 & 17.936 & 2.974 & 0.414 & 0.049\\
      6.0-7.0 & 0.170 & 0.916 & 4.249 & 17.936 & 52.187 & 8.527 & 1.138 & 0.061\\
      7.0-8.5 & 0.049 & 0.225 & 0.707 & 2.974 & 8.527 & 15.067 & 1.952 & 0.104\\
      8.5-10.0 & 0.020 & 0.043 & 0.145 & 0.414 & 1.138 & 1.952 & 4.444 & 0.153\\
      10.0-20.0 & 0.002 & 0.006 & 0.016 & 0.049 & 0.061 & 0.104 & 0.153 & 0.190
    \end{tabular}
  }
  \label{tbl:statnucleonplasticmuon_pb}
\end{table}
\begin{table}
  \centering
  \renewcommand{\arraystretch}{1.15}
  \caption{Systematic covariance matrix of measured cross section in $p_{\mu}$ on scintillator, in units of $10^{-84}$ $(\text{cm}^2$/GeV/c/nucleon$)^2$}
  \resizebox{\textwidth}{!}{
    \begin{tabular}{c||cccccccc}
      Bin edges (GeV/c) & 1.5-3.0 & 3.0-4.0 & 4.0-5.0 & 5.0-6.0 & 6.0-7.0 & 7.0-8.5 & 8.5-10.0 & 10.0-20.0\\
      \hline
      \hline
      1.5-3.0 & 82.125 & 213.956 & 314.097 & 329.205 & 240.409 & 106.454 & 32.919 & 5.696\\
      3.0-4.0 & 213.956 & 590.438 & 854.841 & 906.030 & 662.006 & 293.496 & 94.406 & 15.869\\
      4.0-5.0 & 314.097 & 854.841 & 1292.343 & 1361.955 & 1020.144 & 458.089 & 144.301 & 24.963\\
      5.0-6.0 & 329.205 & 906.030 & 1361.955 & 1461.454 & 1099.429 & 498.336 & 156.016 & 26.910\\
      6.0-7.0 & 240.409 & 662.006 & 1020.144 & 1099.429 & 859.179 & 394.495 & 122.519 & 21.348\\
      7.0-8.5 & 106.454 & 293.496 & 458.089 & 498.336 & 394.495 & 189.709 & 57.596 & 10.355\\
      8.5-10.0 & 32.919 & 94.406 & 144.301 & 156.016 & 122.519 & 57.596 & 19.425 & 3.242\\
      10.0-20.0 & 5.696 & 15.869 & 24.963 & 26.910 & 21.348 & 10.355 & 3.242 & 0.618
    \end{tabular}
  }
  \label{tbl:sysnucleonplasticmuon_pc}
\end{table}
\pagebreak
\begin{turnpage}
\end{turnpage}
\begin{table}
  \centering
  \renewcommand{\arraystretch}{1.15}
  \caption{Measured cross section as function of $\theta_{\mu}$ on scintillator, in units of $10^{-42}$ $\text{cm}^2$/degree/nucleon, and the absolute and fractional cross section uncertainties}
    \begin{tabular}{c||c|c||c|c|c}
      Bin edges (degree) &  $\sigma$  &  Abs. Unc.  &  Frac. Stat. Unc.  &  Frac. Sys. Unc.  &  Frac. Flux Unc. \\
      \hline
      \hline
      0-2 & 93.64 & 6.19 & 0.023 & 0.062 & 0.033\\
      2-3 & 226.67 & 15.25 & 0.025 & 0.063 & 0.035\\
      3-4 & 276.53 & 18.89 & 0.025 & 0.063 & 0.035\\
      4-5 & 278.12 & 22.02 & 0.028 & 0.074 & 0.036\\
      5-6 & 262.12 & 25.46 & 0.031 & 0.092 & 0.038\\
      6-8 & 232.61 & 21.04 & 0.027 & 0.086 & 0.038\\
      8-10 & 167.48 & 16.70 & 0.038 & 0.092 & 0.041\\
      10-13 & 114.14 & 13.91 & 0.047 & 0.112 & 0.046
    \end{tabular}
  \label{tbl:datanucleonplasticmuon_thetaa}
\end{table}
\begin{table}
  \centering
  \renewcommand{\arraystretch}{1.15}
  \caption{Statistical covariance matrix of measured cross section as function of $\theta_{\mu}$ on scintillator, in units of $10^{-84}$ $(\text{cm}^2$/degree/nucleon$)^2$}
  \resizebox{\textwidth}{!}{
    \begin{tabular}{c||cccccccc}
      Bin edges (degree) & 0-2 & 2-3 & 3-4 & 4-5 & 5-6 & 6-8 & 8-10 & 10-13\\
      \hline
      \hline
      0-2 & 4.693 & -0.300 & -0.494 & -0.035 & -0.001 & -0.002 & -0.002 & -0.002\\
      2-3 & -0.300 & 31.566 & 0.489 & -1.641 & -0.173 & -0.013 & -0.015 & -0.013\\
      3-4 & -0.494 & 0.489 & 49.367 & 1.222 & -2.542 & -0.137 & 0.002 & -0.004\\
      4-5 & -0.035 & -1.641 & 1.222 & 60.451 & 2.522 & -1.709 & -0.064 & 0.008\\
      5-6 & -0.001 & -0.173 & -2.542 & 2.522 & 68.064 & -0.152 & -1.074 & -0.021\\
      6-8 & -0.002 & -0.013 & -0.137 & -1.709 & -0.152 & 39.575 & -0.871 & -0.601\\
      8-10 & -0.002 & -0.015 & 0.002 & -0.064 & -1.074 & -0.871 & 39.542 & -1.187\\
      10-13 & -0.002 & -0.013 & -0.004 & 0.008 & -0.021 & -0.601 & -1.187 & 28.649
    \end{tabular}
  }
  \label{tbl:statnucleonplasticmuon_thetab}
\end{table}
\begin{table}
  \centering
  \renewcommand{\arraystretch}{1.15}
  \caption{Systematic covariance matrix of measured cross section in $\theta_{\mu}$ on scintillator, in units of $10^{-84}$ $(\text{cm}^2$/degree/nucleon$)^2$}
  \resizebox{\textwidth}{!}{
    \begin{tabular}{c||cccccccc}
      Bin edges (degree) & 0-2 & 2-3 & 3-4 & 4-5 & 5-6 & 6-8 & 8-10 & 10-13\\
      \hline
      \hline
      0-2 & 33.684 & 78.058 & 87.540 & 87.329 & 86.942 & 68.156 & 50.842 & 37.409\\
      2-3 & 78.058 & 200.937 & 237.301 & 255.413 & 267.730 & 210.228 & 154.423 & 119.171\\
      3-4 & 87.540 & 237.301 & 307.318 & 345.850 & 375.759 & 306.688 & 229.018 & 179.044\\
      4-5 & 87.329 & 255.413 & 345.850 & 424.408 & 479.626 & 386.851 & 290.976 & 229.426\\
      5-6 & 86.942 & 267.730 & 375.759 & 479.626 & 580.386 & 461.821 & 343.697 & 276.712\\
      6-8 & 68.156 & 210.228 & 306.688 & 386.851 & 461.821 & 403.285 & 300.163 & 244.821\\
      8-10 & 50.842 & 154.423 & 229.018 & 290.976 & 343.697 & 300.163 & 239.503 & 193.833\\
      10-13 & 37.409 & 119.171 & 179.044 & 229.426 & 276.712 & 244.821 & 193.833 & 164.717
    \end{tabular}
  }
  \label{tbl:sysnucleonplasticmuon_thetac}
\end{table}
\pagebreak
\begin{turnpage}
\end{turnpage}
\begin{table}
  \centering
  \renewcommand{\arraystretch}{1.15}
  \caption{Measured cross section as function of $p_{\mu,||}$ on scintillator, in units of $10^{-42}$ $\text{cm}^2$/GeV/c/nucleon, and the absolute and fractional cross section uncertainties}
    \begin{tabular}{c||c|c||c|c|c}
      Bin edges (GeV/c) &  $\sigma$  &  Abs. Unc.  &  Frac. Stat. Unc.  &  Frac. Sys. Unc.  &  Frac. Flux Unc. \\
      \hline
      \hline
      1.5-3.0 & 137.27 & 10.42 & 0.030 & 0.070 & 0.037\\
      3.0-4.0 & 347.62 & 25.93 & 0.022 & 0.071 & 0.037\\
      4.0-5.0 & 494.41 & 37.52 & 0.018 & 0.074 & 0.037\\
      5.0-6.0 & 518.20 & 38.78 & 0.017 & 0.073 & 0.037\\
      6.0-7.0 & 386.60 & 29.60 & 0.018 & 0.074 & 0.036\\
      7.0-8.5 & 173.45 & 14.06 & 0.022 & 0.078 & 0.035\\
      8.5-10.0 & 52.71 & 4.76 & 0.040 & 0.081 & 0.033\\
      10.0-20.0 & 9.44 & 0.89 & 0.046 & 0.082 & 0.035
    \end{tabular}
  \label{tbl:datanucleonplasticmuon_pza}
\end{table}
\begin{table}
  \centering
  \renewcommand{\arraystretch}{1.15}
  \caption{Statistical covariance matrix of measured cross section as function of $p_{\mu,||}$ on scintillator, in units of $10^{-84}$ $(\text{cm}^2$/GeV/c/nucleon$)^2$}
  \resizebox{\textwidth}{!}{
    \begin{tabular}{c||cccccccc}
      Bin edges (GeV/c) & 1.5-3.0 & 3.0-4.0 & 4.0-5.0 & 5.0-6.0 & 6.0-7.0 & 7.0-8.5 & 8.5-10.0 & 10.0-20.0\\
      \hline
      \hline
      1.5-3.0 & 17.281 & 4.316 & 0.981 & 0.386 & 0.175 & 0.047 & 0.019 & 0.002\\
      3.0-4.0 & 4.316 & 59.783 & 13.377 & 3.199 & 0.890 & 0.225 & 0.042 & 0.006\\
      4.0-5.0 & 0.981 & 13.377 & 83.368 & 19.478 & 4.159 & 0.688 & 0.142 & 0.016\\
      5.0-6.0 & 0.386 & 3.199 & 19.478 & 76.300 & 17.415 & 2.887 & 0.399 & 0.049\\
      6.0-7.0 & 0.175 & 0.890 & 4.159 & 17.415 & 50.714 & 8.346 & 1.129 & 0.062\\
      7.0-8.5 & 0.047 & 0.225 & 0.688 & 2.887 & 8.346 & 14.498 & 1.871 & 0.101\\
      8.5-10.0 & 0.019 & 0.042 & 0.142 & 0.399 & 1.129 & 1.871 & 4.343 & 0.147\\
      10.0-20.0 & 0.002 & 0.006 & 0.016 & 0.049 & 0.062 & 0.101 & 0.147 & 0.188
    \end{tabular}
  }
  \label{tbl:statnucleonplasticmuon_pzb}
\end{table}
\begin{table}
  \centering
  \renewcommand{\arraystretch}{1.15}
  \caption{Systematic covariance matrix of measured cross section in $p_{\mu,||}$ on scintillator, in units of $10^{-84}$ $(\text{cm}^2$/GeV/c/nucleon$)^2$}
  \resizebox{\textwidth}{!}{
    \begin{tabular}{c||cccccccc}
      Bin edges (GeV/c) & 1.5-3.0 & 3.0-4.0 & 4.0-5.0 & 5.0-6.0 & 6.0-7.0 & 7.0-8.5 & 8.5-10.0 & 10.0-20.0\\
      \hline
      \hline
      1.5-3.0 & 91.226 & 229.520 & 338.349 & 347.709 & 251.637 & 112.813 & 34.549 & 6.067\\
      3.0-4.0 & 229.520 & 612.679 & 877.172 & 909.191 & 657.101 & 291.505 & 93.008 & 15.804\\
      4.0-5.0 & 338.349 & 877.172 & 1324.084 & 1362.526 & 1012.135 & 457.507 & 142.290 & 25.029\\
      5.0-6.0 & 347.709 & 909.191 & 1362.526 & 1427.562 & 1062.163 & 483.211 & 149.666 & 26.138\\
      6.0-7.0 & 251.637 & 657.101 & 1012.135 & 1062.163 & 825.422 & 380.578 & 117.151 & 20.653\\
      7.0-8.5 & 112.813 & 291.505 & 457.507 & 483.211 & 380.578 & 183.242 & 55.337 & 10.013\\
      8.5-10.0 & 34.549 & 93.008 & 142.290 & 149.666 & 117.151 & 55.337 & 18.344 & 3.110\\
      10.0-20.0 & 6.067 & 15.804 & 25.029 & 26.138 & 20.653 & 10.013 & 3.110 & 0.598
    \end{tabular}
  }
  \label{tbl:sysnucleonplasticmuon_pzc}
\end{table}
\pagebreak
\begin{turnpage}
\end{turnpage}
\begin{table}
  \centering
  \renewcommand{\arraystretch}{1.15}
  \caption{Measured cross section as function of $p_{\mu,T}$ on scintillator, in units of $10^{-42}$ $\text{cm}^2$/GeV/c/nucleon, and the absolute and fractional cross section uncertainties}
    \begin{tabular}{c||c|c||c|c|c}
      Bin edges (GeV/c) &  $\sigma$  &  Abs. Unc.  &  Frac. Stat. Unc.  &  Frac. Sys. Unc.  &  Frac. Flux Unc. \\
      \hline
      \hline
      0.00-0.15 & 646.46 & 44.23 & 0.030 & 0.062 & 0.033\\
      0.15-0.30 & 2212.06 & 141.95 & 0.019 & 0.061 & 0.033\\
      0.30-0.45 & 3421.09 & 240.61 & 0.019 & 0.068 & 0.035\\
      0.45-0.60 & 3545.67 & 308.07 & 0.022 & 0.084 & 0.036\\
      0.60-0.75 & 2728.36 & 278.16 & 0.028 & 0.098 & 0.037\\
      0.75-0.90 & 1665.26 & 200.09 & 0.041 & 0.113 & 0.041\\
      0.90-1.25 & 559.10 & 97.31 & 0.065 & 0.161 & 0.045\\
      1.25-2.50 & 19.99 & 5.98 & 0.221 & 0.202 & 0.045
    \end{tabular}
  \label{tbl:datanucleonplasticmuon_pta}
\end{table}
\pagebreak
\begin{turnpage}
\begin{table}
  \centering
  \renewcommand{\arraystretch}{1.15}
  \caption{Statistical covariance matrix of measured cross section as function of $p_{\mu,T}$ on scintillator, in units of $10^{-84}$ $(\text{cm}^2$/GeV/c/nucleon$)^2$}
  \resizebox{\textheight}{!}{
    \begin{tabular}{c||cccccccc}
      Bin edges (GeV/c) & 0.00-0.15 & 0.15-0.30 & 0.30-0.45 & 0.45-0.60 & 0.60-0.75 & 0.75-0.90 & 0.90-1.25 & 1.25-2.50\\
      \hline
      \hline
      0.00-0.15 & 375.058 & -5.551 & -47.201 & -7.308 & -0.821 & -0.902 & -0.344 & -0.030\\
      0.15-0.30 & -5.551 & 1821.647 & 31.115 & -156.758 & -39.058 & -9.338 & -2.217 & -0.043\\
      0.30-0.45 & -47.201 & 31.115 & 4348.247 & 269.040 & -313.708 & -92.462 & -17.984 & -0.914\\
      0.45-0.60 & -7.308 & -156.758 & 269.040 & 5947.956 & 700.721 & -342.839 & -78.672 & -2.713\\
      0.60-0.75 & -0.821 & -39.058 & -313.708 & 700.721 & 5878.758 & 1013.662 & -145.751 & -2.860\\
      0.75-0.90 & -0.902 & -9.338 & -92.462 & -342.839 & 1013.662 & 4679.844 & 465.915 & -9.180\\
      0.90-1.25 & -0.344 & -2.217 & -17.984 & -78.672 & -145.751 & 465.915 & 1335.605 & 34.903\\
      1.25-2.50 & -0.030 & -0.043 & -0.914 & -2.713 & -2.860 & -9.180 & 34.903 & 19.497
    \end{tabular}
  }
  \label{tbl:statnucleonplasticmuon_ptb}
\end{table}
\end{turnpage}
\clearpage
\pagebreak
\begin{turnpage}
\begin{table}
  \centering
  \renewcommand{\arraystretch}{1.15}
  \caption{Systematic covariance matrix of measured cross section in $p_{\mu,T}$ on scintillator, in units of $10^{-84}$ $(\text{cm}^2$/GeV/c/nucleon$)^2$}
  \resizebox{\textheight}{!}{
    \begin{tabular}{c||cccccccc}
      Bin edges (GeV/c) & 0.00-0.15 & 0.15-0.30 & 0.30-0.45 & 0.45-0.60 & 0.60-0.75 & 0.75-0.90 & 0.90-1.25 & 1.25-2.50\\
      \hline
      \hline
      0.00-0.15 & 1581.186 & 5096.119 & 6401.031 & 6050.531 & 4214.151 & 2732.787 & 955.250 & 36.236\\
      0.15-0.30 & 5096.119 & 18327.195 & 26728.571 & 28116.824 & 20742.179 & 13158.005 & 4370.207 & 161.226\\
      0.30-0.45 & 6401.031 & 26728.571 & 53542.567 & 65838.516 & 54164.519 & 34609.724 & 11743.643 & 442.813\\
      0.45-0.60 & 6050.531 & 28116.824 & 65838.516 & 88957.669 & 77342.326 & 50764.557 & 18249.458 & 689.458\\
      0.60-0.75 & 4214.151 & 20742.179 & 54164.519 & 77342.326 & 71495.961 & 48206.936 & 18687.460 & 724.443\\
      0.75-0.90 & 2732.787 & 13158.005 & 34609.724 & 50764.557 & 48206.936 & 35354.928 & 15091.296 & 591.143\\
      0.90-1.25 & 955.250 & 4370.207 & 11743.643 & 18249.458 & 18687.460 & 15091.296 & 8133.526 & 341.419\\
      1.25-2.50 & 36.236 & 161.226 & 442.813 & 689.458 & 724.443 & 591.143 & 341.419 & 16.250
    \end{tabular}
  }
  \label{tbl:sysnucleonplasticmuon_ptc}
\end{table}
\end{turnpage}
\pagebreak
\begin{turnpage}
\end{turnpage}
\begin{table}
  \centering
  \renewcommand{\arraystretch}{1.15}
  \caption{Measured cross section as function of $Q^2$ on scintillator, in units of $10^{-42}$ $\text{cm}^2$/GeV${}^2/c{}^2$/nucleon, and the absolute and fractional cross section uncertainties}
    \begin{tabular}{c||c|c||c|c|c}
      Bin edges (GeV${}^2/c{}^2$) &  $\sigma$  &  Abs. Unc.  &  Frac. Stat. Unc.  &  Frac. Sys. Unc.  &  Frac. Flux Unc. \\
      \hline
      \hline
      0.000-0.025 & 3825.00 & 261.09 & 0.029 & 0.062 & 0.033\\
      0.025-0.050 & 4428.16 & 286.72 & 0.028 & 0.059 & 0.033\\
      0.050-0.100 & 4535.82 & 301.52 & 0.023 & 0.062 & 0.034\\
      0.100-0.200 & 4059.36 & 285.14 & 0.021 & 0.067 & 0.035\\
      0.200-0.300 & 3452.35 & 286.17 & 0.024 & 0.079 & 0.036\\
      0.300-0.400 & 2733.16 & 247.43 & 0.028 & 0.086 & 0.036\\
      0.400-0.500 & 2103.42 & 214.43 & 0.033 & 0.096 & 0.037\\
      0.500-0.700 & 1412.01 & 153.16 & 0.034 & 0.103 & 0.038\\
      0.700-1.000 & 718.94 & 89.74 & 0.046 & 0.116 & 0.041\\
      1.000-1.300 & 331.82 & 54.29 & 0.071 & 0.147 & 0.044\\
      1.300-2.000 & 90.74 & 24.36 & 0.137 & 0.231 & 0.051\\
      2.000-3.000 & 21.03 & 7.04 & 0.278 & 0.186 & 0.039
    \end{tabular}
  \label{tbl:datanucleonplasticq2a}
\end{table}
\pagebreak
\begin{turnpage}
\begin{table}
  \centering
  \renewcommand{\arraystretch}{1.15}
  \caption{Statistical covariance matrix of measured cross section as function of $Q^2$ on scintillator, in units of $10^{-84}$ $(\text{cm}^2$/GeV${}^2/c{}^2$/nucleon$)^2$}
  \resizebox{\textheight}{!}{
    \begin{tabular}{c||cccccccccccc}
      Bin edges (GeV${}^2/c{}^2$) & 0.000-0.025 & 0.025-0.050 & 0.050-0.100 & 0.100-0.200 & 0.200-0.300 & 0.300-0.400 & 0.400-0.500 & 0.500-0.700 & 0.700-1.000 & 1.000-1.300 & 1.300-2.000 & 2.000-3.000\\
      \hline
      \hline
      0.000-0.025 & 12588.365 & 997.907 & -805.851 & -175.415 & -12.014 & -3.840 & -2.954 & -1.013 & -1.935 & -0.723 & -0.531 & -0.175\\
      0.025-0.050 & 997.907 & 14929.823 & 1929.772 & -589.284 & -214.404 & -36.350 & -8.937 & -5.562 & -3.890 & -1.779 & -0.493 & -0.030\\
      0.050-0.100 & -805.851 & 1929.772 & 10965.271 & 674.926 & -508.532 & -190.296 & -62.976 & -16.436 & -3.694 & -3.023 & -1.259 & -0.104\\
      0.100-0.200 & -175.415 & -589.284 & 674.926 & 7118.253 & 814.386 & -378.057 & -250.650 & -76.736 & -19.479 & -5.880 & -2.364 & -0.485\\
      0.200-0.300 & -12.014 & -214.404 & -508.532 & 814.386 & 6771.264 & 1416.831 & -132.953 & -214.847 & -56.635 & -19.106 & -2.782 & -1.670\\
      0.300-0.400 & -3.840 & -36.350 & -190.296 & -378.057 & 1416.831 & 5865.008 & 1736.357 & -24.934 & -128.954 & -45.772 & -7.149 & -2.457\\
      0.400-0.500 & -2.954 & -8.937 & -62.976 & -250.650 & -132.953 & 1736.357 & 4823.638 & 887.183 & -77.544 & -55.772 & -11.482 & 0.465\\
      0.500-0.700 & -1.013 & -5.562 & -16.436 & -76.736 & -214.847 & -24.934 & 887.183 & 2338.753 & 352.481 & -49.685 & -20.726 & -1.956\\
      0.700-1.000 & -1.935 & -3.890 & -3.694 & -19.479 & -56.635 & -128.954 & -77.544 & 352.481 & 1092.546 & 239.344 & 3.234 & -4.504\\
      1.000-1.300 & -0.723 & -1.779 & -3.023 & -5.880 & -19.106 & -45.772 & -55.772 & -49.685 & 239.344 & 562.636 & 91.453 & 3.066\\
      1.300-2.000 & -0.531 & -0.493 & -1.259 & -2.364 & -2.782 & -7.149 & -11.482 & -20.726 & 3.234 & 91.453 & 153.493 & 11.032\\
      2.000-3.000 & -0.175 & -0.030 & -0.104 & -0.485 & -1.670 & -2.457 & 0.465 & -1.956 & -4.504 & 3.066 & 11.032 & 34.274
    \end{tabular}
  }
  \label{tbl:statnucleonplasticq2b}
\end{table}
\end{turnpage}
\clearpage
\pagebreak
\begin{turnpage}
\begin{table}
  \centering
  \renewcommand{\arraystretch}{1.15}
  \caption{Systematic covariance matrix of measured cross section in $Q^2$ on scintillator, in units of $10^{-84}$ $(\text{cm}^2$/GeV${}^2/c{}^2$/nucleon$)^2$}
  \resizebox{\textheight}{!}{
    \begin{tabular}{c||cccccccccccc}
      Bin edges (GeV${}^2/c{}^2$) & 0.000-0.025 & 0.025-0.050 & 0.050-0.100 & 0.100-0.200 & 0.200-0.300 & 0.300-0.400 & 0.400-0.500 & 0.500-0.700 & 0.700-1.000 & 1.000-1.300 & 1.300-2.000 & 2.000-3.000\\
      \hline
      \hline
      0.000-0.025 & 55578.445 & 58860.557 & 60378.311 & 45220.647 & 34618.319 & 26584.728 & 20025.404 & 12986.738 & 6460.467 & 3981.279 & 829.538 & 205.817\\
      0.025-0.050 & 58860.557 & 67278.874 & 71209.690 & 60105.923 & 50535.710 & 39818.233 & 31808.493 & 20735.959 & 10139.678 & 5164.600 & 1487.729 & 261.792\\
      0.050-0.100 & 60378.311 & 71209.690 & 79947.393 & 69924.610 & 60147.726 & 47554.008 & 38176.351 & 25061.644 & 12064.101 & 6143.531 & 1774.373 & 308.808\\
      0.100-0.200 & 45220.647 & 60105.923 & 69924.610 & 74185.188 & 71691.029 & 59199.546 & 49384.624 & 33309.494 & 16489.765 & 7198.661 & 2728.158 & 399.253\\
      0.200-0.300 & 34618.319 & 50535.710 & 60147.726 & 71691.029 & 75124.673 & 63525.200 & 54225.183 & 37359.079 & 19135.220 & 8013.084 & 3474.311 & 468.633\\
      0.300-0.400 & 26584.728 & 39818.233 & 47554.008 & 59199.546 & 63525.200 & 55354.698 & 46951.905 & 33113.395 & 17206.412 & 7264.286 & 3260.826 & 458.856\\
      0.400-0.500 & 20025.404 & 31808.493 & 38176.351 & 49384.624 & 54225.183 & 46951.905 & 41157.215 & 28869.769 & 15207.015 & 6487.562 & 2953.839 & 418.405\\
\end{tabular}
  }
  \label{tbl:sysnucleonplasticq2c}
\end{table}
\end{turnpage}
\pagebreak
\begin{turnpage}
\end{turnpage}
\begin{table}
  \centering
  \renewcommand{\arraystretch}{1.15}
  \caption{Measured cross section as function of $W_{exp}$ on scintillator, in units of $10^{-42}$ $\text{cm}^2$/GeV/c${}^2$/nucleon, and the absolute and fractional cross section uncertainties}
    \begin{tabular}{c||c|c||c|c|c}
      Bin edges (GeV/c${}^2$) &  $\sigma$  &  Abs. Unc.  &  Frac. Stat. Unc.  &  Frac. Sys. Unc.  &  Frac. Flux Unc. \\
      \hline
      \hline
      0.9-1.1 & 380.21 & 89.00 & 0.100 & 0.212 & 0.059\\
      1.1-1.2 & 4969.82 & 394.14 & 0.026 & 0.075 & 0.038\\
      1.2-1.3 & 10893.98 & 791.75 & 0.015 & 0.071 & 0.034\\
      1.3-1.4 & 7534.12 & 786.69 & 0.031 & 0.100 & 0.033
    \end{tabular}
  \label{tbl:datanucleonplasticWa}
\end{table}
\begin{table}
  \centering
  \renewcommand{\arraystretch}{1.15}
  \caption{Statistical covariance matrix of measured cross section as function of $W_{exp}$ on scintillator, in units of $10^{-84}$ $(\text{cm}^2$/GeV/c${}^2$/nucleon$)^2$}
    \begin{tabular}{c||cccc}
      Bin edges (GeV/c${}^2$) & 0.9-1.1 & 1.1-1.2 & 1.2-1.3 & 1.3-1.4\\
      \hline
      \hline
      0.9-1.1 & 1434.912 & 298.236 & -1959.309 & -15.500\\
      1.1-1.2 & 298.236 & 16108.522 & -6199.017 & -16343.752\\
      1.2-1.3 & -1959.309 & -6199.017 & 28235.499 & 17223.717\\
      1.3-1.4 & -15.500 & -16343.752 & 17223.717 & 53971.062
    \end{tabular}
  \label{tbl:statnucleonplasticWb}
\end{table}
\begin{table}
  \centering
  \renewcommand{\arraystretch}{1.15}
  \caption{Systematic covariance matrix of measured cross section in $W_{exp}$ on scintillator, in units of $10^{-84}$ $(\text{cm}^2$/GeV/c${}^2$/nucleon$)^2$}
    \begin{tabular}{c||cccc}
      Bin edges (GeV/c${}^2$) & 0.9-1.1 & 1.1-1.2 & 1.2-1.3 & 1.3-1.4\\
      \hline
      \hline
      0.9-1.1 & 6486.707 & 15615.022 & 35973.498 & 36397.745\\
      1.1-1.2 & 15615.022 & 139234.924 & 211164.417 & 136341.941\\
      1.2-1.3 & 35973.498 & 211164.417 & 598624.922 & 523114.324\\
      1.3-1.4 & 36397.745 & 136341.941 & 523114.324 & 564914.576
    \end{tabular}
  \label{tbl:sysnucleonplasticWc}
\end{table}
\pagebreak
\begin{turnpage}
\end{turnpage}
\begin{table}
  \centering
  \renewcommand{\arraystretch}{1.15}
  \caption{Measured cross section as function of $T_{\pi}$ on scintillator, in units of $10^{-42}$ $\text{cm}^2$/GeV/nucleon, and the absolute and fractional cross section uncertainties}
    \begin{tabular}{c||c|c||c|c|c}
      Bin edges (GeV) &  $\sigma$  &  Abs. Unc.  &  Frac. Stat. Unc.  &  Frac. Sys. Unc.  &  Frac. Flux Unc. \\
      \hline
      \hline
      0.035-0.100 & 13123.56 & 1025.94 & 0.016 & 0.076 & 0.037\\
      0.100-0.150 & 9902.55 & 655.87 & 0.015 & 0.064 & 0.035\\
      0.150-0.200 & 6746.23 & 521.56 & 0.023 & 0.074 & 0.035\\
      0.200-0.350 & 3574.69 & 367.65 & 0.032 & 0.098 & 0.037
    \end{tabular}
  \label{tbl:datanucleonplasticpion_ekina}
\end{table}
\begin{table}
  \centering
  \renewcommand{\arraystretch}{1.15}
  \caption{Statistical covariance matrix of measured cross section as function of $T_{\pi}$ on scintillator, in units of $10^{-84}$ $(\text{cm}^2$/GeV/nucleon$)^2$}
    \begin{tabular}{c||cccc}
      Bin edges (GeV) & 0.035-0.100 & 0.100-0.150 & 0.150-0.200 & 0.200-0.350\\
      \hline
      \hline
      0.035-0.100 & 46118.291 & 870.889 & -2121.179 & -1189.136\\
      0.100-0.150 & 870.889 & 23295.488 & 2478.311 & -458.209\\
      0.150-0.200 & -2121.179 & 2478.311 & 24009.621 & 3296.240\\
      0.200-0.350 & -1189.136 & -458.209 & 3296.240 & 13381.849
    \end{tabular}
  \label{tbl:statnucleonplasticpion_ekinb}
\end{table}
\begin{table}
  \centering
  \renewcommand{\arraystretch}{1.15}
  \caption{Systematic covariance matrix of measured cross section in $T_{\pi}$ on scintillator, in units of $10^{-84}$ $(\text{cm}^2$/GeV/nucleon$)^2$}
    \begin{tabular}{c||cccc}
      Bin edges (GeV) & 0.035-0.100 & 0.100-0.150 & 0.150-0.200 & 0.200-0.350\\
      \hline
      \hline
      0.035-0.100 & 1006425.993 & 600004.017 & 464834.038 & 298853.945\\
      0.100-0.150 & 600004.017 & 406873.221 & 299782.750 & 190420.988\\
      0.150-0.200 & 464834.038 & 299782.750 & 248015.149 & 165499.617\\
      0.200-0.350 & 298853.945 & 190420.988 & 165499.617 & 121786.467
    \end{tabular}
  \label{tbl:sysnucleonplasticpion_ekinc}
\end{table}
\pagebreak
\begin{turnpage}
\end{turnpage}
\begin{table}
  \centering
  \renewcommand{\arraystretch}{1.15}
  \caption{Measured cross section as function of $\theta_{\pi}$ on scintillator, in units of $10^{-42}$ $\text{cm}^2$/degree/nucleon, and the absolute and fractional cross section uncertainties}
    \begin{tabular}{c||c|c||c|c|c}
      Bin edges (degree) &  $\sigma$  &  Abs. Unc.  &  Frac. Stat. Unc.  &  Frac. Sys. Unc.  &  Frac. Flux Unc. \\
      \hline
      \hline
      0-10 & 3.46 & 0.45 & 0.079 & 0.102 & 0.039\\
      10-20 & 10.35 & 1.05 & 0.041 & 0.092 & 0.037\\
      20-30 & 15.83 & 1.50 & 0.032 & 0.089 & 0.036\\
      30-40 & 19.13 & 1.65 & 0.027 & 0.082 & 0.037\\
      40-50 & 21.54 & 1.81 & 0.026 & 0.080 & 0.038\\
      50-60 & 23.20 & 1.83 & 0.026 & 0.074 & 0.037\\
      60-120 & 19.20 & 1.53 & 0.019 & 0.077 & 0.035\\
      120-135 & 9.30 & 0.69 & 0.027 & 0.069 & 0.033\\
      135-150 & 6.21 & 0.46 & 0.031 & 0.068 & 0.034\\
      150-165 & 3.20 & 0.30 & 0.049 & 0.079 & 0.035\\
      165-180 & 0.78 & 0.10 & 0.110 & 0.074 & 0.037
    \end{tabular}
  \label{tbl:datanucleonplasticpion_thetaa}
\end{table}
\pagebreak
\begin{turnpage}
\begin{table}
  \centering
  \renewcommand{\arraystretch}{1.15}
  \caption{Statistical covariance matrix of measured cross section as function of $\theta_{\pi}$ on scintillator, in units of $10^{-84}$ $(\text{cm}^2$/degree/nucleon$)^2$}
  \resizebox{\textheight}{!}{
    \begin{tabular}{c||ccccccccccc}
      Bin edges (degree) & 0-10 & 10-20 & 20-30 & 30-40 & 40-50 & 50-60 & 60-120 & 120-135 & 135-150 & 150-165 & 165-180\\
      \hline
      \hline
      0-10 & 0.074 & 0.022 & -0.002 & -0.003 & -0.001 & -0.001 & -0.000 & -0.001 & -0.000 & -0.000 & 0.005\\
      10-20 & 0.022 & 0.183 & 0.029 & -0.007 & -0.004 & -0.004 & -0.003 & -0.002 & -0.002 & 0.002 & 0.005\\
      20-30 & -0.002 & 0.029 & 0.250 & 0.034 & -0.009 & -0.006 & -0.003 & -0.004 & -0.003 & 0.005 & -0.000\\
      30-40 & -0.003 & -0.007 & 0.034 & 0.276 & 0.040 & -0.011 & -0.005 & -0.006 & 0.002 & 0.002 & -0.000\\
      40-50 & -0.001 & -0.004 & -0.009 & 0.040 & 0.323 & 0.044 & -0.006 & -0.004 & 0.002 & -0.003 & -0.001\\
      50-60 & -0.001 & -0.004 & -0.006 & -0.011 & 0.044 & 0.376 & 0.020 & 0.002 & -0.006 & -0.002 & -0.000\\
      60-120 & -0.000 & -0.003 & -0.003 & -0.005 & -0.006 & 0.020 & 0.133 & 0.007 & -0.002 & 0.000 & 0.000\\
      120-135 & -0.001 & -0.002 & -0.004 & -0.006 & -0.004 & 0.002 & 0.007 & 0.064 & 0.009 & -0.001 & 0.000\\
      135-150 & -0.000 & -0.002 & -0.003 & 0.002 & 0.002 & -0.006 & -0.002 & 0.009 & 0.036 & 0.008 & 0.001\\
      150-165 & -0.000 & 0.002 & 0.005 & 0.002 & -0.003 & -0.002 & 0.000 & -0.001 & 0.008 & 0.024 & 0.001\\
      165-180 & 0.005 & 0.005 & -0.000 & -0.000 & -0.001 & -0.000 & 0.000 & 0.000 & 0.001 & 0.001 & 0.007
    \end{tabular}
  }
  \label{tbl:statnucleonplasticpion_thetab}
\end{table}
\end{turnpage}
\clearpage
\pagebreak
\begin{turnpage}
\begin{table}
  \centering
  \renewcommand{\arraystretch}{1.15}
  \caption{Systematic covariance matrix of measured cross section in $\theta_{\pi}$ on scintillator, in units of $10^{-84}$ $(\text{cm}^2$/degree/nucleon$)^2$}
  \resizebox{\textheight}{!}{
    \begin{tabular}{c||ccccccccccc}
      Bin edges (degree) & 0-10 & 10-20 & 20-30 & 30-40 & 40-50 & 50-60 & 60-120 & 120-135 & 135-150 & 150-165 & 165-180\\
      \hline
      \hline
      0-10 & 0.125 & 0.325 & 0.470 & 0.510 & 0.522 & 0.494 & 0.301 & 0.146 & 0.097 & 0.054 & 0.014\\
      10-20 & 0.325 & 0.913 & 1.275 & 1.384 & 1.399 & 1.314 & 0.860 & 0.434 & 0.280 & 0.161 & 0.038\\
      20-30 & 0.470 & 1.275 & 1.989 & 2.183 & 2.306 & 2.214 & 1.631 & 0.653 & 0.431 & 0.258 & 0.059\\
      30-40 & 0.510 & 1.384 & 2.183 & 2.450 & 2.624 & 2.532 & 1.914 & 0.750 & 0.500 & 0.299 & 0.071\\
      40-50 & 0.522 & 1.399 & 2.306 & 2.624 & 2.936 & 2.888 & 2.220 & 0.856 & 0.575 & 0.326 & 0.078\\
      50-60 & 0.494 & 1.314 & 2.214 & 2.532 & 2.888 & 2.972 & 2.272 & 0.903 & 0.616 & 0.349 & 0.079\\
      60-120 & 0.301 & 0.860 & 1.631 & 1.914 & 2.220 & 2.272 & 2.193 & 0.675 & 0.463 & 0.301 & 0.059\\
      120-135 & 0.146 & 0.434 & 0.653 & 0.750 & 0.856 & 0.903 & 0.675 & 0.407 & 0.261 & 0.118 & 0.025\\
      135-150 & 0.097 & 0.280 & 0.431 & 0.500 & 0.575 & 0.616 & 0.463 & 0.261 & 0.176 & 0.085 & 0.018\\
      150-165 & 0.054 & 0.161 & 0.258 & 0.299 & 0.326 & 0.349 & 0.301 & 0.118 & 0.085 & 0.064 & 0.012\\
      165-180 & 0.014 & 0.038 & 0.059 & 0.071 & 0.078 & 0.079 & 0.059 & 0.025 & 0.018 & 0.012 & 0.003
    \end{tabular}
  }
  \label{tbl:sysnucleonplasticpion_thetac}
\end{table}
\end{turnpage}
\pagebreak
\begin{turnpage}
\end{turnpage}
\begin{table}
  \centering
  \renewcommand{\arraystretch}{1.15}
  \caption{Measured cross section as function of $p_{\mu}$ on carbon, in units of $10^{-42}$ $\text{cm}^2$/GeV/c/nucleon, and the absolute and fractional cross section uncertainties}
    \begin{tabular}{c||c|c||c|c|c}
      Bin edges (GeV/c) &  $\sigma$  &  Abs. Unc.  &  Frac. Stat. Unc.  &  Frac. Sys. Unc.  &  Frac. Flux Unc. \\
      \hline
      \hline
      1.5-3.0 & 143.43 & 53.59 & 0.348 & 0.137 & 0.038\\
      3.0-4.0 & 400.66 & 96.71 & 0.207 & 0.124 & 0.038\\
      4.0-5.0 & 489.84 & 119.50 & 0.183 & 0.162 & 0.039\\
      5.0-6.0 & 489.52 & 115.11 & 0.170 & 0.163 & 0.037\\
      6.0-7.0 & 310.21 & 84.33 & 0.194 & 0.191 & 0.037\\
      7.0-8.5 & 135.12 & 44.89 & 0.257 & 0.210 & 0.038\\
      8.5-10.0 & 36.97 & 16.68 & 0.364 & 0.267 & 0.041\\
      10.0-20.0 & 11.32 & 5.78 & 0.446 & 0.247 & 0.038
    \end{tabular}
  \label{tbl:datanucleonnuc6muon_pa}
\end{table}
\begin{table}
  \centering
  \renewcommand{\arraystretch}{1.15}
  \caption{Statistical covariance matrix of measured cross section as function of $p_{\mu}$ on carbon, in units of $10^{-84}$ $(\text{cm}^2$/GeV/c/nucleon$)^2$}
  \resizebox{\textwidth}{!}{
    \begin{tabular}{c||cccccccc}
      Bin edges (GeV/c) & 1.5-3.0 & 3.0-4.0 & 4.0-5.0 & 5.0-6.0 & 6.0-7.0 & 7.0-8.5 & 8.5-10.0 & 10.0-20.0\\
      \hline
      \hline
      1.5-3.0 & 2487.306 & 783.530 & 123.415 & 78.780 & 75.516 & 20.167 & 1.410 & 1.759\\
      3.0-4.0 & 783.530 & 6883.567 & 1843.185 & 398.312 & 105.427 & 9.190 & 0.161 & 0.000\\
      4.0-5.0 & 123.415 & 1843.185 & 7998.915 & 2532.020 & 359.777 & 55.429 & 7.510 & 0.000\\
      5.0-6.0 & 78.780 & 398.312 & 2532.020 & 6906.017 & 2408.268 & 351.079 & 61.864 & 4.991\\
      6.0-7.0 & 75.516 & 105.427 & 359.777 & 2408.268 & 3610.355 & 1147.420 & 152.970 & 34.063\\
      7.0-8.5 & 20.167 & 9.190 & 55.429 & 351.079 & 1147.420 & 1207.419 & 270.781 & 15.516\\
      8.5-10.0 & 1.410 & 0.161 & 7.510 & 61.864 & 152.970 & 270.781 & 180.813 & 22.407\\
      10.0-20.0 & 1.759 & 0.000 & 0.000 & 4.991 & 34.063 & 15.516 & 22.407 & 25.537
    \end{tabular}
  }
  \label{tbl:statnucleonnuc6muon_pb}
\end{table}
\begin{table}
  \centering
  \renewcommand{\arraystretch}{1.15}
  \caption{Systematic covariance matrix of measured cross section in $p_{\mu}$ on carbon, in units of $10^{-84}$ $(\text{cm}^2$/GeV/c/nucleon$)^2$}
  \resizebox{\textwidth}{!}{
    \begin{tabular}{c||cccccccc}
      Bin edges (GeV/c) & 1.5-3.0 & 3.0-4.0 & 4.0-5.0 & 5.0-6.0 & 6.0-7.0 & 7.0-8.5 & 8.5-10.0 & 10.0-20.0\\
      \hline
      \hline
      1.5-3.0 & 384.799 & 789.008 & 1193.866 & 1166.558 & 826.990 & 220.270 & 87.278 & 38.919\\
      3.0-4.0 & 789.008 & 2468.666 & 3297.934 & 3441.037 & 2430.366 & 827.244 & 278.953 & 104.621\\
      4.0-5.0 & 1193.866 & 3297.934 & 6281.017 & 5857.713 & 4182.090 & 1528.720 & 419.419 & 150.395\\
      5.0-6.0 & 1166.558 & 3441.037 & 5857.713 & 6343.377 & 4461.196 & 1667.815 & 497.086 & 168.352\\
      6.0-7.0 & 826.990 & 2430.366 & 4182.090 & 4461.196 & 3502.013 & 1401.819 & 439.792 & 128.544\\
      7.0-8.5 & 220.270 & 827.244 & 1528.720 & 1667.815 & 1401.819 & 807.588 & 248.509 & 47.072\\
      8.5-10.0 & 87.278 & 278.953 & 419.419 & 497.086 & 439.792 & 248.509 & 97.422 & 19.134\\
      10.0-20.0 & 38.919 & 104.621 & 150.395 & 168.352 & 128.544 & 47.072 & 19.134 & 7.842
    \end{tabular}
  }
  \label{tbl:sysnucleonnuc6muon_pc}
\end{table}
\pagebreak
\begin{turnpage}
\end{turnpage}
\begin{table}
  \centering
  \renewcommand{\arraystretch}{1.15}
  \caption{Measured cross section as function of $\theta_{\mu}$ on carbon, in units of $10^{-42}$ $\text{cm}^2$/degree/nucleon, and the absolute and fractional cross section uncertainties}
    \begin{tabular}{c||c|c||c|c|c}
      Bin edges (degree) &  $\sigma$  &  Abs. Unc.  &  Frac. Stat. Unc.  &  Frac. Sys. Unc.  &  Frac. Flux Unc. \\
      \hline
      \hline
      0-2 & 69.93 & 21.86 & 0.297 & 0.096 & 0.034\\
      2-3 & 293.81 & 74.18 & 0.226 & 0.113 & 0.033\\
      3-4 & 232.51 & 71.15 & 0.268 & 0.147 & 0.033\\
      4-5 & 243.12 & 83.24 & 0.275 & 0.204 & 0.037\\
      5-6 & 104.44 & 64.79 & 0.510 & 0.353 & 0.046\\
      6-8 & 165.82 & 82.86 & 0.411 & 0.285 & 0.039\\
      8-10 & 272.11 & 109.74 & 0.346 & 0.207 & 0.035\\
      10-13 & 210.36 & 91.15 & 0.384 & 0.202 & 0.039
    \end{tabular}
  \label{tbl:datanucleonnuc6muon_thetaa}
\end{table}
\begin{table}
  \centering
  \renewcommand{\arraystretch}{1.15}
  \caption{Statistical covariance matrix of measured cross section as function of $\theta_{\mu}$ on carbon, in units of $10^{-84}$ $(\text{cm}^2$/degree/nucleon$)^2$}
  \resizebox{\textwidth}{!}{
    \begin{tabular}{c||cccccccc}
      Bin edges (degree) & 0-2 & 2-3 & 3-4 & 4-5 & 5-6 & 6-8 & 8-10 & 10-13\\
      \hline
      \hline
      0-2 & 432.545 & -42.190 & -60.428 & -10.381 & 0.256 & 0.206 & -2.325 & -2.718\\
      2-3 & -42.190 & 4402.124 & 227.802 & -355.022 & -55.531 & -0.703 & -21.100 & -2.610\\
      3-4 & -60.428 & 227.802 & 3889.313 & 556.447 & -267.531 & -59.679 & -5.042 & -9.166\\
      4-5 & -10.381 & -355.022 & 556.447 & 4479.508 & 604.254 & -323.825 & -38.103 & 3.448\\
      5-6 & 0.256 & -55.531 & -267.531 & 604.254 & 2836.385 & 323.719 & -183.679 & -8.398\\
      6-8 & 0.206 & -0.703 & -59.679 & -323.825 & 323.719 & 4640.159 & -126.552 & -144.285\\
      8-10 & -2.325 & -21.100 & -5.042 & -38.103 & -183.679 & -126.552 & 8881.099 & -343.995\\
      10-13 & -2.718 & -2.610 & -9.166 & 3.448 & -8.398 & -144.285 & -343.995 & 6510.178
    \end{tabular}
  }
  \label{tbl:statnucleonnuc6muon_thetab}
\end{table}
\begin{table}
  \centering
  \renewcommand{\arraystretch}{1.15}
  \caption{Systematic covariance matrix of measured cross section in $\theta_{\mu}$ on carbon, in units of $10^{-84}$ $(\text{cm}^2$/degree/nucleon$)^2$}
  \resizebox{\textwidth}{!}{
    \begin{tabular}{c||cccccccc}
      Bin edges (degree) & 0-2 & 2-3 & 3-4 & 4-5 & 5-6 & 6-8 & 8-10 & 10-13\\
      \hline
      \hline
      0-2 & 45.099 & 92.411 & 103.281 & 114.507 & 74.645 & 105.557 & 149.681 & 52.143\\
      2-3 & 92.411 & 1100.609 & 841.832 & 1222.754 & 769.793 & 961.137 & 1118.738 & 1053.933\\
      3-4 & 103.281 & 841.832 & 1172.797 & 1542.172 & 1062.769 & 1138.866 & 1384.286 & 1145.364\\
      4-5 & 114.507 & 1222.754 & 1542.172 & 2449.024 & 1531.587 & 1737.054 & 2156.829 & 1604.998\\
      5-6 & 74.645 & 769.793 & 1062.769 & 1531.587 & 1361.346 & 1240.504 & 1502.525 & 1102.167\\
      6-8 & 105.557 & 961.137 & 1138.866 & 1737.054 & 1240.504 & 2226.414 & 1880.519 & 1195.813\\
      8-10 & 149.681 & 1118.738 & 1384.286 & 2156.829 & 1502.525 & 1880.519 & 3161.621 & 1418.018\\
      10-13 & 52.143 & 1053.933 & 1145.364 & 1604.998 & 1102.167 & 1195.813 & 1418.018 & 1798.854
    \end{tabular}
  }
  \label{tbl:sysnucleonnuc6muon_thetac}
\end{table}
\pagebreak
\begin{turnpage}
\end{turnpage}
\begin{table}
  \centering
  \renewcommand{\arraystretch}{1.15}
  \caption{Measured cross section as function of $p_{\mu,||}$ on carbon, in units of $10^{-42}$ $\text{cm}^2$/GeV/c/nucleon, and the absolute and fractional cross section uncertainties}
    \begin{tabular}{c||c|c||c|c|c}
      Bin edges (GeV/c) &  $\sigma$  &  Abs. Unc.  &  Frac. Stat. Unc.  &  Frac. Sys. Unc.  &  Frac. Flux Unc. \\
      \hline
      \hline
      1.5-3.0 & 142.81 & 54.37 & 0.353 & 0.142 & 0.038\\
      3.0-4.0 & 414.13 & 100.15 & 0.207 & 0.125 & 0.037\\
      4.0-5.0 & 495.24 & 120.33 & 0.181 & 0.162 & 0.039\\
      5.0-6.0 & 470.54 & 113.51 & 0.174 & 0.167 & 0.038\\
      6.0-7.0 & 316.82 & 84.97 & 0.193 & 0.186 & 0.037\\
      7.0-8.5 & 131.56 & 43.90 & 0.258 & 0.211 & 0.038\\
      8.5-10.0 & 36.08 & 16.51 & 0.370 & 0.270 & 0.041\\
      10.0-20.0 & 11.27 & 5.74 & 0.446 & 0.247 & 0.038
    \end{tabular}
  \label{tbl:datanucleonnuc6muon_pza}
\end{table}
\begin{table}
  \centering
  \renewcommand{\arraystretch}{1.15}
  \caption{Statistical covariance matrix of measured cross section as function of $p_{\mu,||}$ on carbon, in units of $10^{-84}$ $(\text{cm}^2$/GeV/c/nucleon$)^2$}
  \resizebox{\textwidth}{!}{
    \begin{tabular}{c||cccccccc}
      Bin edges (GeV/c) & 1.5-3.0 & 3.0-4.0 & 4.0-5.0 & 5.0-6.0 & 6.0-7.0 & 7.0-8.5 & 8.5-10.0 & 10.0-20.0\\
      \hline
      \hline
      1.5-3.0 & 2545.244 & 728.621 & 120.277 & 77.608 & 72.794 & 19.444 & 1.364 & 1.702\\
      3.0-4.0 & 728.621 & 7370.998 & 1832.106 & 356.992 & 93.965 & 8.392 & 0.052 & 0.000\\
      4.0-5.0 & 120.277 & 1832.106 & 8057.985 & 2544.775 & 374.057 & 53.227 & 7.161 & 0.000\\
      5.0-6.0 & 77.608 & 356.992 & 2544.775 & 6684.440 & 2431.505 & 348.631 & 58.569 & 4.882\\
      6.0-7.0 & 72.794 & 93.965 & 374.057 & 2431.505 & 3735.052 & 1088.284 & 140.039 & 33.362\\
      7.0-8.5 & 19.444 & 8.392 & 53.227 & 348.631 & 1088.284 & 1156.326 & 260.739 & 15.362\\
      8.5-10.0 & 1.364 & 0.052 & 7.161 & 58.569 & 140.039 & 260.739 & 177.793 & 22.270\\
      10.0-20.0 & 1.702 & 0.000 & 0.000 & 4.882 & 33.362 & 15.362 & 22.270 & 25.211
    \end{tabular}
  }
  \label{tbl:statnucleonnuc6muon_pzb}
\end{table}
\begin{table}
  \centering
  \renewcommand{\arraystretch}{1.15}
  \caption{Systematic covariance matrix of measured cross section in $p_{\mu,||}$ on carbon, in units of $10^{-84}$ $(\text{cm}^2$/GeV/c/nucleon$)^2$}
  \resizebox{\textwidth}{!}{
    \begin{tabular}{c||cccccccc}
      Bin edges (GeV/c) & 1.5-3.0 & 3.0-4.0 & 4.0-5.0 & 5.0-6.0 & 6.0-7.0 & 7.0-8.5 & 8.5-10.0 & 10.0-20.0\\
      \hline
      \hline
      1.5-3.0 & 410.793 & 807.898 & 1244.792 & 1207.063 & 865.339 & 222.378 & 85.770 & 40.987\\
      3.0-4.0 & 807.898 & 2659.590 & 3370.825 & 3460.466 & 2460.014 & 783.208 & 264.051 & 103.430\\
      4.0-5.0 & 1244.792 & 3370.825 & 6421.428 & 5800.897 & 4234.157 & 1488.178 & 408.526 & 149.058\\
      5.0-6.0 & 1207.063 & 3460.466 & 5800.897 & 6199.385 & 4432.492 & 1594.078 & 488.891 & 166.781\\
      6.0-7.0 & 865.339 & 2460.014 & 4234.157 & 4432.492 & 3484.686 & 1332.120 & 428.733 & 129.620\\
      7.0-8.5 & 222.378 & 783.208 & 1488.178 & 1594.078 & 1332.120 & 770.937 & 239.169 & 44.936\\
      8.5-10.0 & 85.770 & 264.051 & 408.526 & 488.891 & 428.733 & 239.169 & 94.943 & 18.573\\
      10.0-20.0 & 40.987 & 103.430 & 149.058 & 166.781 & 129.620 & 44.936 & 18.573 & 7.764
    \end{tabular}
  }
  \label{tbl:sysnucleonnuc6muon_pzc}
\end{table}
\pagebreak
\begin{turnpage}
\end{turnpage}
\begin{table}
  \centering
  \renewcommand{\arraystretch}{1.15}
  \caption{Measured cross section as function of $p_{\mu,T}$ on carbon, in units of $10^{-42}$ $\text{cm}^2$/GeV/c/nucleon, and the absolute and fractional cross section uncertainties}
    \begin{tabular}{c||c|c||c|c|c}
      Bin edges (GeV/c) &  $\sigma$  &  Abs. Unc.  &  Frac. Stat. Unc.  &  Frac. Sys. Unc.  &  Frac. Flux Unc. \\
      \hline
      \hline
      0.00-0.15 & 677.76 & 225.45 & 0.321 & 0.087 & 0.034\\
      0.15-0.30 & 1738.89 & 427.93 & 0.225 & 0.100 & 0.034\\
      0.30-0.45 & 3334.80 & 784.27 & 0.200 & 0.124 & 0.035\\
      0.45-0.60 & 3255.16 & 968.88 & 0.227 & 0.193 & 0.037\\
      0.60-0.75 & 2354.67 & 914.93 & 0.305 & 0.241 & 0.037\\
      0.75-0.90 & 1095.46 & 966.59 & 0.687 & 0.554 & 0.049\\
      0.90-1.25 & 732.87 & 537.65 & 0.581 & 0.448 & 0.048\\
      1.25-2.50 & 41.20 & 64.42 & 1.409 & 0.678 & 0.036
    \end{tabular}
  \label{tbl:datanucleonnuc6muon_pta}
\end{table}
\pagebreak
\begin{turnpage}
\begin{table}
  \centering
  \renewcommand{\arraystretch}{1.15}
  \caption{Statistical covariance matrix of measured cross section as function of $p_{\mu,T}$ on carbon, in units of $10^{-84}$ $(\text{cm}^2$/GeV/c/nucleon$)^2$}
  \resizebox{\textheight}{!}{
    \begin{tabular}{c||cccccccc}
      Bin edges (GeV/c) & 0.00-0.15 & 0.15-0.30 & 0.30-0.45 & 0.45-0.60 & 0.60-0.75 & 0.75-0.90 & 0.90-1.25 & 1.25-2.50\\
      \hline
      \hline
      0.00-0.15 & 47345.844 & 489.515 & -9627.274 & -1191.565 & -50.835 & 695.864 & -132.553 & -2.979\\
      0.15-0.30 & 489.515 & 152597.493 & 15878.159 & -24434.942 & -4415.122 & -2630.600 & -338.305 & -26.218\\
      0.30-0.45 & -9627.274 & 15878.159 & 442805.481 & 56960.571 & -47584.361 & -18432.176 & -3106.294 & -512.456\\
      0.45-0.60 & -1191.565 & -24434.942 & 56960.571 & 545345.745 & 82151.510 & -38191.909 & -22415.638 & 566.213\\
      0.60-0.75 & -50.835 & -4415.122 & -47584.361 & 82151.510 & 514382.100 & 160232.420 & -4248.074 & -2161.159\\
      0.75-0.90 & 695.864 & -2630.600 & -18432.176 & -38191.909 & 160232.420 & 565653.640 & 116267.363 & -433.914\\
      0.90-1.25 & -132.553 & -338.305 & -3106.294 & -22415.638 & -4248.074 & 116267.363 & 181284.881 & -260.193\\
      1.25-2.50 & -2.979 & -26.218 & -512.456 & 566.213 & -2161.159 & -433.914 & -260.193 & 3369.229
    \end{tabular}
  }
  \label{tbl:statnucleonnuc6muon_ptb}
\end{table}
\end{turnpage}
\clearpage
\pagebreak
\begin{turnpage}
\begin{table}
  \centering
  \renewcommand{\arraystretch}{1.15}
  \caption{Systematic covariance matrix of measured cross section in $p_{\mu,T}$ on carbon, in units of $10^{-84}$ $(\text{cm}^2$/GeV/c/nucleon$)^2$}
  \resizebox{\textheight}{!}{
    \begin{tabular}{c||cccccccc}
      Bin edges (GeV/c) & 0.00-0.15 & 0.15-0.30 & 0.30-0.45 & 0.45-0.60 & 0.60-0.75 & 0.75-0.90 & 0.90-1.25 & 1.25-2.50\\
      \hline
      \hline
      0.00-0.15 & 3483.817 & 7652.058 & 8866.626 & 10183.622 & 11849.515 & 6912.592 & 2468.776 & 265.614\\
      0.15-0.30 & 7652.058 & 30530.330 & 51863.562 & 66554.567 & 71581.100 & 50684.480 & 18484.932 & 1839.939\\
      0.30-0.45 & 8866.626 & 51863.562 & 172270.470 & 234351.239 & 208147.109 & 144317.649 & 56989.645 & 6228.014\\
      0.45-0.60 & 10183.622 & 66554.567 & 234351.239 & 393376.163 & 322202.135 & 212486.586 & 91525.906 & 9190.991\\
      0.60-0.75 & 11849.515 & 71581.100 & 208147.109 & 322202.135 & 322712.829 & 223779.297 & 80646.181 & 9352.017\\
      0.75-0.90 & 6912.592 & 50684.480 & 144317.649 & 212486.586 & 223779.297 & 368643.495 & 152778.165 & 12440.530\\
      0.90-1.25 & 2468.776 & 18484.932 & 56989.645 & 91525.906 & 80646.181 & 152778.165 & 107781.804 & 5769.398\\
      1.25-2.50 & 265.614 & 1839.939 & 6228.014 & 9190.991 & 9352.017 & 12440.530 & 5769.398 & 780.369
    \end{tabular}
  }
  \label{tbl:sysnucleonnuc6muon_ptc}
\end{table}
\end{turnpage}
\pagebreak
\begin{turnpage}
\end{turnpage}
\begin{table}
  \centering
  \renewcommand{\arraystretch}{1.15}
  \caption{Measured cross section as function of $Q^2$ on carbon, in units of $10^{-42}$ $\text{cm}^2$/GeV${}^2/c{}^2$/nucleon, and the absolute and fractional cross section uncertainties}
    \begin{tabular}{c||c|c||c|c|c}
      Bin edges (GeV${}^2/c{}^2$) &  $\sigma$  &  Abs. Unc.  &  Frac. Stat. Unc.  &  Frac. Sys. Unc.  &  Frac. Flux Unc. \\
      \hline
      \hline
      0.000-0.025 & 4110.92 & 1313.92 & 0.309 & 0.082 & 0.034\\
      0.025-0.050 & 3738.85 & 1125.78 & 0.289 & 0.086 & 0.035\\
      0.050-0.100 & 3372.90 & 968.01 & 0.257 & 0.127 & 0.033\\
      0.100-0.200 & 4499.60 & 1037.27 & 0.198 & 0.119 & 0.035\\
      0.200-0.300 & 2654.05 & 855.20 & 0.271 & 0.175 & 0.036\\
      0.300-0.400 & 2401.01 & 861.33 & 0.280 & 0.224 & 0.039\\
      0.400-0.500 & 1622.89 & 738.87 & 0.352 & 0.289 & 0.042\\
      0.500-0.700 & 934.89 & 545.84 & 0.465 & 0.353 & 0.040\\
      0.700-1.000 & 786.73 & 497.09 & 0.515 & 0.366 & 0.039\\
      1.000-1.300 & 508.34 & 308.25 & 0.541 & 0.274 & 0.040\\
      1.300-2.000 & 104.02 & 162.03 & 1.255 & 0.923 & 0.059\\
      2.000-3.000 & 43.44 & 76.15 & 1.619 & 0.673 & 0.050
    \end{tabular}
  \label{tbl:datanucleonnuc6q2a}
\end{table}
\pagebreak
\begin{turnpage}
\begin{table}
  \centering
  \renewcommand{\arraystretch}{1.15}
  \caption{Statistical covariance matrix of measured cross section as function of $Q^2$ on carbon, in units of $10^{-84}$ $(\text{cm}^2$/GeV${}^2/c{}^2$/nucleon$)^2$}
  \resizebox{\textheight}{!}{
    \begin{tabular}{c||cccccccccccc}
      Bin edges (GeV${}^2/c{}^2$) & 0.000-0.025 & 0.025-0.050 & 0.050-0.100 & 0.100-0.200 & 0.200-0.300 & 0.300-0.400 & 0.400-0.500 & 0.500-0.700 & 0.700-1.000 & 1.000-1.300 & 1.300-2.000 & 2.000-3.000\\
      \hline
      \hline
      0.000-0.025 & 1612047.844 & 133164.747 & -109292.560 & -38037.712 & -210.854 & -289.388 & -431.586 & -381.587 & 1410.826 & -116.093 & -134.401 & -15.298\\
      0.025-0.050 & 133164.747 & 1163631.354 & 296929.122 & -89658.591 & -29868.450 & -4855.439 & -2137.794 & -1577.626 & -2178.074 & -485.529 & -223.131 & 12.313\\
      0.050-0.100 & -109292.560 & 296929.122 & 753462.588 & 147506.891 & -58070.191 & -21930.282 & -6742.827 & -1558.373 & 317.631 & -580.345 & -609.480 & -76.275\\
      0.100-0.200 & -38037.712 & -89658.591 & 147506.891 & 790275.558 & 111043.502 & -56840.469 & -38915.673 & -5121.184 & -3538.403 & -2738.221 & -473.804 & -314.496\\
      0.200-0.300 & -210.854 & -29868.450 & -58070.191 & 111043.502 & 516648.021 & 193831.617 & -9914.202 & -26131.027 & -9527.948 & -4605.887 & -2429.364 & -225.566\\
      0.300-0.400 & -289.388 & -4855.439 & -21930.282 & -56840.469 & 193831.617 & 453257.205 & 226912.682 & 727.533 & -25604.708 & -10358.162 & 2497.816 & -92.988\\
      0.400-0.500 & -431.586 & -2137.794 & -6742.827 & -38915.673 & -9914.202 & 226912.682 & 325672.916 & 114465.431 & -2753.532 & -9681.002 & -1603.063 & -511.652\\
\end{tabular}
  }
  \label{tbl:statnucleonnuc6q2b}
\end{table}
\end{turnpage}
\clearpage
\pagebreak
\begin{turnpage}
\begin{table}
  \centering
  \renewcommand{\arraystretch}{1.15}
  \caption{Systematic covariance matrix of measured cross section in $Q^2$ on carbon, in units of $10^{-84}$ $(\text{cm}^2$/GeV${}^2/c{}^2$/nucleon$)^2$}
  \resizebox{\textheight}{!}{
    \begin{tabular}{c||cccccccccccc}
      Bin edges (GeV${}^2/c{}^2$) & 0.000-0.025 & 0.025-0.050 & 0.050-0.100 & 0.100-0.200 & 0.200-0.300 & 0.300-0.400 & 0.400-0.500 & 0.500-0.700 & 0.700-1.000 & 1.000-1.300 & 1.300-2.000 & 2.000-3.000\\
      \hline
      \hline
      0.000-0.025 & 114340.749 & 63203.907 & 97883.668 & 69702.441 & 50788.032 & 44606.610 & 50119.145 & 27594.492 & 15002.767 & 9359.854 & 2927.239 & 1444.850\\
      0.025-0.050 & 63203.907 & 103757.689 & 84392.474 & 109039.744 & 61996.950 & 53071.693 & 63755.278 & 53044.149 & 28270.909 & 14167.656 & 3437.170 & 2164.095\\
      0.050-0.100 & 97883.668 & 84392.474 & 183575.702 & 170724.672 & 142519.813 & 153118.040 & 139533.540 & 99833.908 & 67543.256 & 30334.664 & 12124.226 & 4509.597\\
      0.100-0.200 & 69702.441 & 109039.744 & 170724.672 & 285646.828 & 223066.738 & 226804.698 & 203762.849 & 146664.829 & 78950.233 & 38488.714 & 19109.338 & 6346.318\\
      0.200-0.300 & 50788.032 & 61996.950 & 142519.813 & 223066.738 & 214718.388 & 225859.884 & 194853.625 & 120137.142 & 64661.826 & 32947.723 & 16705.759 & 4606.379\\
      0.300-0.400 & 44606.610 & 53071.693 & 153118.040 & 226804.698 & 225859.884 & 288638.745 & 230674.942 & 132653.732 & 72541.441 & 34292.653 & 21857.936 & 5416.617\\
      0.400-0.500 & 50119.145 & 63755.278 & 139533.540 & 203762.849 & 194853.625 & 230674.942 & 220250.698 & 124449.181 & 63894.843 & 32171.484 & 19459.809 & 5250.944\\
\end{tabular}
  }
  \label{tbl:sysnucleonnuc6q2c}
\end{table}
\end{turnpage}
\pagebreak
\begin{turnpage}
\end{turnpage}
\begin{table}
  \centering
  \renewcommand{\arraystretch}{1.15}
  \caption{Measured cross section as function of $W_{exp}$ on carbon, in units of $10^{-42}$ $\text{cm}^2$/GeV/c${}^2$/nucleon, and the absolute and fractional cross section uncertainties}
    \begin{tabular}{c||c|c||c|c|c}
      Bin edges (GeV/c${}^2$) &  $\sigma$  &  Abs. Unc.  &  Frac. Stat. Unc.  &  Frac. Sys. Unc.  &  Frac. Flux Unc. \\
      \hline
      \hline
      0.9-1.1 & 935.92 & 613.41 & 0.627 & 0.191 & 0.043\\
      1.1-1.2 & 5238.76 & 1031.06 & 0.175 & 0.091 & 0.036\\
      1.2-1.3 & 8434.70 & 2309.31 & 0.200 & 0.187 & 0.033\\
      1.3-1.4 & 4744.74 & 3791.22 & 0.529 & 0.599 & 0.033
    \end{tabular}
  \label{tbl:datanucleonnuc6Wa}
\end{table}
\begin{table}
  \centering
  \renewcommand{\arraystretch}{1.15}
  \caption{Statistical covariance matrix of measured cross section as function of $W_{exp}$ on carbon, in units of $10^{-84}$ $(\text{cm}^2$/GeV/c${}^2$/nucleon$)^2$}
    \begin{tabular}{c||cccc}
      Bin edges (GeV/c${}^2$) & 0.9-1.1 & 1.1-1.2 & 1.2-1.3 & 1.3-1.4\\
      \hline
      \hline
      0.9-1.1 & 344280.921 & -5955.534 & -433736.932 & -328835.505\\
      1.1-1.2 & -5955.534 & 837000.556 & 82998.926 & -910335.711\\
      1.2-1.3 & -433736.932 & 82998.926 & 2848202.794 & 1532250.771\\
      1.3-1.4 & -328835.505 & -910335.711 & 1532250.771 & 6306221.739
    \end{tabular}
  \label{tbl:statnucleonnuc6Wb}
\end{table}
\begin{table}
  \centering
  \renewcommand{\arraystretch}{1.15}
  \caption{Systematic covariance matrix of measured cross section in $W_{exp}$ on carbon, in units of $10^{-84}$ $(\text{cm}^2$/GeV/c${}^2$/nucleon$)^2$}
    \begin{tabular}{c||cccc}
      Bin edges (GeV/c${}^2$) & 0.9-1.1 & 1.1-1.2 & 1.2-1.3 & 1.3-1.4\\
      \hline
      \hline
      0.9-1.1 & 31991.340 & 34754.971 & 26908.976 & 39260.342\\
      1.1-1.2 & 34754.971 & 226074.839 & 269792.794 & 290485.892\\
      1.2-1.3 & 26908.976 & 269792.794 & 2484692.008 & 4127535.293\\
      1.3-1.4 & 39260.342 & 290485.892 & 4127535.293 & 8067146.466
    \end{tabular}
  \label{tbl:sysnucleonnuc6Wc}
\end{table}
\pagebreak
\begin{turnpage}
\end{turnpage}
\begin{table}
  \centering
  \renewcommand{\arraystretch}{1.15}
  \caption{Measured cross section as function of $T_{\pi}$ on carbon, in units of $10^{-42}$ $\text{cm}^2$/GeV/nucleon, and the absolute and fractional cross section uncertainties}
    \begin{tabular}{c||c|c||c|c|c}
      Bin edges (GeV) &  $\sigma$  &  Abs. Unc.  &  Frac. Stat. Unc.  &  Frac. Sys. Unc.  &  Frac. Flux Unc. \\
      \hline
      \hline
      0.035-0.100 & 11544.77 & 2671.16 & 0.176 & 0.151 & 0.036\\
      0.100-0.150 & 9366.27 & 1783.83 & 0.132 & 0.137 & 0.034\\
      0.150-0.200 & 5705.08 & 1763.73 & 0.226 & 0.211 & 0.033\\
      0.200-0.350 & 2730.47 & 1583.72 & 0.425 & 0.394 & 0.040
    \end{tabular}
  \label{tbl:datanucleonnuc6pion_ekina}
\end{table}
\begin{table}
  \centering
  \renewcommand{\arraystretch}{1.15}
  \caption{Statistical covariance matrix of measured cross section as function of $T_{\pi}$ on carbon, in units of $10^{-84}$ $(\text{cm}^2$/GeV/nucleon$)^2$}
    \begin{tabular}{c||cccc}
      Bin edges (GeV) & 0.035-0.100 & 0.100-0.150 & 0.150-0.200 & 0.200-0.350\\
      \hline
      \hline
      0.035-0.100 & 4110120.545 & 450633.521 & -366102.636 & -6362.214\\
      0.100-0.150 & 450633.521 & 1529252.308 & 596865.376 & 246606.388\\
      0.150-0.200 & -366102.636 & 596865.376 & 1657591.298 & 826260.188\\
      0.200-0.350 & -6362.214 & 246606.388 & 826260.188 & 1349320.553
    \end{tabular}
  \label{tbl:statnucleonnuc6pion_ekinb}
\end{table}
\begin{table}
  \centering
  \renewcommand{\arraystretch}{1.15}
  \caption{Systematic covariance matrix of measured cross section in $T_{\pi}$ on carbon, in units of $10^{-84}$ $(\text{cm}^2$/GeV/nucleon$)^2$}
    \begin{tabular}{c||cccc}
      Bin edges (GeV) & 0.035-0.100 & 0.100-0.150 & 0.150-0.200 & 0.200-0.350\\
      \hline
      \hline
      0.035-0.100 & 3024981.510 & 1978891.970 & 1702989.640 & 1319045.699\\
      0.100-0.150 & 1978891.970 & 1652801.001 & 1398786.930 & 954689.298\\
      0.150-0.200 & 1702989.640 & 1398786.930 & 1453140.559 & 1071815.858\\
      0.200-0.350 & 1319045.699 & 954689.298 & 1071815.858 & 1158852.632
    \end{tabular}
  \label{tbl:sysnucleonnuc6pion_ekinc}
\end{table}
\pagebreak
\begin{turnpage}
\end{turnpage}
\begin{table}
  \centering
  \renewcommand{\arraystretch}{1.15}
  \caption{Measured cross section as function of $\theta_{\pi}$ on carbon, in units of $10^{-42}$ $\text{cm}^2$/degree/nucleon, and the absolute and fractional cross section uncertainties}
    \begin{tabular}{c||c|c||c|c|c}
      Bin edges (degree) &  $\sigma$  &  Abs. Unc.  &  Frac. Stat. Unc.  &  Frac. Sys. Unc.  &  Frac. Flux Unc. \\
      \hline
      \hline
      0-10 & 3.68 & 5.99 & 1.483 & 0.671 & 0.045\\
      10-20 & 9.04 & 6.38 & 0.627 & 0.325 & 0.040\\
      20-30 & 14.15 & 5.91 & 0.339 & 0.244 & 0.035\\
      30-40 & 16.26 & 5.52 & 0.279 & 0.194 & 0.037\\
      40-50 & 21.86 & 6.14 & 0.228 & 0.163 & 0.036\\
      50-60 & 18.12 & 5.63 & 0.256 & 0.176 & 0.034\\
      60-120 & 17.45 & 4.32 & 0.207 & 0.135 & 0.035\\
      120-135 & 9.75 & 3.15 & 0.291 & 0.141 & 0.033\\
      135-150 & 8.14 & 2.79 & 0.310 & 0.146 & 0.033\\
      150-165 & 5.01 & 3.21 & 0.556 & 0.319 & 0.039\\
      165-180 & 1.05 & 0.91 & 0.768 & 0.395 & 0.036
    \end{tabular}
  \label{tbl:datanucleonnuc6pion_thetaa}
\end{table}
\pagebreak
\begin{turnpage}
\begin{table}
  \centering
  \renewcommand{\arraystretch}{1.15}
  \caption{Statistical covariance matrix of measured cross section as function of $\theta_{\pi}$ on carbon, in units of $10^{-84}$ $(\text{cm}^2$/degree/nucleon$)^2$}
  \resizebox{\textheight}{!}{
    \begin{tabular}{c||ccccccccccc}
      Bin edges (degree) & 0-10 & 10-20 & 20-30 & 30-40 & 40-50 & 50-60 & 60-120 & 120-135 & 135-150 & 150-165 & 165-180\\
      \hline
      \hline
      0-10 & 29.766 & 12.455 & 4.832 & -1.353 & 2.463 & -0.442 & -0.606 & -0.302 & -0.460 & -0.539 & 0.102\\
      10-20 & 12.455 & 32.119 & 11.299 & -0.769 & -0.780 & -1.218 & -0.439 & 0.010 & -0.135 & 0.286 & 2.196\\
      20-30 & 4.832 & 11.299 & 23.034 & 5.013 & -1.963 & -1.259 & -0.814 & -0.799 & -0.565 & 0.421 & 0.728\\
      30-40 & -1.353 & -0.769 & 5.013 & 20.496 & 7.417 & -0.553 & -0.845 & -0.311 & 0.136 & -0.222 & -0.201\\
      40-50 & 2.463 & -0.780 & -1.963 & 7.417 & 24.897 & 6.425 & -0.456 & -0.672 & -0.583 & -1.089 & -0.313\\
      50-60 & -0.442 & -1.218 & -1.259 & -0.553 & 6.425 & 21.543 & 4.262 & -0.433 & -0.423 & -0.570 & -0.167\\
      60-120 & -0.606 & -0.439 & -0.814 & -0.845 & -0.456 & 4.262 & 13.098 & 1.406 & -0.111 & -0.194 & -0.021\\
      120-135 & -0.302 & 0.010 & -0.799 & -0.311 & -0.672 & -0.433 & 1.406 & 8.040 & 1.280 & 0.071 & 0.060\\
      135-150 & -0.460 & -0.135 & -0.565 & 0.136 & -0.583 & -0.423 & -0.111 & 1.280 & 6.378 & 4.357 & 0.394\\
      150-165 & -0.539 & 0.286 & 0.421 & -0.222 & -1.089 & -0.570 & -0.194 & 0.071 & 4.357 & 7.762 & 0.096\\
      165-180 & 0.102 & 2.196 & 0.728 & -0.201 & -0.313 & -0.167 & -0.021 & 0.060 & 0.394 & 0.096 & 0.653
    \end{tabular}
  }
  \label{tbl:statnucleonnuc6pion_thetab}
\end{table}
\end{turnpage}
\clearpage
\pagebreak
\begin{turnpage}
\begin{table}
  \centering
  \renewcommand{\arraystretch}{1.15}
  \caption{Systematic covariance matrix of measured cross section in $\theta_{\pi}$ on carbon, in units of $10^{-84}$ $(\text{cm}^2$/degree/nucleon$)^2$}
  \resizebox{\textheight}{!}{
    \begin{tabular}{c||ccccccccccc}
      Bin edges (degree) & 0-10 & 10-20 & 20-30 & 30-40 & 40-50 & 50-60 & 60-120 & 120-135 & 135-150 & 150-165 & 165-180\\
      \hline
      \hline
      0-10 & 6.099 & 2.112 & 3.025 & 2.738 & 2.224 & 2.087 & 1.429 & 1.276 & 1.283 & 1.506 & 0.034\\
      10-20 & 2.112 & 8.617 & 9.250 & 7.393 & 8.503 & 7.705 & 4.256 & 2.782 & 2.277 & 1.852 & 0.710\\
      20-30 & 3.025 & 9.250 & 11.938 & 9.913 & 10.228 & 8.821 & 5.209 & 3.110 & 2.750 & 1.823 & 0.670\\
      30-40 & 2.738 & 7.393 & 9.913 & 9.939 & 8.952 & 7.789 & 4.739 & 2.801 & 2.587 & 1.857 & 0.574\\
      40-50 & 2.224 & 8.503 & 10.228 & 8.952 & 12.744 & 10.942 & 6.933 & 3.974 & 3.154 & 1.512 & 0.677\\
      50-60 & 2.087 & 7.705 & 8.821 & 7.789 & 10.942 & 10.147 & 6.381 & 3.840 & 2.837 & 1.708 & 0.712\\
      60-120 & 1.429 & 4.256 & 5.209 & 4.739 & 6.933 & 6.381 & 5.548 & 2.655 & 1.714 & 0.957 & 0.559\\
      120-135 & 1.276 & 2.782 & 3.110 & 2.801 & 3.974 & 3.840 & 2.655 & 1.883 & 1.275 & 1.207 & 0.335\\
      135-150 & 1.283 & 2.277 & 2.750 & 2.587 & 3.154 & 2.837 & 1.714 & 1.275 & 1.417 & 0.745 & 0.186\\
      150-165 & 1.506 & 1.852 & 1.823 & 1.857 & 1.512 & 1.708 & 0.957 & 1.207 & 0.745 & 2.546 & 0.259\\
      165-180 & 0.034 & 0.710 & 0.670 & 0.574 & 0.677 & 0.712 & 0.559 & 0.335 & 0.186 & 0.259 & 0.173
    \end{tabular}
  }
  \label{tbl:sysnucleonnuc6pion_thetac}
\end{table}
\end{turnpage}
\pagebreak
\begin{turnpage}
\end{turnpage}
\begin{table}
  \centering
  \renewcommand{\arraystretch}{1.15}
  \caption{Measured cross section as function of $p_{\mu}$ on water, in units of $10^{-42}$ $\text{cm}^2$/GeV/c/nucleon, and the absolute and fractional cross section uncertainties}
    \begin{tabular}{c||c|c||c|c|c}
      Bin edges (GeV/c) &  $\sigma$  &  Abs. Unc.  &  Frac. Stat. Unc.  &  Frac. Sys. Unc.  &  Frac. Flux Unc. \\
      \hline
      \hline
      1.5-3.0 & 117.95 & 52.87 & 0.410 & 0.180 & 0.035\\
      3.0-4.0 & 185.58 & 88.01 & 0.386 & 0.276 & 0.048\\
      4.0-5.0 & 457.01 & 118.99 & 0.211 & 0.153 & 0.041\\
      5.0-6.0 & 476.40 & 113.68 & 0.187 & 0.148 & 0.037\\
      6.0-7.0 & 340.82 & 84.27 & 0.207 & 0.135 & 0.035\\
      7.0-8.5 & 162.09 & 46.27 & 0.233 & 0.165 & 0.032\\
      8.5-10.0 & 46.21 & 16.77 & 0.317 & 0.176 & 0.031\\
      10.0-20.0 & 6.09 & 2.91 & 0.438 & 0.192 & 0.037
    \end{tabular}
  \label{tbl:datanucleonnuc10muon_pa}
\end{table}
\begin{table}
  \centering
  \renewcommand{\arraystretch}{1.15}
  \caption{Statistical covariance matrix of measured cross section as function of $p_{\mu}$ on water, in units of $10^{-84}$ $(\text{cm}^2$/GeV/c/nucleon$)^2$}
  \resizebox{\textwidth}{!}{
    \begin{tabular}{c||cccccccc}
      Bin edges (GeV/c) & 1.5-3.0 & 3.0-4.0 & 4.0-5.0 & 5.0-6.0 & 6.0-7.0 & 7.0-8.5 & 8.5-10.0 & 10.0-20.0\\
      \hline
      \hline
      1.5-3.0 & 2344.004 & 445.008 & 69.019 & 37.629 & 28.309 & 10.974 & 8.907 & 0.011\\
      3.0-4.0 & 445.008 & 5127.450 & 2079.943 & 557.499 & 60.656 & 35.261 & 3.708 & 1.132\\
      4.0-5.0 & 69.019 & 2079.943 & 9264.813 & 3453.632 & 698.630 & 141.989 & 30.908 & 2.745\\
      5.0-6.0 & 37.629 & 557.499 & 3453.632 & 7926.040 & 3136.162 & 481.482 & 55.213 & 7.682\\
      6.0-7.0 & 28.309 & 60.656 & 698.630 & 3136.162 & 4992.546 & 1267.690 & 85.475 & 1.391\\
      7.0-8.5 & 10.974 & 35.261 & 141.989 & 481.482 & 1267.690 & 1425.398 & 217.459 & 10.233\\
      8.5-10.0 & 8.907 & 3.708 & 30.908 & 55.213 & 85.475 & 217.459 & 215.068 & 20.762\\
      10.0-20.0 & 0.011 & 1.132 & 2.745 & 7.682 & 1.391 & 10.233 & 20.762 & 7.127
    \end{tabular}
  }
  \label{tbl:statnucleonnuc10muon_pb}
\end{table}
\begin{table}
  \centering
  \renewcommand{\arraystretch}{1.15}
  \caption{Systematic covariance matrix of measured cross section in $p_{\mu}$ on water, in units of $10^{-84}$ $(\text{cm}^2$/GeV/c/nucleon$)^2$}
  \resizebox{\textwidth}{!}{
    \begin{tabular}{c||cccccccc}
      Bin edges (GeV/c) & 1.5-3.0 & 3.0-4.0 & 4.0-5.0 & 5.0-6.0 & 6.0-7.0 & 7.0-8.5 & 8.5-10.0 & 10.0-20.0\\
      \hline
      \hline
      1.5-3.0 & 451.412 & 564.022 & 898.038 & 799.031 & 682.157 & 419.447 & 119.619 & 9.347\\
      3.0-4.0 & 564.022 & 2619.122 & 2753.930 & 2802.366 & 1580.375 & 770.858 & 208.530 & 26.336\\
      4.0-5.0 & 898.038 & 2753.930 & 4893.632 & 4338.470 & 2681.647 & 1244.144 & 288.103 & 47.295\\
      5.0-6.0 & 799.031 & 2802.366 & 4338.470 & 4998.182 & 2697.240 & 1079.987 & 305.276 & 52.912\\
      6.0-7.0 & 682.157 & 1580.375 & 2681.647 & 2697.240 & 2109.131 & 974.516 & 247.648 & 36.237\\
      7.0-8.5 & 419.447 & 770.858 & 1244.144 & 1079.987 & 974.516 & 715.744 & 182.693 & 18.250\\
      8.5-10.0 & 119.619 & 208.530 & 288.103 & 305.276 & 247.648 & 182.693 & 66.308 & 4.025\\
      10.0-20.0 & 9.347 & 26.336 & 47.295 & 52.912 & 36.237 & 18.250 & 4.025 & 1.361
    \end{tabular}
  }
  \label{tbl:sysnucleonnuc10muon_pc}
\end{table}
\pagebreak
\begin{turnpage}
\end{turnpage}
\begin{table}
  \centering
  \renewcommand{\arraystretch}{1.15}
  \caption{Measured cross section as function of $\theta_{\mu}$ on water, in units of $10^{-42}$ $\text{cm}^2$/degree/nucleon, and the absolute and fractional cross section uncertainties}
    \begin{tabular}{c||c|c||c|c|c}
      Bin edges (degree) &  $\sigma$  &  Abs. Unc.  &  Frac. Stat. Unc.  &  Frac. Sys. Unc.  &  Frac. Flux Unc. \\
      \hline
      \hline
      0-2 & 63.72 & 18.14 & 0.260 & 0.117 & 0.031\\
      2-3 & 217.48 & 63.11 & 0.271 & 0.103 & 0.035\\
      3-4 & 246.44 & 72.74 & 0.267 & 0.125 & 0.033\\
      4-5 & 207.57 & 88.04 & 0.382 & 0.184 & 0.034\\
      5-6 & 136.66 & 83.70 & 0.536 & 0.296 & 0.037\\
      6-8 & 236.70 & 102.60 & 0.366 & 0.233 & 0.040\\
      8-10 & 290.88 & 114.06 & 0.340 & 0.195 & 0.039\\
      10-13 & 14.67 & 57.44 & 3.399 & 1.945 & 0.181
    \end{tabular}
  \label{tbl:datanucleonnuc10muon_thetaa}
\end{table}
\begin{table}
  \centering
  \renewcommand{\arraystretch}{1.15}
  \caption{Statistical covariance matrix of measured cross section as function of $\theta_{\mu}$ on water, in units of $10^{-84}$ $(\text{cm}^2$/degree/nucleon$)^2$}
  \resizebox{\textwidth}{!}{
    \begin{tabular}{c||cccccccc}
      Bin edges (degree) & 0-2 & 2-3 & 3-4 & 4-5 & 5-6 & 6-8 & 8-10 & 10-13\\
      \hline
      \hline
      0-2 & 273.470 & 54.585 & -101.543 & -17.484 & -0.727 & -4.685 & -0.455 & 0.011\\
      2-3 & 54.585 & 3478.295 & 496.736 & -513.517 & -80.607 & -13.769 & -12.044 & 0.139\\
      3-4 & -101.543 & 496.736 & 4336.745 & 942.105 & -394.960 & -115.405 & 5.655 & 0.149\\
      4-5 & -17.484 & -513.517 & 942.105 & 6294.647 & 1188.026 & -463.504 & -39.117 & 0.892\\
      5-6 & -0.727 & -80.607 & -394.960 & 1188.026 & 5373.660 & 607.700 & -342.608 & -2.822\\
      6-8 & -4.685 & -13.769 & -115.405 & -463.504 & 607.700 & 7487.508 & -398.111 & -72.215\\
      8-10 & -0.455 & -12.044 & 5.655 & -39.117 & -342.608 & -398.111 & 9799.885 & 587.996\\
      10-13 & 0.011 & 0.139 & 0.149 & 0.892 & -2.822 & -72.215 & 587.996 & 2485.525
    \end{tabular}
  }
  \label{tbl:statnucleonnuc10muon_thetab}
\end{table}
\begin{table}
  \centering
  \renewcommand{\arraystretch}{1.15}
  \caption{Systematic covariance matrix of measured cross section in $\theta_{\mu}$ on water, in units of $10^{-84}$ $(\text{cm}^2$/degree/nucleon$)^2$}
  \resizebox{\textwidth}{!}{
    \begin{tabular}{c||cccccccc}
      Bin edges (degree) & 0-2 & 2-3 & 3-4 & 4-5 & 5-6 & 6-8 & 8-10 & 10-13\\
      \hline
      \hline
      0-2 & 55.609 & 108.738 & 139.883 & 81.848 & 35.719 & 28.759 & 40.840 & -6.741\\
      2-3 & 108.738 & 504.494 & 552.733 & 601.157 & 570.449 & 631.701 & 410.197 & 287.381\\
      3-4 & 139.883 & 552.733 & 954.832 & 802.914 & 660.357 & 1116.479 & 944.109 & 450.452\\
      4-5 & 81.848 & 601.157 & 802.914 & 1456.929 & 1320.732 & 1478.837 & 1299.382 & 751.433\\
      5-6 & 35.719 & 570.449 & 660.357 & 1320.732 & 1632.620 & 1328.831 & 1269.663 & 881.853\\
      6-8 & 28.759 & 631.701 & 1116.479 & 1478.837 & 1328.831 & 3040.213 & 1943.068 & 1180.314\\
      8-10 & 40.840 & 410.197 & 944.109 & 1299.382 & 1269.663 & 1943.068 & 3209.779 & 1043.855\\
      10-13 & -6.741 & 287.381 & 450.452 & 751.433 & 881.853 & 1180.314 & 1043.855 & 814.089
    \end{tabular}
  }
  \label{tbl:sysnucleonnuc10muon_thetac}
\end{table}
\pagebreak
\begin{turnpage}
\end{turnpage}
\begin{table}
  \centering
  \renewcommand{\arraystretch}{1.15}
  \caption{Measured cross section as function of $p_{\mu,||}$ on water, in units of $10^{-42}$ $\text{cm}^2$/GeV/c/nucleon, and the absolute and fractional cross section uncertainties}
    \begin{tabular}{c||c|c||c|c|c}
      Bin edges (GeV/c) &  $\sigma$  &  Abs. Unc.  &  Frac. Stat. Unc.  &  Frac. Sys. Unc.  &  Frac. Flux Unc. \\
      \hline
      \hline
      1.5-3.0 & 136.29 & 60.24 & 0.389 & 0.210 & 0.035\\
      3.0-4.0 & 209.53 & 91.74 & 0.357 & 0.253 & 0.046\\
      4.0-5.0 & 467.09 & 122.37 & 0.211 & 0.155 & 0.041\\
      5.0-6.0 & 454.55 & 110.72 & 0.192 & 0.149 & 0.037\\
      6.0-7.0 & 332.04 & 82.82 & 0.211 & 0.134 & 0.034\\
      7.0-8.5 & 150.87 & 44.38 & 0.239 & 0.171 & 0.032\\
      8.5-10.0 & 44.53 & 16.29 & 0.319 & 0.180 & 0.031\\
      10.0-20.0 & 6.13 & 2.91 & 0.436 & 0.189 & 0.037
    \end{tabular}
  \label{tbl:datanucleonnuc10muon_pza}
\end{table}
\begin{table}
  \centering
  \renewcommand{\arraystretch}{1.15}
  \caption{Statistical covariance matrix of measured cross section as function of $p_{\mu,||}$ on water, in units of $10^{-84}$ $(\text{cm}^2$/GeV/c/nucleon$)^2$}
  \resizebox{\textwidth}{!}{
    \begin{tabular}{c||cccccccc}
      Bin edges (GeV/c) & 1.5-3.0 & 3.0-4.0 & 4.0-5.0 & 5.0-6.0 & 6.0-7.0 & 7.0-8.5 & 8.5-10.0 & 10.0-20.0\\
      \hline
      \hline
      1.5-3.0 & 2806.437 & 504.890 & 73.486 & 37.165 & 29.685 & 10.703 & 9.867 & 0.011\\
      3.0-4.0 & 504.890 & 5594.629 & 2192.711 & 520.469 & 52.881 & 35.865 & 3.809 & 1.132\\
      4.0-5.0 & 73.486 & 2192.711 & 9743.624 & 3427.026 & 667.524 & 136.501 & 30.407 & 2.822\\
      5.0-6.0 & 37.165 & 520.469 & 3427.026 & 7644.453 & 2938.764 & 447.993 & 53.709 & 7.692\\
      6.0-7.0 & 29.685 & 52.881 & 667.524 & 2938.764 & 4887.686 & 1195.760 & 81.566 & 1.344\\
      7.0-8.5 & 10.703 & 35.865 & 136.501 & 447.993 & 1195.760 & 1302.963 & 209.148 & 9.855\\
      8.5-10.0 & 9.867 & 3.809 & 30.407 & 53.709 & 81.566 & 209.148 & 201.152 & 20.494\\
      10.0-20.0 & 0.011 & 1.132 & 2.822 & 7.692 & 1.344 & 9.855 & 20.494 & 7.146
    \end{tabular}
  }
  \label{tbl:statnucleonnuc10muon_pzb}
\end{table}
\begin{table}
  \centering
  \renewcommand{\arraystretch}{1.15}
  \caption{Systematic covariance matrix of measured cross section in $p_{\mu,||}$ on water, in units of $10^{-84}$ $(\text{cm}^2$/GeV/c/nucleon$)^2$}
  \resizebox{\textwidth}{!}{
    \begin{tabular}{c||cccccccc}
      Bin edges (GeV/c) & 1.5-3.0 & 3.0-4.0 & 4.0-5.0 & 5.0-6.0 & 6.0-7.0 & 7.0-8.5 & 8.5-10.0 & 10.0-20.0\\
      \hline
      \hline
      1.5-3.0 & 822.289 & 462.724 & 1028.704 & 801.765 & 665.651 & 470.679 & 127.720 & 9.731\\
      3.0-4.0 & 462.724 & 2820.764 & 2891.273 & 2824.462 & 1605.847 & 795.496 & 223.430 & 27.984\\
      4.0-5.0 & 1028.704 & 2891.273 & 5230.407 & 4401.956 & 2668.438 & 1195.644 & 289.928 & 49.303\\
      5.0-6.0 & 801.765 & 2824.462 & 4401.956 & 4615.427 & 2549.872 & 1005.696 & 294.413 & 50.238\\
      6.0-7.0 & 665.651 & 1605.847 & 2668.438 & 2549.872 & 1972.173 & 891.953 & 230.546 & 34.929\\
      7.0-8.5 & 470.679 & 795.496 & 1195.644 & 1005.696 & 891.953 & 666.928 & 168.923 & 17.475\\
      8.5-10.0 & 127.720 & 223.430 & 289.928 & 294.413 & 230.546 & 168.923 & 64.052 & 3.808\\
      10.0-20.0 & 9.731 & 27.984 & 49.303 & 50.238 & 34.929 & 17.475 & 3.808 & 1.345
    \end{tabular}
  }
  \label{tbl:sysnucleonnuc10muon_pzc}
\end{table}
\pagebreak
\begin{turnpage}
\end{turnpage}
\begin{table}
  \centering
  \renewcommand{\arraystretch}{1.15}
  \caption{Measured cross section as function of $p_{\mu,T}$ on water, in units of $10^{-42}$ $\text{cm}^2$/GeV/c/nucleon, and the absolute and fractional cross section uncertainties}
    \begin{tabular}{c||c|c||c|c|c}
      Bin edges (GeV/c) &  $\sigma$  &  Abs. Unc.  &  Frac. Stat. Unc.  &  Frac. Sys. Unc.  &  Frac. Flux Unc. \\
      \hline
      \hline
      0.00-0.15 & 427.07 & 156.83 & 0.350 & 0.110 & 0.031\\
      0.15-0.30 & 1552.75 & 434.21 & 0.252 & 0.122 & 0.030\\
      0.30-0.45 & 3698.80 & 892.81 & 0.212 & 0.115 & 0.035\\
      0.45-0.60 & 3113.42 & 1107.87 & 0.288 & 0.209 & 0.035\\
      0.60-0.75 & 762.06 & 764.44 & 0.739 & 0.678 & 0.047\\
      0.75-0.90 & 2313.61 & 1115.44 & 0.419 & 0.239 & 0.032\\
      0.90-1.25 & 1015.62 & 585.68 & 0.541 & 0.201 & 0.038\\
      1.25-2.50 & 61.79 & 101.64 & 0.926 & 1.360 & 0.066
    \end{tabular}
  \label{tbl:datanucleonnuc10muon_pta}
\end{table}
\pagebreak
\begin{turnpage}
\begin{table}
  \centering
  \renewcommand{\arraystretch}{1.15}
  \caption{Statistical covariance matrix of measured cross section as function of $p_{\mu,T}$ on water, in units of $10^{-84}$ $(\text{cm}^2$/GeV/c/nucleon$)^2$}
  \resizebox{\textheight}{!}{
    \begin{tabular}{c||cccccccc}
      Bin edges (GeV/c) & 0.00-0.15 & 0.15-0.30 & 0.30-0.45 & 0.45-0.60 & 0.60-0.75 & 0.75-0.90 & 0.90-1.25 & 1.25-2.50\\
      \hline
      \hline
      0.00-0.15 & 22384.290 & 309.360 & -7344.185 & -1230.080 & -94.202 & -98.135 & -6.266 & 1.649\\
      0.15-0.30 & 309.360 & 152790.960 & 14256.803 & -30091.672 & -6430.353 & -2133.393 & -3061.313 & -32.386\\
      0.30-0.45 & -7344.185 & 14256.803 & 617215.664 & 92003.011 & -20825.426 & -38780.595 & -3925.249 & -1075.423\\
      0.45-0.60 & -1230.080 & -30091.672 & 92003.011 & 805716.499 & 171697.794 & -44210.017 & -7286.457 & 778.975\\
      0.60-0.75 & -94.202 & -6430.353 & -20825.426 & 171697.794 & 317423.898 & 241253.824 & 4577.656 & -1266.640\\
      0.75-0.90 & -98.135 & -2133.393 & -38780.595 & -44210.017 & 241253.824 & 938404.156 & 35594.495 & -1501.990\\
      0.90-1.25 & -6.266 & -3061.313 & -3925.249 & -7286.457 & 4577.656 & 35594.495 & 301365.386 & 99.880\\
      1.25-2.50 & 1.649 & -32.386 & -1075.423 & 778.975 & -1266.640 & -1501.990 & 99.880 & 3273.325
    \end{tabular}
  }
  \label{tbl:statnucleonnuc10muon_ptb}
\end{table}
\end{turnpage}
\clearpage
\pagebreak
\begin{turnpage}
\begin{table}
  \centering
  \renewcommand{\arraystretch}{1.15}
  \caption{Systematic covariance matrix of measured cross section in $p_{\mu,T}$ on water, in units of $10^{-84}$ $(\text{cm}^2$/GeV/c/nucleon$)^2$}
  \resizebox{\textheight}{!}{
    %
  }
  \label{tbl:sysnucleonnuc10muon_ptc}
\end{table}
\end{turnpage}
\pagebreak
\begin{turnpage}
\end{turnpage}
\begin{table}
  \centering
  \renewcommand{\arraystretch}{1.15}
  \caption{Measured cross section as function of $Q^2$ on water, in units of $10^{-42}$ $\text{cm}^2$/GeV${}^2/c{}^2$/nucleon, and the absolute and fractional cross section uncertainties}
    %
  \label{tbl:datanucleonnuc10q2a}
\end{table}
\pagebreak
\begin{turnpage}
\begin{table}
  \centering
  \renewcommand{\arraystretch}{1.15}
  \caption{Statistical covariance matrix of measured cross section as function of $Q^2$ on water, in units of $10^{-84}$ $(\text{cm}^2$/GeV${}^2/c{}^2$/nucleon$)^2$}
  \resizebox{\textheight}{!}{
    %
  }
  \label{tbl:statnucleonnuc10q2b}
\end{table}
\end{turnpage}
\clearpage
\pagebreak
\begin{turnpage}
\begin{table}
  \centering
  \renewcommand{\arraystretch}{1.15}
  \caption{Systematic covariance matrix of measured cross section in $Q^2$ on water, in units of $10^{-84}$ $(\text{cm}^2$/GeV${}^2/c{}^2$/nucleon$)^2$}
  \resizebox{\textheight}{!}{
    %
  }
  \label{tbl:sysnucleonnuc10q2c}
\end{table}
\end{turnpage}
\pagebreak
\begin{turnpage}
\end{turnpage}
\begin{table}
  \centering
  \renewcommand{\arraystretch}{1.15}
  \caption{Measured cross section as function of $W_{exp}$ on water, in units of $10^{-42}$ $\text{cm}^2$/GeV/c${}^2$/nucleon, and the absolute and fractional cross section uncertainties}
    \begin{tabular}{c||c|c||c|c|c}
      Bin edges (GeV/c${}^2$) &  $\sigma$  &  Abs. Unc.  &  Frac. Stat. Unc.  &  Frac. Sys. Unc.  &  Frac. Flux Unc. \\
      \hline
      \hline
      0.9-1.1 & 1604.24 & 1413.62 & 0.847 & 0.242 & 0.037\\
      1.1-1.2 & 7771.74 & 1887.69 & 0.206 & 0.129 & 0.034\\
      1.2-1.3 & 5069.96 & 2456.47 & 0.368 & 0.315 & 0.033\\
      1.3-1.4 & 1635.65 & 1510.53 & 0.665 & 0.641 & 0.035
    \end{tabular}
  \label{tbl:datanucleonnuc10Wa}
\end{table}
\begin{table}
  \centering
  \renewcommand{\arraystretch}{1.15}
  \caption{Statistical covariance matrix of measured cross section as function of $W_{exp}$ on water, in units of $10^{-84}$ $(\text{cm}^2$/GeV/c${}^2$/nucleon$)^2$}
    %
  \label{tbl:statnucleonnuc10Wb}
\end{table}
\begin{table}
  \centering
  \renewcommand{\arraystretch}{1.15}
  \caption{Systematic covariance matrix of measured cross section in $W_{exp}$ on water, in units of $10^{-84}$ $(\text{cm}^2$/GeV/c${}^2$/nucleon$)^2$}
    %
  \label{tbl:sysnucleonnuc10Wc}
\end{table}
\pagebreak
\begin{turnpage}
\end{turnpage}
\begin{table}
  \centering
  \renewcommand{\arraystretch}{1.15}
  \caption{Measured cross section as function of $T_{\pi}$ on water, in units of $10^{-42}$ $\text{cm}^2$/GeV/nucleon, and the absolute and fractional cross section uncertainties}
    %
  \label{tbl:datanucleonnuc10pion_ekina}
\end{table}
\begin{table}
  \centering
  \renewcommand{\arraystretch}{1.15}
  \caption{Statistical covariance matrix of measured cross section as function of $T_{\pi}$ on water, in units of $10^{-84}$ $(\text{cm}^2$/GeV/nucleon$)^2$}
    %
  \label{tbl:statnucleonnuc10pion_ekinb}
\end{table}
\begin{table}
  \centering
  \renewcommand{\arraystretch}{1.15}
  \caption{Systematic covariance matrix of measured cross section in $T_{\pi}$ on water, in units of $10^{-84}$ $(\text{cm}^2$/GeV/nucleon$)^2$}
    %
  \label{tbl:sysnucleonnuc10pion_ekinc}
\end{table}
\pagebreak
\begin{turnpage}
\end{turnpage}
\begin{table}
  \centering
  \renewcommand{\arraystretch}{1.15}
  \caption{Measured cross section as function of $\theta_{\pi}$ on water, in units of $10^{-42}$ $\text{cm}^2$/degree/nucleon, and the absolute and fractional cross section uncertainties}
    %
  \label{tbl:datanucleonnuc10pion_thetaa}
\end{table}
\pagebreak
\begin{turnpage}
\begin{table}
  \centering
  \renewcommand{\arraystretch}{1.15}
  \caption{Statistical covariance matrix of measured cross section as function of $\theta_{\pi}$ on water, in units of $10^{-84}$ $(\text{cm}^2$/degree/nucleon$)^2$}
  \resizebox{\textheight}{!}{
    %
  }
  \label{tbl:statnucleonnuc10pion_thetab}
\end{table}
\end{turnpage}
\clearpage
\pagebreak
\begin{turnpage}
\begin{table}
  \centering
  \renewcommand{\arraystretch}{1.15}
  \caption{Systematic covariance matrix of measured cross section in $\theta_{\pi}$ on water, in units of $10^{-84}$ $(\text{cm}^2$/degree/nucleon$)^2$}
  \resizebox{\textheight}{!}{
    %
  }
  \label{tbl:sysnucleonnuc10pion_thetac}
\end{table}
\end{turnpage}
\pagebreak
\begin{turnpage}
\end{turnpage}
\begin{table}
  \centering
  \renewcommand{\arraystretch}{1.15}
  \caption{Measured cross section as function of $p_{\mu}$ on iron, in units of $10^{-42}$ $\text{cm}^2$/GeV/c/nucleon, and the absolute and fractional cross section uncertainties}
    %
  \label{tbl:datanucleonnuc26muon_pa}
\end{table}
\begin{table}
  \centering
  \renewcommand{\arraystretch}{1.15}
  \caption{Statistical covariance matrix of measured cross section as function of $p_{\mu}$ on iron, in units of $10^{-84}$ $(\text{cm}^2$/GeV/c/nucleon$)^2$}
  \resizebox{\textwidth}{!}{
    %
  }
  \label{tbl:statnucleonnuc26muon_pb}
\end{table}
\begin{table}
  \centering
  \renewcommand{\arraystretch}{1.15}
  \caption{Systematic covariance matrix of measured cross section in $p_{\mu}$ on iron, in units of $10^{-84}$ $(\text{cm}^2$/GeV/c/nucleon$)^2$}
  \resizebox{\textwidth}{!}{
    %
  }
  \label{tbl:sysnucleonnuc26muon_pc}
\end{table}
\pagebreak
\begin{turnpage}
\end{turnpage}
\begin{table}
  \centering
  \renewcommand{\arraystretch}{1.15}
  \caption{Measured cross section as function of $\theta_{\mu}$ on iron, in units of $10^{-42}$ $\text{cm}^2$/degree/nucleon, and the absolute and fractional cross section uncertainties}
    %
  \label{tbl:datanucleonnuc26muon_thetaa}
\end{table}
\begin{table}
  \centering
  \renewcommand{\arraystretch}{1.15}
  \caption{Statistical covariance matrix of measured cross section as function of $\theta_{\mu}$ on iron, in units of $10^{-84}$ $(\text{cm}^2$/degree/nucleon$)^2$}
  \resizebox{\textwidth}{!}{
    %
  }
  \label{tbl:statnucleonnuc26muon_thetab}
\end{table}
\begin{table}
  \centering
  \renewcommand{\arraystretch}{1.15}
  \caption{Systematic covariance matrix of measured cross section in $\theta_{\mu}$ on iron, in units of $10^{-84}$ $(\text{cm}^2$/degree/nucleon$)^2$}
  \resizebox{\textwidth}{!}{
    %
  }
  \label{tbl:sysnucleonnuc26muon_thetac}
\end{table}
\pagebreak
\begin{turnpage}
\end{turnpage}
\begin{table}
  \centering
  \renewcommand{\arraystretch}{1.15}
  \caption{Measured cross section as function of $p_{\mu,||}$ on iron, in units of $10^{-42}$ $\text{cm}^2$/GeV/c/nucleon, and the absolute and fractional cross section uncertainties}
    %
  \label{tbl:datanucleonnuc26muon_pza}
\end{table}
\begin{table}
  \centering
  \renewcommand{\arraystretch}{1.15}
  \caption{Statistical covariance matrix of measured cross section as function of $p_{\mu,||}$ on iron, in units of $10^{-84}$ $(\text{cm}^2$/GeV/c/nucleon$)^2$}
  \resizebox{\textwidth}{!}{
    %
  }
  \label{tbl:statnucleonnuc26muon_pzb}
\end{table}
\begin{table}
  \centering
  \renewcommand{\arraystretch}{1.15}
  \caption{Systematic covariance matrix of measured cross section in $p_{\mu,||}$ on iron, in units of $10^{-84}$ $(\text{cm}^2$/GeV/c/nucleon$)^2$}
  \resizebox{\textwidth}{!}{
    %
  }
  \label{tbl:sysnucleonnuc26muon_pzc}
\end{table}
\pagebreak
\begin{turnpage}
\end{turnpage}
\begin{table}
  \centering
  \renewcommand{\arraystretch}{1.15}
  \caption{Measured cross section as function of $p_{\mu,T}$ on iron, in units of $10^{-42}$ $\text{cm}^2$/GeV/c/nucleon, and the absolute and fractional cross section uncertainties}
    %
  \label{tbl:datanucleonnuc26muon_pta}
\end{table}
\pagebreak
\begin{turnpage}
\begin{table}
  \centering
  \renewcommand{\arraystretch}{1.15}
  \caption{Statistical covariance matrix of measured cross section as function of $p_{\mu,T}$ on iron, in units of $10^{-84}$ $(\text{cm}^2$/GeV/c/nucleon$)^2$}
  \resizebox{\textheight}{!}{
    %
  }
  \label{tbl:statnucleonnuc26muon_ptb}
\end{table}
\end{turnpage}
\clearpage
\pagebreak
\begin{turnpage}
\begin{table}
  \centering
  \renewcommand{\arraystretch}{1.15}
  \caption{Systematic covariance matrix of measured cross section in $p_{\mu,T}$ on iron, in units of $10^{-84}$ $(\text{cm}^2$/GeV/c/nucleon$)^2$}
  \resizebox{\textheight}{!}{
    %
  }
  \label{tbl:sysnucleonnuc26muon_ptc}
\end{table}
\end{turnpage}
\pagebreak
\begin{turnpage}
\end{turnpage}
\begin{table}
  \centering
  \renewcommand{\arraystretch}{1.15}
  \caption{Measured cross section as function of $Q^2$ on iron, in units of $10^{-42}$ $\text{cm}^2$/GeV${}^2/c{}^2$/nucleon, and the absolute and fractional cross section uncertainties}
    %
  \label{tbl:datanucleonnuc26q2a}
\end{table}
\pagebreak
\begin{turnpage}
\begin{table}
  \centering
  \renewcommand{\arraystretch}{1.15}
  \caption{Statistical covariance matrix of measured cross section as function of $Q^2$ on iron, in units of $10^{-84}$ $(\text{cm}^2$/GeV${}^2/c{}^2$/nucleon$)^2$}
  \resizebox{\textheight}{!}{
    %
  }
  \label{tbl:statnucleonnuc26q2b}
\end{table}
\end{turnpage}
\clearpage
\pagebreak
\begin{turnpage}
\begin{table}
  \centering
  \renewcommand{\arraystretch}{1.15}
  \caption{Systematic covariance matrix of measured cross section in $Q^2$ on iron, in units of $10^{-84}$ $(\text{cm}^2$/GeV${}^2/c{}^2$/nucleon$)^2$}
  \resizebox{\textheight}{!}{
    %
  }
  \label{tbl:sysnucleonnuc26q2c}
\end{table}
\end{turnpage}
\pagebreak
\begin{turnpage}
\end{turnpage}
\begin{table}
  \centering
  \renewcommand{\arraystretch}{1.15}
  \caption{Measured cross section as function of $W_{exp}$ on iron, in units of $10^{-42}$ $\text{cm}^2$/GeV/c${}^2$/nucleon, and the absolute and fractional cross section uncertainties}
    %
  \label{tbl:datanucleonnuc26Wa}
\end{table}
\begin{table}
  \centering
  \renewcommand{\arraystretch}{1.15}
  \caption{Statistical covariance matrix of measured cross section as function of $W_{exp}$ on iron, in units of $10^{-84}$ $(\text{cm}^2$/GeV/c${}^2$/nucleon$)^2$}
    %
  \label{tbl:statnucleonnuc26Wb}
\end{table}
\begin{table}
  \centering
  \renewcommand{\arraystretch}{1.15}
  \caption{Systematic covariance matrix of measured cross section in $W_{exp}$ on iron, in units of $10^{-84}$ $(\text{cm}^2$/GeV/c${}^2$/nucleon$)^2$}
    %
  \label{tbl:sysnucleonnuc26Wc}
\end{table}
\pagebreak
\begin{turnpage}
\end{turnpage}
\begin{table}
  \centering
  \renewcommand{\arraystretch}{1.15}
  \caption{Measured cross section as function of $T_{\pi}$ on iron, in units of $10^{-42}$ $\text{cm}^2$/GeV/nucleon, and the absolute and fractional cross section uncertainties}
    %
  \label{tbl:datanucleonnuc26pion_ekina}
\end{table}
\begin{table}
  \centering
  \renewcommand{\arraystretch}{1.15}
  \caption{Statistical covariance matrix of measured cross section as function of $T_{\pi}$ on iron, in units of $10^{-84}$ $(\text{cm}^2$/GeV/nucleon$)^2$}
    %
  \label{tbl:statnucleonnuc26pion_ekinb}
\end{table}
\begin{table}
  \centering
  \renewcommand{\arraystretch}{1.15}
  \caption{Systematic covariance matrix of measured cross section in $T_{\pi}$ on iron, in units of $10^{-84}$ $(\text{cm}^2$/GeV/nucleon$)^2$}
    %
  \label{tbl:sysnucleonnuc26pion_ekinc}
\end{table}
\pagebreak
\begin{turnpage}
\end{turnpage}
\begin{table}
  \centering
  \renewcommand{\arraystretch}{1.15}
  \caption{Measured cross section as function of $\theta_{\pi}$ on iron, in units of $10^{-42}$ $\text{cm}^2$/degree/nucleon, and the absolute and fractional cross section uncertainties}
    %
  \label{tbl:datanucleonnuc26pion_thetaa}
\end{table}
\pagebreak
\begin{turnpage}
\begin{table}
  \centering
  \renewcommand{\arraystretch}{1.15}
  \caption{Statistical covariance matrix of measured cross section as function of $\theta_{\pi}$ on iron, in units of $10^{-84}$ $(\text{cm}^2$/degree/nucleon$)^2$}
  \resizebox{\textheight}{!}{
    %
  }
  \label{tbl:statnucleonnuc26pion_thetab}
\end{table}
\end{turnpage}
\clearpage
\pagebreak
\begin{turnpage}
\begin{table}
  \centering
  \renewcommand{\arraystretch}{1.15}
  \caption{Systematic covariance matrix of measured cross section in $\theta_{\pi}$ on iron, in units of $10^{-84}$ $(\text{cm}^2$/degree/nucleon$)^2$}
  \resizebox{\textheight}{!}{
    %
  }
  \label{tbl:sysnucleonnuc26pion_thetac}
\end{table}
\end{turnpage}
\pagebreak
\begin{turnpage}
\end{turnpage}
\begin{table}
  \centering
  \renewcommand{\arraystretch}{1.15}
  \caption{Measured cross section as function of $p_{\mu}$ on lead, in units of $10^{-42}$ $\text{cm}^2$/GeV/c/nucleon, and the absolute and fractional cross section uncertainties}
    %
  \label{tbl:datanucleonnuc82muon_pa}
\end{table}
\begin{table}
  \centering
  \renewcommand{\arraystretch}{1.15}
  \caption{Statistical covariance matrix of measured cross section as function of $p_{\mu}$ on lead, in units of $10^{-84}$ $(\text{cm}^2$/GeV/c/nucleon$)^2$}
  \resizebox{\textwidth}{!}{
    %
  }
  \label{tbl:statnucleonnuc82muon_pb}
\end{table}
\begin{table}
  \centering
  \renewcommand{\arraystretch}{1.15}
  \caption{Systematic covariance matrix of measured cross section in $p_{\mu}$ on lead, in units of $10^{-84}$ $(\text{cm}^2$/GeV/c/nucleon$)^2$}
  \resizebox{\textwidth}{!}{
    %
  }
  \label{tbl:sysnucleonnuc82muon_pc}
\end{table}
\pagebreak
\begin{turnpage}
\end{turnpage}
\begin{table}
  \centering
  \renewcommand{\arraystretch}{1.15}
  \caption{Measured cross section as function of $\theta_{\mu}$ on lead, in units of $10^{-42}$ $\text{cm}^2$/degree/nucleon, and the absolute and fractional cross section uncertainties}
    %
  \label{tbl:datanucleonnuc82muon_thetaa}
\end{table}
\begin{table}
  \centering
  \renewcommand{\arraystretch}{1.15}
  \caption{Statistical covariance matrix of measured cross section as function of $\theta_{\mu}$ on lead, in units of $10^{-84}$ $(\text{cm}^2$/degree/nucleon$)^2$}
  \resizebox{\textwidth}{!}{
    %
  }
  \label{tbl:statnucleonnuc82muon_thetab}
\end{table}
\begin{table}
  \centering
  \renewcommand{\arraystretch}{1.15}
  \caption{Systematic covariance matrix of measured cross section in $\theta_{\mu}$ on lead, in units of $10^{-84}$ $(\text{cm}^2$/degree/nucleon$)^2$}
  \resizebox{\textwidth}{!}{
    %
  }
  \label{tbl:sysnucleonnuc82muon_thetac}
\end{table}
\pagebreak
\begin{turnpage}
\end{turnpage}
\begin{table}
  \centering
  \renewcommand{\arraystretch}{1.15}
  \caption{Measured cross section as function of $p_{\mu,||}$ on lead, in units of $10^{-42}$ $\text{cm}^2$/GeV/c/nucleon, and the absolute and fractional cross section uncertainties}
    %
  \label{tbl:datanucleonnuc82muon_pza}
\end{table}
\begin{table}
  \centering
  \renewcommand{\arraystretch}{1.15}
  \caption{Statistical covariance matrix of measured cross section as function of $p_{\mu,||}$ on lead, in units of $10^{-84}$ $(\text{cm}^2$/GeV/c/nucleon$)^2$}
  \resizebox{\textwidth}{!}{
    %
  }
  \label{tbl:statnucleonnuc82muon_pzb}
\end{table}
\begin{table}
  \centering
  \renewcommand{\arraystretch}{1.15}
  \caption{Systematic covariance matrix of measured cross section in $p_{\mu,||}$ on lead, in units of $10^{-84}$ $(\text{cm}^2$/GeV/c/nucleon$)^2$}
  \resizebox{\textwidth}{!}{
    %
  }
  \label{tbl:sysnucleonnuc82muon_pzc}
\end{table}
\pagebreak
\begin{turnpage}
\end{turnpage}
\begin{table}
  \centering
  \renewcommand{\arraystretch}{1.15}
  \caption{Measured cross section as function of $p_{\mu,T}$ on lead, in units of $10^{-42}$ $\text{cm}^2$/GeV/c/nucleon, and the absolute and fractional cross section uncertainties}
    %
  \label{tbl:datanucleonnuc82muon_pta}
\end{table}
\pagebreak
\begin{turnpage}
\begin{table}
  \centering
  \renewcommand{\arraystretch}{1.15}
  \caption{Statistical covariance matrix of measured cross section as function of $p_{\mu,T}$ on lead, in units of $10^{-84}$ $(\text{cm}^2$/GeV/c/nucleon$)^2$}
  \resizebox{\textheight}{!}{
    %
  }
  \label{tbl:statnucleonnuc82muon_ptb}
\end{table}
\end{turnpage}
\clearpage
\pagebreak
\begin{turnpage}
\begin{table}
  \centering
  \renewcommand{\arraystretch}{1.15}
  \caption{Systematic covariance matrix of measured cross section in $p_{\mu,T}$ on lead, in units of $10^{-84}$ $(\text{cm}^2$/GeV/c/nucleon$)^2$}
  \resizebox{\textheight}{!}{
    %
  }
  \label{tbl:sysnucleonnuc82muon_ptc}
\end{table}
\end{turnpage}
\pagebreak
\begin{turnpage}
\end{turnpage}
\begin{table}
  \centering
  \renewcommand{\arraystretch}{1.15}
  \caption{Measured cross section as function of $Q^2$ on lead, in units of $10^{-42}$ $\text{cm}^2$/GeV${}^2/c{}^2$/nucleon, and the absolute and fractional cross section uncertainties}
    %
  \label{tbl:datanucleonnuc82q2a}
\end{table}
\pagebreak
\begin{turnpage}
\begin{table}
  \centering
  \renewcommand{\arraystretch}{1.15}
  \caption{Statistical covariance matrix of measured cross section as function of $Q^2$ on lead, in units of $10^{-84}$ $(\text{cm}^2$/GeV${}^2/c{}^2$/nucleon$)^2$}
  \resizebox{\textheight}{!}{
    %
  }
  \label{tbl:statnucleonnuc82q2b}
\end{table}
\end{turnpage}
\clearpage
\pagebreak
\begin{turnpage}
\begin{table}
  \centering
  \renewcommand{\arraystretch}{1.15}
  \caption{Systematic covariance matrix of measured cross section in $Q^2$ on lead, in units of $10^{-84}$ $(\text{cm}^2$/GeV${}^2/c{}^2$/nucleon$)^2$}
  \resizebox{\textheight}{!}{
    %
  }
  \label{tbl:sysnucleonnuc82q2c}
\end{table}
\end{turnpage}
\pagebreak
\begin{turnpage}
\end{turnpage}
\begin{table}
  \centering
  \renewcommand{\arraystretch}{1.15}
  \caption{Measured cross section as function of $W_{exp}$ on lead, in units of $10^{-42}$ $\text{cm}^2$/GeV/c${}^2$/nucleon, and the absolute and fractional cross section uncertainties}
    %
  \label{tbl:datanucleonnuc82Wa}
\end{table}
\begin{table}
  \centering
  \renewcommand{\arraystretch}{1.15}
  \caption{Statistical covariance matrix of measured cross section as function of $W_{exp}$ on lead, in units of $10^{-84}$ $(\text{cm}^2$/GeV/c${}^2$/nucleon$)^2$}
    %
  \label{tbl:statnucleonnuc82Wb}
\end{table}
\begin{table}
  \centering
  \renewcommand{\arraystretch}{1.15}
  \caption{Systematic covariance matrix of measured cross section in $W_{exp}$ on lead, in units of $10^{-84}$ $(\text{cm}^2$/GeV/c${}^2$/nucleon$)^2$}
    %
  \label{tbl:sysnucleonnuc82Wc}
\end{table}
\pagebreak
\begin{turnpage}
\end{turnpage}
\begin{table}
  \centering
  \renewcommand{\arraystretch}{1.15}
  \caption{Measured cross section as function of $T_{\pi}$ on lead, in units of $10^{-42}$ $\text{cm}^2$/GeV/nucleon, and the absolute and fractional cross section uncertainties}
    %
  \label{tbl:datanucleonnuc82pion_ekina}
\end{table}
\begin{table}
  \centering
  \renewcommand{\arraystretch}{1.15}
  \caption{Statistical covariance matrix of measured cross section as function of $T_{\pi}$ on lead, in units of $10^{-84}$ $(\text{cm}^2$/GeV/nucleon$)^2$}
    %
  \label{tbl:statnucleonnuc82pion_ekinb}
\end{table}
\begin{table}
  \centering
  \renewcommand{\arraystretch}{1.15}
  \caption{Systematic covariance matrix of measured cross section in $T_{\pi}$ on lead, in units of $10^{-84}$ $(\text{cm}^2$/GeV/nucleon$)^2$}
    %
  \label{tbl:sysnucleonnuc82pion_ekinc}
\end{table}
\pagebreak
\begin{turnpage}
\end{turnpage}
\begin{table}
  \centering
  \renewcommand{\arraystretch}{1.15}
  \caption{Measured cross section as function of $\theta_{\pi}$ on lead, in units of $10^{-42}$ $\text{cm}^2$/degree/nucleon, and the absolute and fractional cross section uncertainties}
    %
  \label{tbl:datanucleonnuc82pion_thetaa}
\end{table}
\pagebreak
\begin{turnpage}
\begin{table}
  \centering
  \renewcommand{\arraystretch}{1.15}
  \caption{Statistical covariance matrix of measured cross section as function of $\theta_{\pi}$ on lead, in units of $10^{-84}$ $(\text{cm}^2$/degree/nucleon$)^2$}
  \resizebox{\textheight}{!}{
    %
  }
  \label{tbl:statnucleonnuc82pion_thetab}
\end{table}
\end{turnpage}
\clearpage
\pagebreak
\begin{turnpage}
\begin{table}
  \centering
  \renewcommand{\arraystretch}{1.15}
  \caption{Systematic covariance matrix of measured cross section in $\theta_{\pi}$ on lead, in units of $10^{-84}$ $(\text{cm}^2$/degree/nucleon$)^2$}
  \resizebox{\textheight}{!}{
    %
  }
  \label{tbl:sysnucleonnuc82pion_thetac}
\end{table}
\end{turnpage}
\pagebreak
\begin{turnpage}
\end{turnpage}

%% file: includeTex/PRL_Table_Var_Ratio.tex
\setlength\tabcolsep{0.5em}
\begin{table}
  \centering
  \renewcommand{\arraystretch}{1.15}
  \caption{Measured cross section ratio of carbon over scintillator in $p_{\mu}$, and the absolute and fractional cross section ratio uncertainties}
    %
  \label{tbl:rationucleonnuc6muon_pa}
\end{table}
\begin{table}
  \centering
  \renewcommand{\arraystretch}{1.15}
  \caption{Statistical covariance matrix of measured cross section ratio of carbon over scintillator in $p_{\mu}$}
  \resizebox{\textwidth}{!}{
    %
  }
  \label{tbl:rationucleonnuc6muon_pb}
\end{table}
\begin{table}
  \centering
  \renewcommand{\arraystretch}{1.15}
  \caption{Systematic covariance matrix of measured cross section ratio of carbon over scintillator in $p_{\mu}$}
  \resizebox{\textwidth}{!}{
    %
  }
  \label{tbl:rationucleonnuc6muon_pc}
\end{table}
\begin{table}
  \centering
  \renewcommand{\arraystretch}{1.15}
  \caption{Measured cross section ratio of carbon over scintillator in $\theta_{\mu}$, and the absolute and fractional cross section ratio uncertainties}
    %
  \label{tbl:rationucleonnuc6muon_thetaa}
\end{table}
\begin{table}
  \centering
  \renewcommand{\arraystretch}{1.15}
  \caption{Statistical covariance matrix of measured cross section ratio of carbon over scintillator in $\theta_{\mu}$}
  \resizebox{\textwidth}{!}{
    %
  }
  \label{tbl:rationucleonnuc6muon_thetab}
\end{table}
\begin{table}
  \centering
  \renewcommand{\arraystretch}{1.15}
  \caption{Systematic covariance matrix of measured cross section ratio of carbon over scintillator in $\theta_{\mu}$}
  \resizebox{\textwidth}{!}{
    %
  }
  \label{tbl:rationucleonnuc6muon_thetac}
\end{table}
\begin{table}
  \centering
  \renewcommand{\arraystretch}{1.15}
  \caption{Measured cross section ratio of carbon over scintillator in $p_{\mu,||}$, and the absolute and fractional cross section ratio uncertainties}
    %
  \label{tbl:rationucleonnuc6muon_pza}
\end{table}
\begin{table}
  \centering
  \renewcommand{\arraystretch}{1.15}
  \caption{Statistical covariance matrix of measured cross section ratio of carbon over scintillator in $p_{\mu,||}$}
  \resizebox{\textwidth}{!}{
    %
  }
  \label{tbl:rationucleonnuc6muon_pzb}
\end{table}
\begin{table}
  \centering
  \renewcommand{\arraystretch}{1.15}
  \caption{Systematic covariance matrix of measured cross section ratio of carbon over scintillator in $p_{\mu,||}$}
  \resizebox{\textwidth}{!}{
    %
  }
  \label{tbl:rationucleonnuc6muon_pzc}
\end{table}
\begin{table}
  \centering
  \renewcommand{\arraystretch}{1.15}
  \caption{Measured cross section ratio of carbon over scintillator in $p_{\mu,T}$, and the absolute and fractional cross section ratio uncertainties}
    %
  \label{tbl:rationucleonnuc6muon_pta}
\end{table}
\pagebreak
\begin{turnpage}
\begin{table}
  \centering
  \renewcommand{\arraystretch}{1.15}
  \caption{Statistical covariance matrix of measured cross section ratio of carbon over scintillator in $p_{\mu,T}$}
  \resizebox{\textheight}{!}{
    %
  }
  \label{tbl:rationucleonnuc6muon_ptc}
\end{table}
\end{turnpage}
\begin{table}
  \centering
  \renewcommand{\arraystretch}{1.15}
  \caption{Measured cross section ratio of carbon over scintillator in $Q^2$, and the absolute and fractional cross section ratio uncertainties}
    %
  \label{tbl:rationucleonnuc6q2a}
\end{table}
\pagebreak
\begin{turnpage}
\begin{table}
  \centering
  \renewcommand{\arraystretch}{1.15}
  \caption{Statistical covariance matrix of measured cross section ratio of carbon over scintillator in $Q^2$}
  \resizebox{\textheight}{!}{
    %
  }
  \label{tbl:rationucleonnuc6q2c}
\end{table}
\end{turnpage}
\begin{table}
  \centering
  \renewcommand{\arraystretch}{1.15}
  \caption{Measured cross section ratio of carbon over scintillator in $W_{exp}$, and the absolute and fractional cross section ratio uncertainties}
    %
  \label{tbl:rationucleonnuc6Wa}
\end{table}
\begin{table}
  \centering
  \renewcommand{\arraystretch}{1.15}
  \caption{Statistical covariance matrix of measured cross section ratio of carbon over scintillator in $W_{exp}$}
    %
  \label{tbl:rationucleonnuc6pion_thetaa}
\end{table}
\pagebreak
\begin{turnpage}
\begin{table}
  \centering
  \renewcommand{\arraystretch}{1.15}
  \caption{Statistical covariance matrix of measured cross section ratio of carbon over scintillator in $\theta_{\pi}$}
  \resizebox{\textheight}{!}{
    %
  }
  \label{tbl:rationucleonnuc6pion_thetac}
\end{table}
\end{turnpage}
\begin{table}
  \centering
  \renewcommand{\arraystretch}{1.15}
  \caption{Measured cross section ratio of water over scintillator in $p_{\mu}$, and the absolute and fractional cross section ratio uncertainties}
    %
  \label{tbl:rationucleonnuc10muon_pa}
\end{table}
\begin{table}
  \centering
  \renewcommand{\arraystretch}{1.15}
  \caption{Statistical covariance matrix of measured cross section ratio of water over scintillator in $p_{\mu}$}
  \resizebox{\textwidth}{!}{
    %
  }
  \label{tbl:rationucleonnuc10muon_pc}
\end{table}
\begin{table}
  \centering
  \renewcommand{\arraystretch}{1.15}
  \caption{Measured cross section ratio of water over scintillator in $\theta_{\mu}$, and the absolute and fractional cross section ratio uncertainties}
    %
  \label{tbl:rationucleonnuc10muon_thetaa}
\end{table}
\begin{table}
  \centering
  \renewcommand{\arraystretch}{1.15}
  \caption{Statistical covariance matrix of measured cross section ratio of water over scintillator in $\theta_{\mu}$}
  \resizebox{\textwidth}{!}{
    %
  }
  \label{tbl:rationucleonnuc10muon_thetac}
\end{table}
\begin{table}
  \centering
  \renewcommand{\arraystretch}{1.15}
  \caption{Measured cross section ratio of water over scintillator in $p_{\mu,||}$, and the absolute and fractional cross section ratio uncertainties}
    %
  \label{tbl:rationucleonnuc10muon_pza}
\end{table}
\begin{table}
  \centering
  \renewcommand{\arraystretch}{1.15}
  \caption{Statistical covariance matrix of measured cross section ratio of water over scintillator in $p_{\mu,||}$}
  \resizebox{\textwidth}{!}{
    %
  }
  \label{tbl:rationucleonnuc10muon_pzc}
\end{table}
\begin{table}
  \centering
  \renewcommand{\arraystretch}{1.15}
  \caption{Measured cross section ratio of water over scintillator in $p_{\mu,T}$, and the absolute and fractional cross section ratio uncertainties}
    %
  \label{tbl:rationucleonnuc10muon_pta}
\end{table}
\pagebreak
\begin{turnpage}
\begin{table}
  \centering
  \renewcommand{\arraystretch}{1.15}
  \caption{Statistical covariance matrix of measured cross section ratio of water over scintillator in $p_{\mu,T}$}
  \resizebox{\textheight}{!}{
    %
  }
  \label{tbl:rationucleonnuc10muon_ptc}
\end{table}
\end{turnpage}
\begin{table}
  \centering
  \renewcommand{\arraystretch}{1.15}
  \caption{Measured cross section ratio of water over scintillator in $Q^2$, and the absolute and fractional cross section ratio uncertainties}
    %
  \label{tbl:rationucleonnuc10q2a}
\end{table}
\pagebreak
\begin{turnpage}
\begin{table}
  \centering
  \renewcommand{\arraystretch}{1.15}
  \caption{Statistical covariance matrix of measured cross section ratio of water over scintillator in $Q^2$}
  \resizebox{\textheight}{!}{
    %
  }
  \label{tbl:rationucleonnuc10q2c}
\end{table}
\end{turnpage}
\begin{table}
  \centering
  \renewcommand{\arraystretch}{1.15}
  \caption{Measured cross section ratio of water over scintillator in $W_{exp}$, and the absolute and fractional cross section ratio uncertainties}
    %
  \label{tbl:rationucleonnuc10pion_thetaa}
\end{table}
\pagebreak
\begin{turnpage}
\begin{table}
  \centering
  \renewcommand{\arraystretch}{1.15}
  \caption{Statistical covariance matrix of measured cross section ratio of water over scintillator in $\theta_{\pi}$}
  \resizebox{\textheight}{!}{
    %
  }
  \label{tbl:rationucleonnuc10pion_thetac}
\end{table}
\end{turnpage}
\begin{table}
  \centering
  \renewcommand{\arraystretch}{1.15}
  \caption{Measured cross section ratio of iron over scintillator in $p_{\mu}$, and the absolute and fractional cross section ratio uncertainties}
    %
  \label{tbl:rationucleonnuc26muon_pa}
\end{table}
\begin{table}
  \centering
  \renewcommand{\arraystretch}{1.15}
  \caption{Statistical covariance matrix of measured cross section ratio of iron over scintillator in $p_{\mu}$}
  \resizebox{\textwidth}{!}{
    %
  }
  \label{tbl:rationucleonnuc26muon_pc}
\end{table}
\begin{table}
  \centering
  \renewcommand{\arraystretch}{1.15}
  \caption{Measured cross section ratio of iron over scintillator in $\theta_{\mu}$, and the absolute and fractional cross section ratio uncertainties}
    %
  \label{tbl:rationucleonnuc26muon_thetaa}
\end{table}
\begin{table}
  \centering
  \renewcommand{\arraystretch}{1.15}
  \caption{Statistical covariance matrix of measured cross section ratio of iron over scintillator in $\theta_{\mu}$}
  \resizebox{\textwidth}{!}{
    %
  }
  \label{tbl:rationucleonnuc26muon_thetac}
\end{table}
\begin{table}
  \centering
  \renewcommand{\arraystretch}{1.15}
  \caption{Measured cross section ratio of iron over scintillator in $p_{\mu,||}$, and the absolute and fractional cross section ratio uncertainties}
    %
  \label{tbl:rationucleonnuc26muon_pza}
\end{table}
\begin{table}
  \centering
  \renewcommand{\arraystretch}{1.15}
  \caption{Statistical covariance matrix of measured cross section ratio of iron over scintillator in $p_{\mu,||}$}
  \resizebox{\textwidth}{!}{
    %
  }
  \label{tbl:rationucleonnuc26muon_pzc}
\end{table}
\begin{table}
  \centering
  \renewcommand{\arraystretch}{1.15}
  \caption{Measured cross section ratio of iron over scintillator in $p_{\mu,T}$, and the absolute and fractional cross section ratio uncertainties}
    %
  \label{tbl:rationucleonnuc26muon_pta}
\end{table}
\pagebreak
\begin{turnpage}
\begin{table}
  \centering
  \renewcommand{\arraystretch}{1.15}
  \caption{Statistical covariance matrix of measured cross section ratio of iron over scintillator in $p_{\mu,T}$}
  \resizebox{\textheight}{!}{
    %
  }
  \label{tbl:rationucleonnuc26muon_ptc}
\end{table}
\end{turnpage}
\begin{table}
  \centering
  \renewcommand{\arraystretch}{1.15}
  \caption{Measured cross section ratio of iron over scintillator in $Q^2$, and the absolute and fractional cross section ratio uncertainties}
    %
  \label{tbl:rationucleonnuc26q2a}
\end{table}
\pagebreak
\begin{turnpage}
\begin{table}
  \centering
  \renewcommand{\arraystretch}{1.15}
  \caption{Statistical covariance matrix of measured cross section ratio of iron over scintillator in $Q^2$}
  \resizebox{\textheight}{!}{
    %
  }
  \label{tbl:rationucleonnuc26q2c}
\end{table}
\end{turnpage}
\begin{table}
  \centering
  \renewcommand{\arraystretch}{1.15}
  \caption{Measured cross section ratio of iron over scintillator in $W_{exp}$, and the absolute and fractional cross section ratio uncertainties}
    %
  \label{tbl:rationucleonnuc26pion_thetaa}
\end{table}
\pagebreak
\begin{turnpage}
\begin{table}
  \centering
  \renewcommand{\arraystretch}{1.15}
  \caption{Statistical covariance matrix of measured cross section ratio of iron over scintillator in $\theta_{\pi}$}
  \resizebox{\textheight}{!}{
    %
  }
  \label{tbl:rationucleonnuc26pion_thetac}
\end{table}
\end{turnpage}
\begin{table}
  \centering
  \renewcommand{\arraystretch}{1.15}
  \caption{Measured cross section ratio of lead over scintillator in $p_{\mu}$, and the absolute and fractional cross section ratio uncertainties}
    %
  \label{tbl:rationucleonnuc82muon_pa}
\end{table}
\begin{table}
  \centering
  \renewcommand{\arraystretch}{1.15}
  \caption{Statistical covariance matrix of measured cross section ratio of lead over scintillator in $p_{\mu}$}
  \resizebox{\textwidth}{!}{
    %
  }
  \label{tbl:rationucleonnuc82muon_pc}
\end{table}
\begin{table}
  \centering
  \renewcommand{\arraystretch}{1.15}
  \caption{Measured cross section ratio of lead over scintillator in $\theta_{\mu}$, and the absolute and fractional cross section ratio uncertainties}
    %
  \label{tbl:rationucleonnuc82muon_thetaa}
\end{table}
\begin{table}
  \centering
  \renewcommand{\arraystretch}{1.15}
  \caption{Statistical covariance matrix of measured cross section ratio of lead over scintillator in $\theta_{\mu}$}
  \resizebox{\textwidth}{!}{
    %
  }
  \label{tbl:rationucleonnuc82muon_thetac}
\end{table}
\begin{table}
  \centering
  \renewcommand{\arraystretch}{1.15}
  \caption{Measured cross section ratio of lead over scintillator in $p_{\mu,||}$, and the absolute and fractional cross section ratio uncertainties}
    %
  \label{tbl:rationucleonnuc82muon_pza}
\end{table}
\begin{table}
  \centering
  \renewcommand{\arraystretch}{1.15}
  \caption{Statistical covariance matrix of measured cross section ratio of lead over scintillator in $p_{\mu,||}$}
  \resizebox{\textwidth}{!}{
    %
  }
  \label{tbl:rationucleonnuc82muon_pzc}
\end{table}
\begin{table}
  \centering
  \renewcommand{\arraystretch}{1.15}
  \caption{Measured cross section ratio of lead over scintillator in $p_{\mu,T}$, and the absolute and fractional cross section ratio uncertainties}
    %
  \label{tbl:rationucleonnuc82muon_pta}
\end{table}
\pagebreak
\begin{turnpage}
\begin{table}
  \centering
  \renewcommand{\arraystretch}{1.15}
  \caption{Statistical covariance matrix of measured cross section ratio of lead over scintillator in $p_{\mu,T}$}
  \resizebox{\textheight}{!}{
    %
  }
  \label{tbl:rationucleonnuc82muon_ptc}
\end{table}
\end{turnpage}
\begin{table}
  \centering
  \renewcommand{\arraystretch}{1.15}
  \caption{Measured cross section ratio of lead over scintillator in $Q^2$, and the absolute and fractional cross section ratio uncertainties}
    %
  \label{tbl:rationucleonnuc82q2a}
\end{table}
\pagebreak
\begin{turnpage}
\begin{table}
  \centering
  \renewcommand{\arraystretch}{1.15}
  \caption{Statistical covariance matrix of measured cross section ratio of lead over scintillator in $Q^2$}
  \resizebox{\textheight}{!}{
    %
  }
  \label{tbl:rationucleonnuc82q2c}
\end{table}
\end{turnpage}
\begin{table}
  \centering
  \renewcommand{\arraystretch}{1.15}
  \caption{Measured cross section ratio of lead over scintillator in $W_{exp}$, and the absolute and fractional cross section ratio uncertainties}
    %
  \label{tbl:rationucleonnuc82pion_thetaa}
\end{table}
\pagebreak
\begin{turnpage}
\begin{table}
  \centering
  \renewcommand{\arraystretch}{1.15}
  \caption{Statistical covariance matrix of measured cross section ratio of lead over scintillator in $\theta_{\pi}$}
  \resizebox{\textheight}{!}{
    %
  }
  \label{tbl:rationucleonnuc82pion_thetac}
\end{table}
\end{turnpage}